\documentclass[hyperref,amsmath,amssymb,showpacs,floatfix,journal=ancac3,manuscript=article]{achemso}
\usepackage[version=4]{mhchem} % Formula subscripts using \ce{}
\usepackage{graphicx}% Include figure files
\usepackage{bm}% bold math
\usepackage{color}
\usepackage[utf8]{inputenc}

\usepackage{booktabs, caption, makecell}

\usepackage{threeparttable}

\newcommand{\ib}[1]{{\color{black}#1}}

\newcommand{\si}{SI}
\newcommand{\secname}{Section~}
\newcommand{\secsname}{Sections~}

\newcommand{\rad}[1]{\raisebox{-0.0ex}{$\stackrel{\mbox{\tiny{$\bullet $}}}{\mbox{#1}}$}}
\newcommand{\dirad}[1]{\raisebox{-0.0ex}{$\stackrel{\mbox{\tiny{$\bullet \bullet $}}}{\mbox{#1}}$}}

\newcommand{\rdirad}[1]{\raisebox{-0.0ex}{#1$\stackrel{ \mbox{\tiny{$\bullet $}} }{ \mbox{\tiny{$\bullet $}} }$}}

\DeclareGraphicsExtensions{.jpg, .eps, .ps, .png, .pdf}
\graphicspath{{ }{/home/ioan/floppy/astro/data/.}}
\usepackage{mhchem}
\usepackage{morefloats}
\SectionNumbersOn
\author{Ioan B\^aldea}
\email{ioan.baldea@pci.uni-heidelberg.de}
\affiliation{Theoretische Chemie, Universit\"at Heidelberg, Im Neuenheimer Feld 229, D-69120 Heidelberg, Germany}
\alsoaffiliation{Institute of Space Sciences, National Institute of Lasers, Plasma and Radiation Physics,
RO 077125, Bucharest-M\u{a}gurele, Romania}

\title[Alternation of Singlet and Triplet States
  in Carbon-Based Chain Molecules]{Alternation of Singlet and Triplet States
  in Carbon-Based Chain Molecules and Its Astrochemical Implications.
  Results of an Extensive Theoretical Study}

\newcommand{\figname}{Figure~}
\newcommand{\figsname}{Figures~}
\begin{document}
\begin{abstract}
A variety of homologous carbon chains (HCnH, HCnN, CnS, CnO, and OCnO) are found to exhibit an appealing even-odd effect. Chains containing a number of carbon atoms of a certain parity possess singlet ground states, while members of opposite parity have triplet ground states. From a general perspective, it is important that this even–odd effect confounds straightforward chemical intuition. Whether the most stable form is a triplet or a singlet is neither simply related to the fact that the species in question is a ``normal'' (closed-shell, nonradical) molecule nor a (di)radical or to the (e.g., cumulene-type) C-C bond succession across the chain. From a computational perspective, the present results are important also because they demonstrate that electron correlations in carbon-based chains are extremely strong. Whether the ``gold-standard'' CCSD(T) (coupled-cluster expansions with single and double excitations and triple excitations corrections) framework suffices to describe such strongly correlated systems remains an open question that calls for further clarification. Most importantly for astrochemistry, the present results may explain why certain members are not astronomically observed although larger members of the same homologous series are detected; the missing species are exactly those for which the present calculations predict triplet ground states.
\end{abstract}
\section{Introduction}
\label{sec:intro}
Linear carbon chains represent a field of great current interest for fields
ranging from (bio-)molecular electronics to astrochemistry.
Until the advent of nanoelectronics \cite{Datta:05,CuevasScheer:17,Baldea:2015e,Fazzi:16}
such molecules were often of little interest for (terrestrial) laboratory studies and practical applications.
Cyclic or combined ring-chain structures are usually more stable energetically
than linear isomers.\cite{Goulay:09} This explains the scarcity of available information on carbon-based chains.
For the vast majority of the
molecular species to be considered in the present paper the NIST database contains no entries.
A number of molecules comprising carbon chains were observed in the past decades
in cold interstellar and circumstellar clouds.
\cite{Smith:71,Snyder:71,Turner:71,Avery:76,Souza:77,Kroto:78,Broten:78,Matthews:84,Hinkle:88,Bernath:89,Ohishi:91,Guelin:91,Bell:97,Cernicharo:00,Cernicharo:04,Graupner:08}

The identification of the type of ground state (which is the most stable isomer?)
of the various molecular species of interest
represents the most basic information, also needed to correctly
understand a certain extraterrestrial environment and to simulate its astrochemical evolution.
In the present paper, we addressed this issue by examining in detail representative homologous
series of carbon-based linear chains (\ce{HC_nH}, \ce{HC_nN}, \ce{C_n S}, \ce{C_nO}, and \ce{OC_nO}).
We found that, by successively adding atoms to the molecular backbone of all
these astrochemically relevant families of carbon-based chains with an even number of electrons,
the most stable
form systematically switches back and forth between singlet and triplet isomers.
Rephrasing, we will show below that the most stable state of
members of a certain parity (even or odd) of a chain family is a singlet (ground) state,
while members of opposite parity (odd or even, respectively) have a triplet ground state.
One should note at this point that the present finding that triplet states of carbon-based chains
can be surprisingly lower in energy and lie below singlet states 
contradicts some recent studies done in the astronomical/astrophysical community.\cite{Etim:16a,Etim16aImplicitACS}
Those studies 
explicitly \cite{Etim:16b} or implicitly \cite{Etim:16a} claimed (see
{\secname}\ref{sec:etim}
and the {\si}) % \cite{Etim:16a,Etim16aImplicitACS}
that all carbon chains of the type considered in the present paper
possess a singlet ground state.

Noteworthily,
the even-odd singlet-triplet alternation \cite{Fan:89}
extensively discussed in this paper is qualitatively different from all other even-odd effects (\emph{e.g.},
in multilayers,\cite{Wu:16} self-assembled monolayers,\cite{Tao:07}
quantum dot arrays,\cite{Baldea:99a,Baldea:99b,Baldea:2001a,Baldea:2002}
molecular electronic devices \cite{CuevasScheer:17,Baldea:2015e})
known since the early days of quantum mechanics.
\cite{Hueckel:31a,Hueckel:31b,Hueckel:32,London:37}
In those cases, it is merely a certain property rather than the
very nature of the ground state that exhibits alternation.

\section{Methods}
\label{sec:methods}
All quantum chemical calculations done in conjunction with the present study
were performed by running the GAUSSIAN 16 suite of programs \cite{g16} on the bwHPC platform.\cite{bwHPC}
Optimized geometries ($\mathbf{R}_{S,T}$) of all singlet (S) and triplet (T)
carbon-based chains considered in this paper
were obtained from DFT calculations using the B3LYP hybrid exchange functional and, 
unless otherwise stated
(\emph{cf.}~Tables~S1--S4 of the {\si}) the largest Pople 6-311++g(3df, 3pd) basis sets.
In all cases, we checked that all frequencies were real.

For the largest molecular species of each homologous series, we also performed
state-of-the-art calculations based on 
coupled-cluster (CC) expansions with single and double excitations (CCSD) supplemented
by perturbative treatment of triple excitations (CCSD(T)).\cite{Bartlett:78}
For triplet states, we employed the unrestricted and restricted open shell formalism for DFT calculations and 
CC-calculations, respectively. More technical details are presented in {\secname}S1 of the {\si}.

Enthalpies of formations $\Delta_{f} H^{0}$ 
(Tables~\ref{table:H-hcxh}, \ref{table:H-hcxn},
\ref{table:H-cxs}, \ref{table:H-cxo}, and \ref{table:H-ocxo})
were computed using the standard methodology.\cite{Ochterski:00}
For comparison purposes
(\emph{cf.}~{\secname}\ref{sec:etim}),
along with the values obtained within
a DFT/B3LYP/6-311++g(3df, 3pd) approach
we also estimated enthalpies of formation using the 
CBS-QB3 protocol as implemented in GAUSSIAN 16, which are shown in 
Tables~S1--S4 of the {\si}.

A thermochemical analysis may not be sufficient for molecules of interest for astrochemistry,
where single-particle (kinetic) effects also deserve consideration. Therefore, in addition
to enthalpies of formations for both the lowest singlet and triplet electronic ground states,
we also report values estimates for the singlet-triplet separation
energies $\Delta \equiv \mathcal{E}_{T} - \mathcal{E}_{S}$. They 
were obtained as differences of the corresponding total molecular energies
$\mathcal{E}_{S,T}\left(\mathbf{R}\right)$ at the pertaining molecular energies
($\mathbf{R}=\mathbf{R}_{S,T}$). For all the families of carbon chains considered
(Tables~\ref{table:Delta-hcxh}, \ref{table:Delta-hcxn},
\ref{table:Delta-cxs}, \ref{table:Delta-cxo}, and \ref{table:Delta-ocxo}),
we present values of both adiabatic and vertical singlet-triplet separation energies.
The adiabatic value
\begin{equation}
  \label{eq-Delta-adiab}
  \Delta_{adiab} \equiv \mathcal{E}_{T}\left(\mathbf{R}_{T}\right) - \mathcal{E}_{S}\left(\mathbf{R}_{S}\right)
\end{equation}
represents the difference between
    {the triplet (T) and singlet (S)}
energies computed at the 
molecular geometries $\mathbf{R}=\mathbf{R}_{S,T}$ optimized for the singlet
($\mathbf{R}_{S}$) and the triplet ($\mathbf{R}_{T}$) isomers. The two distinct
vertical values $\Delta_{S,T}$
\begin{subequations}
  \label{eq-Delta-S,T}
  \begin{equation}
    \label{eq-Delta-S}
    \Delta_{S} \equiv \mathcal{E}_{T}\left(\mathbf{R}_{S}\right) - \mathcal{E}_{S}\left(\mathbf{R}_{S}\right)
  \end{equation}
  \begin{equation}
    \label{eq-Delta-T}
    \Delta_{T} \equiv \mathcal{E}_{T}\left(\mathbf{R}_{T}\right) - \mathcal{E}_{S}\left(\mathbf{R}_{T}\right)
  \end{equation}
\end{subequations}
correspond to differences between
    {the triplet (T) and singlet (S)}
energies taken at the same molecular
geometry (either singlet $\mathbf{R}_{S}$ or triplet $\mathbf{R}_{T}$). 

       {According to Equation~(\ref{eq-Delta-S,T}), positive $\Delta$-values
imply that singlet isomers are more stable than triplet isomers;
negative $\Delta$-values correspond to triplets more stable than singlets.
The inspection of the various $\Delta$'s presented in the next sections reveals that cases exist
--- \emph{e.g.}, the even-member \ce{HC_{2k} H} chains, \emph{cf.}~Table~\ref{table:Delta-hcxh}
and the odd-member \ce{HC_{2k+1} N} chains, \emph{cf.}~Table~\ref{table:Delta-hcxn},
whose most stable isomers are singlets with polyyne structure --- 
for which the values of $\Delta_{S}$ and $\Delta_{T}$ significantly differ from each other. 
By virtue of Equation~(\ref{eq-Delta-S,T}), for such molecular species the corresponding singlet and
triplet isomers are characterized by significantly different geometries.
In many other cases --- like the longer odd-members \ce{HC_{2k+1} H} chains 
whose most stable isomers are triplets possessing a cumulenic structure, \emph{cf.}~Table~\ref{table:Delta-hcxh} ---
the values of $\Delta_{S}$ and $\Delta_{T}$ are almost equal.}

      {Given the fact discussed in detail below that,
        depending on the parity of the number of atoms, singlet or triplet isomers are more stable,
it makes sense to also consider vertical singlet-triplet splitting energies $\Delta_{m.s.}$
at the geometry of the most stable (acronym $m.s.$) isomer, which are defined as}
\begin{equation}
\hspace*{-5ex} 
\Delta_{m.s.} = \left\{
\begin{array}{lll}
  \Delta_S \equiv \mathcal{E}_T \left(\mathbf{R}_S\right) - \mathcal{E}_S \left(\mathbf{R}_S\right) > 0 &
  \mbox{ if } \mathcal{E}_S \left(\mathbf{R}_S\right) < \mathcal{E}_T \left(\mathbf{R}_S\right) &
  \mbox{(singlet more stable that triplet)}
  \\
  \Delta_T \equiv \mathcal{E}_T \left(\mathbf{R}_T\right) - \mathcal{E}_S \left(\mathbf{R}_T\right) < 0 &
  \mbox{ if } \mathcal{E}_T \left(\mathbf{R}_T\right) < \mathcal{E}_S \left(\mathbf{R}_T\right) &
  \mbox{(triplet more stable that singlet)}
  \\
\end{array}
\right .
\label{eq-Delta-m.s.}
\end{equation}
{Values of $\Delta_{m.s.}$ for the various families of chains considered are depicted in
\figsname\ref{fig:hcxh}c, \ref{fig:hcxn}c, \ref{fig:cxs}c, \ref{fig:cxo}c,
and \ref{fig:ocxo}c.}
  
In addition to DFT/B3LYP estimates of the singlet-triplet separations $\Delta$,
we also report $\Delta$-values deduced
by performing \emph{ab initio} CCSD and state-of-the-art CCSD(T) calculations.
To facilitate comparison with results of other elaborate \emph{ab initio} electronic
structure approaches that could (or should,
\emph{cf.}~{\secname}\ref{sec:ccsd_t})
be applied in subsequent studies, 
the CCSD and CCSD(T) $\Delta$-estimates
reported here (\emph{cf.}~Tables~\ref{table:Delta-hcxh}, \ref{table:Delta-hcxn},
\ref{table:Delta-cxs}, \ref{table:Delta-cxo}, and \ref{table:Delta-ocxo})
do not include
corrections due to zero-point motion.
{One should mention in this context that
  zero-point energy corrections within \emph{ab initio} approaches like CCSD and CCSD(T) are very challenging;
  they require costly numerical frequency calculations. Currently feasible studies of this kind are restricted
  to \emph{closed}-shell (that is, singlet but not triplet) species and smaller molecular sizes.\cite{Doney:18}}
As a trade-off
between accuracy and computationally demanding \emph{ab initio} frequency calculations,
zero-point energy corrections can be applied within the DFT/B3LYP approach.
Such estimates are indicated by label \emph{corr} in Tables~\ref{table:Delta-hcxh}, \ref{table:Delta-hcxn},
\ref{table:Delta-cxs}, \ref{table:Delta-cxo}, and \ref{table:Delta-ocxo}.
{So, values labeled ``B3LYP (corr)'' in those tables
  refer to results of DFT/B3LYP
  calculations including corrections due to zero-point motion while values labeled ``B3LYP''
  were obtained from DFT/B3LYP calculations without zero-point energy corrections.
  The rather minor differences between these two corrected and uncorrected values 
  visible in
  Tables~\ref{table:Delta-hcxh}, \ref{table:Delta-hcxn}, \ref{table:Delta-cxs},
  \ref{table:Delta-cxo}, and \ref{table:Delta-ocxo} suggest that, 
  in spite of the enormous computational effort, the benefit of
  CCSD(T) numerical frequency calculations, even if implemented, would be questionable.}
\section{Results and Discussion}
\label{sec:results}
In the sections that follow we will extensively analyze the most representative families of
carbon-based chains found in interstellar molecular clouds. For each family,
we investigated molecular sizes exceeding the longest chain astronomically observed;
the latter is indicated in the corresponding table caption.
Depending on their chemical composition, the chains considered are either ``normal''
(\emph{i.e.}, closed-shell, nonradical)
or diradical \cite{Abe:13} molecules characterized by polyyne- or cumulene-type structures. 
\subsection{HC$_{n}$H Homologous Series} % \subsection{\ce{HC_{n} H} Homologous Series}
\begin{figure*} % {hbtp}
  \centerline{\includegraphics[width=0.45\textwidth,angle=0]{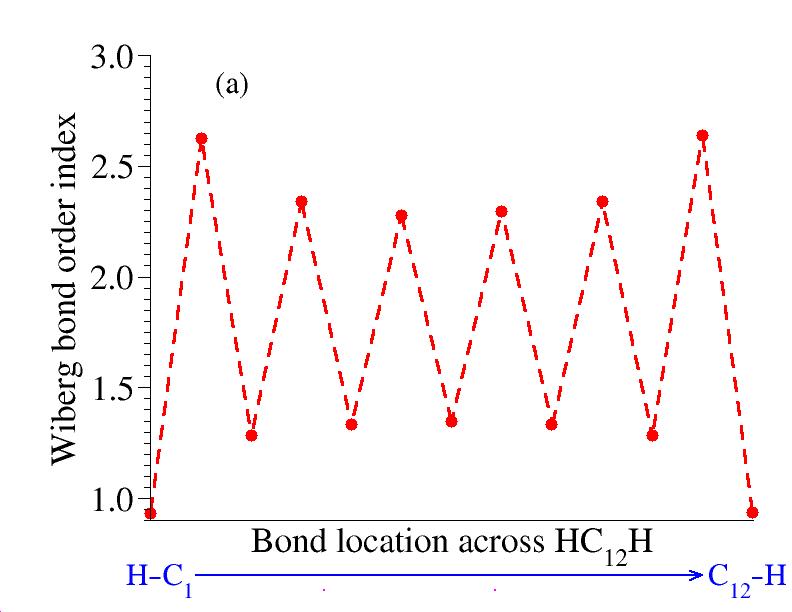}
  \includegraphics[width=0.45\textwidth,angle=0]{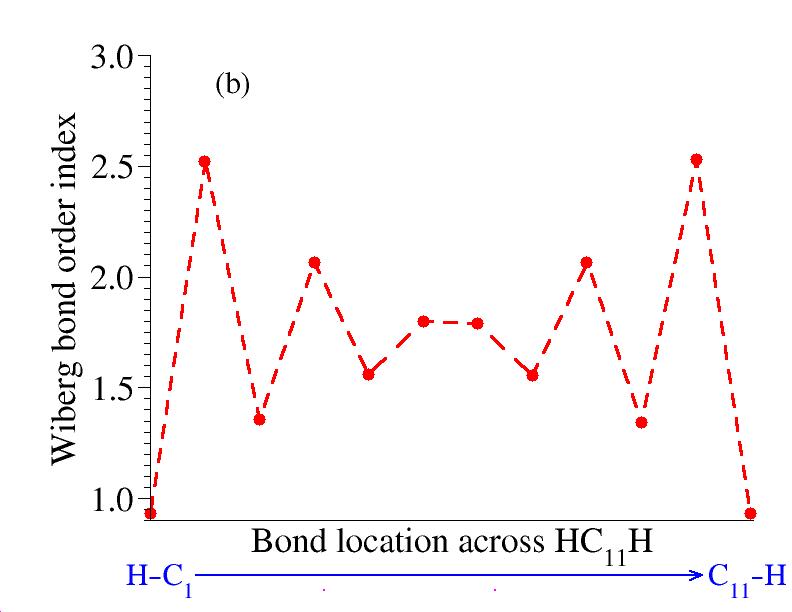}}
  \caption{Wiberg bond order indices for \ce{HC_n H} chains:
    (a) singlet \ce{HC12H} and (b) triplet \ce{HC11H}. The coordinates of these molecules
    at the corresponding energy minima as well as  the HOMO spatial distributions are presented in
    Tables S7 and S6, % Tables~\ref{table:hc12h-xyz-singlet} and \ref{table:hc11h-xyz-triplet},
    and in % \figname\ref{fig:homo-hc12h} and \figname\ref{fig:homo-hc11h},
    {\figsname}S2 and S1, 
    respectively.
  }
  \label{fig:bonds-hc11h-hc12h}
\end{figure*}

\label{sec:hcxh}
The members with an even number ($n=2k$) of carbon atoms
of this family, \ce{HC_{2k} H} ($k=1,2,\ldots$),
which we first consider, are polyynes wherein
single \ce{C\bond{-}C} and triple \ce{C\bond{#}C} carbon-carbon
bonds alternate across the chain backbone \ce{H\bond{1}C\bond{3}C\bond{1}...\bond{3}C\bond{1}H}.
This alternation is illustrated by the results shown in \figname\ref{fig:bonds-hc11h-hc12h}a,
which depicts the Wiberg bond order indices calculated for the singlet \ce{HC12H} chain.
All valence electrons form pairs between adjacent atoms, and this typically renders the singlet
state of the chain to be the most stable form.

\begin{figure*} % {hbtp}
  \centerline{\includegraphics[width=0.3\textwidth,angle=0]{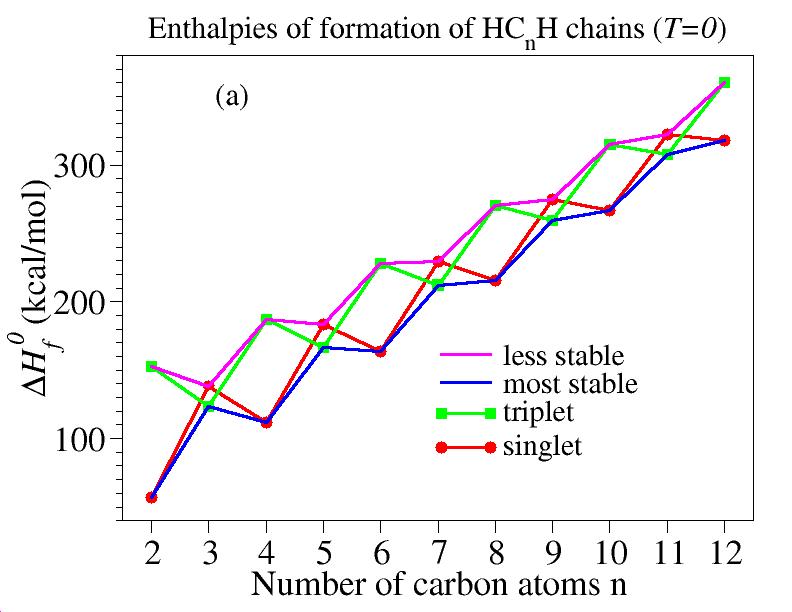}
          \includegraphics[width=0.3\textwidth,angle=0]{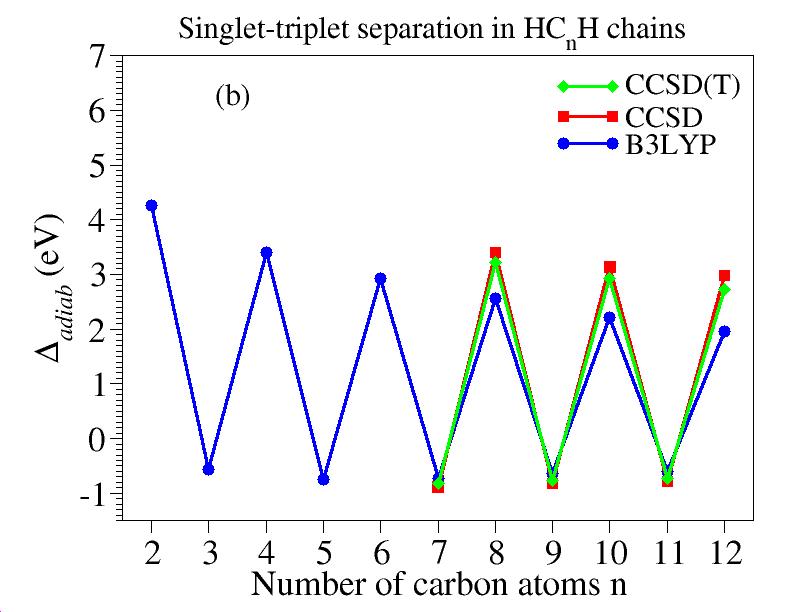}
              \includegraphics[width=0.3\textwidth,angle=0]{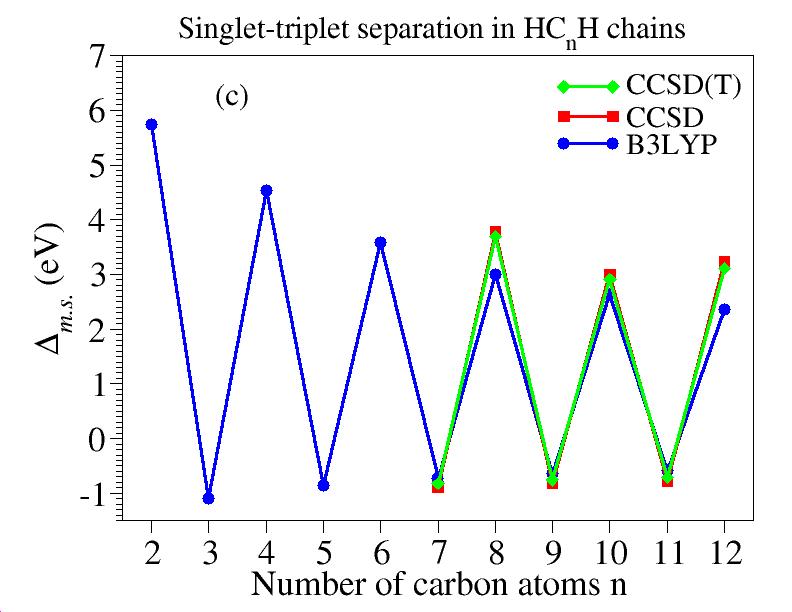}}
  \caption{Results for \ce{HC_n H} chains.
    (a) Enthalpies of formation for singlet and triplet chain isomers.
    (b) Adiabatic singlet-triplet separation energy $\Delta_{adiab}$.
    (c) Vertical singlet-triplet separation at the geometry of the most stable state $\Delta_{m.s.}$
    (namely, singlet for even members and triplet for odd members,
    % \added[remark={\rem{New text added in response to the third comment of the second reviewer}}]
          {\emph{cf.}~Equation~(\ref{eq-Delta-m.s.})}).
    Lines are guide to the eye. The numerical values underlying this figure are presented in Tables~\ref{table:H-hcxh} and \ref{table:Delta-hcxh}.
  }
  \label{fig:hcxh}
\end{figure*}

Table~\ref{table:H-hcxh}, which collects our results for the enthalpies of formation
$\Delta_{f}H^{0} = \Delta_{f} H^{0}(T)$ at zero $\Delta_{f} H^{0}_{0}$ and room temperature
$\Delta_{f} H^{0}_{RT}$ ($T=0$\,K and $T=298.15$\,K, respectively)
and the related \figname\ref{fig:hcxh}a confirm this picture.
As visible there, for the even members \ce{HC_{2k} H},
the values $\left . \Delta_{f}H^{0}\right\vert_{S}$ for singlet ($S$) isomers are smaller than 
the values $\left . \Delta_{f}H^{0}\right\vert_{T}$ for triplet ($T$) isomers.
\begin{table}[htbp] % [h!]
  \scriptsize % \footnotesize % \small \footnotesize % \scriptsize % \footnotesize % \tiny
  \begin{center}
    \begin{threeparttable}
    %%%%%%%%%%%%%%%%%%%%%%%%%%%%
    \begin{tabular*}{0.47\textwidth}{@{\extracolsep{\fill}}lrrrr}
      \hline
      Molec.
      & $\Delta_{f} H^{0}_{0}\vert_{S}$ & $\Delta_{f} H^{0}_{RT}\vert_{S}$
      & $\Delta_{f} H^{0}_{0}\vert_{T}$ & $\Delta_{f} H^{0}_{RT}\vert_{T}$
      \\
      \hline
      % hc2h_optimized_singlet_b3lyp_6311++g_3df3pd_dft_ccsd_ST_split.log MA; 
      % hc2h_optimized_triplet_b3lyp_6311++g_3df3pd_dft_ccsd_ST_split.log MA
      % hc2h_optimized_singlet_b3lyp_6311++g_3df3pd_n_IP_EA_ovgf_ST_split.log, MA; forces ok
      % hc2h_optimized_singlet_b3lyp_6311++g_3df3pd_n_nbo_ST_split_IP_EA_both_dft_and_ccsd_t.log, MA; forces ok
      \ce{HC2H}  &  56.706 & 56.540 & 152.995 & 152.954 \\
      % hc3h_optimized_singlet_b3lyp_6311++g_3df3pd_n_IP_EA_ovgf_ST_split.log, MA; forces ok
      % hc3h_optimized_triplet_b3lyp_6311++g_3df3pd_n_IP_EA_ovgf_ST_split.log, ulm; forces ok
      \ce{HC3H}  & 138.464 & 138.910 & 123.415 & 125.742 \\
      % hc4h_optimized_singlet_b3lyp_6311++g_3df3pd_n_IP_EA_ovgf_ST_split.log, ulm; forces ok
      % hc4h_optimized_triplet_b3lyp_6311++g_3df3pd_n_IP_EA_ovgf_ST_split.log, MA; forces ok
      \ce{HC4H}  & 111.469 & 111.709 & 187.331 & 187.863 \\
% 
      % hc5h_b3lyp_singlet_6311++g_3df3pd_zmat_nbo_IP_EA_ST_split_ovgf.log, MA; forces ok
      % hc5h_b3lyp_triplet_6311++g_3df3pd_zmat_nbo_IP_EA_ST_split_ovgf.log, MA; forces ok
      \ce{HC5H}  & 183.280 & 184.472 & 166.461 & 167.299 \\
      % hc6h_optimized_singlet_b3lyp_6311++g_3df3pd_n_IP_EA_ovgf_ST_split.log, MA; forces ok
      % hc6h_triplet_b3lyp_6311++g_3df3pd_n_IP_EA_ovgf_ST_split.log, MA; forces ok
% 
      % hc5h_triplet_cbs-qb3.log, ulm: t=-191.566534340; zmpt=0.037715; ent=0.044500; hmt=-191.528820; hm=hmt; en=ent; zpm=zpmt
% 
% 
% 
% 
% 
      \ce{HC6H} \tnote{$\ast $}  & 163.773 & 164.660 & 227.837 & 229.260 \\
% 
% 
% 
% 
      % hc7h_optimized_singlet_b3lyp_6311++g_3df3pd_n_IP_EA_ovgf_ST_split.log, ulm; forces ok
      % hc7h_optimized_triplet_b3lyp_6311++g_3df3pd_zmat_n_IP_EA_ST_split_ovgf.log, ulm; forces ok
      \ce{HC7H}  & 229.287 & 231.208 & 212.236 & 213.732 \\
      % hc8h_optimized_singlet_b3lyp_6311++g_3df3pd_n_IP_EA_ovgf_ST_split.log, ulm; forces ok
      % hc8h_optimized_triplet_b3lyp_6311++g_3df3pd_n_IP_EA_ovgf_ST_split.log, ulm; forces ok
% 
% 
% 
% 
% 
% 
% 
      \ce{HC8H}  & 215.390 & 216.935 & 270.313 & 272.668 \\
% 
% 
% 
% 
      % hc9h_optimized_singlet_b3lyp_6311++g_3df3pd_n_IP_EA_ovgf_ST_split.log, ulm; forces ok
      % hc9h_triplet_b3lyp_6311++g_3df3pd_n_IP_EA_ovgf_ST_split.log, MA; forces ok
      \ce{HC9H}  & 275.106 & 277.356 & 259.606 & 261.482 \\
      % hc9h_singlet_cbs-qb3.log, ulm: s=-343.915194016; zpms=0.059596; ens=0.070394; hms=-343.855598; hm=hms; en=ens; zpm=zpms;
      % hc9h_triplet_cbs-qb3.log, ulm: t=-343.939360120; zpmt=0.058627; ent=0.068736; hmt=-343.880733; hm=hmt; en=ent; zpm=zpmt;
% 
% 
      % hc10h_optimized_singlet_b3lyp_6311++g_3df3pd_n_IP_EA_ovgf_ST_split.log, ulm; forces ok
      % hc10h_triplet_b3lyp_6311++g_3df3pd_n_IP_EA_ovgf_ST_split.log, ulm; forces ok
      \ce{HC10H} & 266.706 & 268.921 & 314.956 & 317.451 \\
% 
% 
% 
      % hc11h_optimized_singlet_b3lyp_6311++g_3df3pd_n_ST_split_IP_EA.log, ulm, 
      % hc11h_optimized_triplet_b3lyp_6311++g_3df3pd_n_IP_EA_ovgf_ST_split.log, ulm; forces ok
      \ce{HC11H} & 322.336 & 325.149 & 307.943 & 310.455 \\
% 
% 
% 
      % hc12h_optimized_singlet_b3lyp_6311++g_3df3pd_n_IP_EA_ovgf_ST_split.log, ulm; forces ok
      % hc12h_triplet_b3lyp_6311++g_3df3pd_n_IP_EA_ovgf_ST_split.log, ulm; forces ok
      \ce{HC12H} & 317.923 & 320.797 & 360.526 & 363.516 \\
      \hline
    \end{tabular*}
    \begin{tablenotes}\footnotesize
    \item[$\ast $] Longest chain of this family astronomically observed.\cite{Cernicharo:01}
    \end{tablenotes}
    \end{threeparttable}
     %%%%%%%%%%%%%%%%%%%%%%%%%%%%%%%%%%%%%%%%%%%%%%%%%%%%%%%%%%%%%%%%%%%%%%%%%%%%%%%%%%%%%%
    \caption{Enthalpies of formation of linear \ce{HC_nH} chains at zero and room temperature (subscript $0$ and $RT$, respectively). Notice that for the even members 
      \ce{HC_{2k} H} the values for singlet (label $S$) are smaller than those for triplet (label $T$),
      while for the odd members \ce{HC_{2k+1} H} the values for triplet are smaller than those for singlet. 
    }
    % \ib{Compare with Table~9 and Fig.~10 of ref.~\citenum{Etim:16a}. IR spectrum of the singlet (nonlinear) is significantly different from that of the triplet (triplet is linear).}
    \label{table:H-hcxh}
  \end{center}
\end{table}
\begin{table}[htbp] % [h!]
  \scriptsize % \small % \footnotesize % \scriptsize % \footnotesize % \tiny
  \begin{center}
    \begin{tabular*}{0.47\textwidth}{@{\extracolsep{\fill}}rrrrr}
      \hline
      Molec. & Method & $\Delta_{adiab}$      & $\Delta_{S}$      & $\Delta_{T}$     
      \\
      \hline
      % hc2h_optimized_singlet_b3lyp_6311++g_3df3pd_dft_ccsd_ST_split.log MA
      % hc2h_optimized_triplet_b3lyp_6311++g_3df3pd_dft_ccsd_ST_split.log MA
      \ce{HC2H}  & B3LYP    &   4.269 &  5.739   &  2.285 \\
      \ce{HC2H}  & B3LYP (corr) &   4.176 &  5.645   &  2.191 \\
      %%CCSD
% 
      % hc2h_optimized_triplet_b3lyp_6311++g_3df3pd_n_nbo_ST_split_IP_EA_both_dft_and_ccsd_t.log MA
      % hc2h_optimized_triplet_b3lyp_6311++g_3df3pd_n_nbo_ST_split_IP_EA_both_dft_and_ccsd_t.log MA
% 
      \hline
      % hc3h_optimized_singlet_b3lyp_6311++g_3df3pd_dft_ccsd_ST_split.log MA
      % hc3h_optimized_triplet_b3lyp_6311++g_3df3pd_dft_ccsd_ST_split.log ulm
      \ce{HC3H}  & B3LYP    &  -0.571 & -0.044   & -1.097 \\
      \ce{HC3H}  & B3LYP (corr) &  -0.653 & -0.125   & -1.179 \\
      %%CCSD
% 
      % hc3h_optimized_singlet_b3lyp_6311++g_3df3pd_ccsd_t_n_ST_split_IP_EA.log MA
      % hc3h_optimized_triplet_b3lyp_6311++g_3df3pd_ccsd_t_n_ST_split_IP_EA.log ulm
% 
      \hline      
      % hc4h_optimized_singlet_b3lyp_6311++g_3df3pd_dft_ccsd_ST_split.log ulm
      % hc4h_optimized_triplet_b3lyp_6311++g_3df3pd_dft_ccsd_ST_split.log ulm
      \ce{HC4H}  & B3LYP    &   3.410 &  4.532   &  1.856  \\
      \ce{HC4H}  & B3LYP (corr) &   3.290 &  4.411   &  1.735  \\
      %%CCSD
% 
      % hc4h_optimized_singlet_b3lyp_6311++g_3df3pd_dft_ccsd_t_ST_split_IP_EA.log MA
      % hc4h_optimized_triplet_b3lyp_6311++g_3df3pd_dft_ccsd_t_ST_split_IP_EA.log MA
% 
      \hline      
      % hc5h_optimized_singlet_b3lyp_6311++g_3df3pd_dft_ccsd_ST_split.log ulm
      % hc5h_optimized_triplet_b3lyp_6311++g_3df3pd_dft_ccsd_ST_split.log ulm
      \ce{HC5H}  & B3LYP    &  -0.740 & -0.376   & -0.861 \\
      \ce{HC5H}  & B3LYP (corr) &  -0.729 & -0.366   & -0.850 \\
      % hc5h_optimized_singlet_b3lyp_6311++g_3df3pd_dft_ccsd_ST_split_IP_EA_ccsd.log ulm
      % hc5h_optimized_singlet_b3lyp_6311++g_3df3pd_n_nbo_ST_split_IP_EA_both_dft_and_ccsd_t.log MA
      % hc5h_optimized_triplet_b3lyp_6311++g_3df3pd_dft_ccsd_t_ST_split_IP_EA.log MA
      %%CCSD
% 
% 
      % 
      \hline
      % hc6h_optimized_singlet_b3lyp_6311++g_3df3pd_dft_ccsd_ST_split.log MA
      % hc6h_optimized_triplet_b3lyp_6311++g_3df3pd_dft_ccsd_ST_split.log MA
      \ce{HC6H}  & B3LYP    &  2.930  &  3.579   &  1.809 \\
      \ce{HC6H}  & B3LYP (corr) &  2.778  &  3.427   &  1.657 \\
      %%CCSD
% 
% 
      \hline
      % hc7h_optimized_singlet_b3lyp_6311++g_3df3pd_n_IP_EA_ovgf_ST_split.log, ulm
      % hc7h_optimized_triplet_b3lyp_6311++g_3df3pd_zmat_n_IP_EA_ST_split_ovgf.log, ulm
      \ce{HC7H}  & B3LYP    & -0.730  & -0.729   & -0.730 \\
      \ce{HC7H}  & B3LYP (corr) & -0.740  & -0.739   & -0.740 \\
      % hc7h_optimized_singlet_b3lyp_6311++g_3df3pd_dft_ccsd_t_ST_split_IP_EA.log MA
      % hc7h_optimized_triplet_b3lyp_6311++g_3df3pd_dft_ccsd_t_ST_split_IP_EA.log MA
      %%CCSD
% 
% 
      % hc7h_singlet_b3lyp_6311++g_3df3pd_n_nbo_mayer_ST_split_IP_EA_dft_ccsd_t.log, ulm
      % hc7h_optimized_triplet_rob3lyp_6311++g_3df3pd_roh_n_nbo_mayer_ST_split_IP_EA_dft_ccsd_t.log, ulm
      \ce{HC7H}  & CCSD    & -0.885 & -0.880 & -0.881 \\
      \ce{HC7H}  & CCSD(T) & -0.827 & -0.820 & -0.822 \\
      \hline
      % hc8h_optimized_singlet_b3lyp_6311++g_3df3pd_n_IP_EA_ovgf_ST_split.log ulm
      % hc8h_optimized_triplet_b3lyp_6311++g_3df3pd_n_IP_EA_ovgf_ST_split.log ulm
      \ce{HC8H}  & B3LYP    &  2.552  &  3.004   &  1.895 \\
      \ce{HC8H}  & B3LYP (corr) &  2.382  &  2.834   &  1.725 \\
      % hc8h_optimized_singlet_b3lyp_6311++g_3df3pd_dft_ccsd_t_ST_split_IP_EA.log MA
      % hc8h_optimized_triplet_b3lyp_6311++g_3df3pd_dft_ccsd_t_ST_split_IP_EA.log MA
      %%CCSD
% 
% 
      % hc8h_singlet_b3lyp_6311++g_3df3pd_n_nbo_mayer_ST_split_IP_EA_dft_ccsd_t.log, ulm
      % hc8h_optimized_triplet_rob3lyp_6311++g_3df3pd_roh_n_nbo_mayer_ST_split_IP_EA_dft_ccsd_t.log, ulm
      \ce{HC8H}  & CCSD    & 3.403 & 3.776 & 2.455 \\
      \ce{HC8H}  & CCSD(T) & 3.211 & 3.693 & 2.529 \\
      \hline
      % hc9h_optimized_singlet_b3lyp_6311++g_3df3pd_dft_ccsd_ST_split.log MA
      % hc9h_optimized_triplet_b3lyp_6311++g_3df3pd_dft_ccsd_ST_split.log MA    
      \ce{HC9H}  & B3LYP    & -0.645  & -0.647   & -0.648 \\
      \ce{HC9H}  & B3LYP (corr) & -0.672  & -0.674   & -0.675 \\
      % hc9h_optimized_singlet_b3lyp_6311++g_3df3pd_dft_ccsd_t_ST_split_IP_EA.log MA
      % hc9h_optimized_triplet_b3lyp_6311++g_3df3pd_dft_ccsd_t_ST_split_IP_EA.log MA
      %%CCSD
% 
% 
      % hc9h_singlet_b3lyp_6311++g_3df3pd_n_nbo_mayer_ST_split_IP_EA_dft_ccsd_t.log, ulm
      % hc9h_optimized_triplet_rob3lyp_6311++g_3df3pd_roh_n_nbo_mayer_ST_split_IP_EA_dft_ccsd_t.log, ulm
      \ce{HC9H}  & CCSD    & -0.823 & -0.820 & -0.821 \\
      \ce{HC9H}  & CCSD(T) & -0.759 & -0.754 & -0.755 \\ 
      \hline
      % hc10h_optimized_singlet_b3lyp_6311++g_3df3pd_n_IP_EA_ovgf_ST_split.log, ulm
      % hc10h_triplet_b3lyp_6311++g_3df3pd_n_IP_EA_ovgf_ST_split.log, ulm
      \ce{HC10H} & B3LYP    &  2.210  &  2.628   &  1.804 \\
      \ce{HC10H} & B3LYP (corr) &  2.092  &  2.510   &  1.686 \\
      % hc10h_optimized_singlet_b3lyp_6311++g_3df3pd_ccsd_t_n_ST_split_IP_EA.log ulm
      % hc10h_optimized_triplet_b3lyp_6311++g_3df3pd_ccsd_t_n_ST_split_IP_EA.log ulm
      %%CCSD
% 
% 
% 
% 
      \ce{HC10H} & CCSD     & 3.137 & 2.997 & 2.378 \\
      \ce{HC10H} & CCSD(T)  & 2.928 & 2.902 & 2.426 \\
      \hline
      \ce{HC11H} & B3LYP    & -0.592  & -0.591   & -0.592 \\
      \ce{HC11H} & B3LYP (corr) & -0.624  & -0.624   & -0.625 \\
      % hc11h_optimized_singlet_b3lyp_6311++g_3df3pd_ccsd_t_n_ST_split_IP_EA.log ulm
      % hc11h_optimized_triplet_b3lyp_6311++g_3df3pd_ccsd_t_n_ST_split_IP_EA.log ulm
      %%CCSD
% 
% 
% 
% 
% 
      \ce{HC11H} & CCSD     & -0.787  &  -0.781   & -0.783 \\
      \ce{HC11H} & CCSD(T)  & -0.724  &  -0.716   & -0.717 \\
      \hline
      % hc12h_optimized_singlet_b3lyp_6311++g_3df3pd_n_IP_EA_ovgf_ST_split.log, ulm
      % hc12h_triplet_b3lyp_6311++g_3df3pd_n_IP_EA_ovgf_ST_split.log, ulm
      \ce{HC12H} & B3LYP    &  1.960  &  2.365   &  1.565 \\
      \ce{HC12H} & B3LYP (corr) &  1.847  &  2.253   &  1.453 \\
      % hc12h_optimized_singlet_b3lyp_6311++g_3df3pd_ccsd_t_n_ST_split_IP_EA.log ulm
      % hc12h_optimized_triplet_b3lyp_6311++g_3df3pd_ccsd_t_n_ST_split.log ulm
      %%CCSD
% 
% 
% 
% 
      \ce{HC12H}  & CCSD    &  2.981  &  3.241  &  2.151  \\
      \ce{HC12H}  & CCSD(T) &  2.721  &  3.109  &  2.185  \\
\hline
    \end{tabular*}      
    %%%%%%%%%%%%%%%%%%%%%%%%%%%%%%%%%%%%%%%%%%%%%%%%%%%%%%%%%%%%%%%%%%%%%%%%%%%%%%%%%%%%%%
    %%%%%%%%%%%%%%%%%%%%%%%%%%%%%%%%%%%%%%%%%%%%%%%%%%%%%%%%%%%%%%%%%%%%%%%%%%%%%%%%%%%%%%
    \caption{Adiabatic ($\Delta_{adiab}$) and vertical ($\Delta_{S,T}$) values of the singlet-triplet
      energy separation (in eV) for \ce{HC_{n} H} chains obtained within the methods indicated in the second column.
      The two vertical values shown here correspond to the optimized singlet ($\Delta_{S}$)
      and triplet ($\Delta_{T}$) geometries. Notice that the sign of $\Delta$
      indicates that the most stable isomers are singlets for even members
      ($\Delta > 0$, \ce{HC_{2k} H}) and triplets for odd members ($\Delta < 0$, \ce{HC_{2k+1} H}).
      Corrections due to zero-point motion (label \emph{corr}) were deduced within the DFT/B3LYP
      approach;
      % \added[remark={\rem{New text added in response to the first comment of the second reviewer}}]
            {see the last paragraph of \secname~\ref{sec:methods}.}
            % \added[remark={\rem{New text added in response to the second comment of the second reviewer}}]
                  {The almost equal values $\Delta_S \approx \Delta_T$ for larger odd-members (\ce{HC_{n} H}, $n=7,9,11$)
                    indicate that the singlet and triplet geometries are similar ($\mathbf{R}_S \approx \mathbf{R}_T$,
                    \emph{cf.}~Equation~(\ref{eq-Delta-S,T})).}
    }
    \label{table:Delta-hcxh}
  \end{center}
\end{table}

The alternation between single and triple bonds is incompatible with the
standard rules of valence % chemical formula of
for the
odd members species \ce{HC_{2k+1} H}. Forms that could come into question 
here are either of polyyne- (acetylenic)-type 
(\emph{e.g.}, \ce{H\bond{1}C\bond{3}C\bond{1}...\bond{3}C\bond{1}}\dirad{C}\ce{\bond{1}H})
or of cumulene type
(\emph{e.g.}, \ce{H\bond{1}}\rad{C}\ce{\bond{2}C\bond{2}...\bond{2}C\bond{2}}\rad{C}\ce{\bond{1}H}).
As illustrated by a specific example,
namely the \ce{HC11H} chain depicted in \figname\ref{fig:bonds-hc11h-hc12h}b,
calculations confirm the cumulene structure.
By inspecting the enthalpies of formation (Table~\ref{table:H-hcxh} and \figname\ref{fig:hcxh}a),
one can see that the triplet state rather than the singlet state
is the most stable form of the odd members of this family.

Above, we described the stability of the two subclasses (\ce{HC_{2k+1}H} and \ce{HC_{2k}H})
in a picture based on thermochemistry, which may not be the most adequate for systems of
interest for astrochemistry, where single-molecule kinetic effects may prevail.
Therefore, as a counterpart of this thermochemical analysis, 
in Table~\ref{table:Delta-hcxh} and \figname\ref{fig:hcxh}b and c we also report results for the 
singlet-triplet separation energies $\Delta$ computed as differences between the corresponding total electronic energies 
of a molecule
(\emph{cf.}~{\secname}\ref{sec:methods}).
The negative values ($\Delta \equiv \mathcal{E}_{T} - \mathcal{E}_{S} < 0$)
for odd members indicate that the triplet state rather than the singlet state
is preferably energetically for the odd members of this family, while the opposite
($\Delta > 0$) holds for even members, for which the singlet isomers are the most stable.
\subsection{HC$_{n}$N Homologous Series} % \subsection{\ce{HC_{n} N} Homologous Series}
\label{sec:hcxn}
Let us next consider the related \ce{HC_nN} family.

Out of the various carbon-based homologous series, this is probably the most
numerous family investigated in different contexts in the past.
\cite{Avery:76,Kroto:78,Broten:78,Bell:97,Sommerfeld:05,Graupner:06,Baldea:2019e,Baldea:2019g}
In this case, it is the odd-member
($n=2k+1$) subclass \ce{HC_{2k+1} N} ($k=1, 2, \ldots $)
\ce{H\bond{1}C\bond{3}...\bond{1}C\bond{3}N}
wherein the standard valence rules allow a singlet-triplet alternation of the
carbon-carbon bonds along the chain. This is illustrated by the case of the \ce{HC9N}
chain in \figname\ref{fig:bonds-hc8n-hc9n}a. Calculations confirm that the electronic ground state 
of this \ce{HC_{2k+1} N} cyanopolyyne structure is of ``normal'' type, \emph{i.e.},
a singlet state. 
Indeed, the corresponding enthalpies of formation for singlet are lower than for triplet
($\Delta_{f} H^{0}_{0}\vert_{S} < \Delta_{f} H^{0}_{0}\vert_{T}$),
and the values of the singlet-triplet
energy separation are positive ($\Delta > 0$);
see~Tables~\ref{table:H-hcxn} and \ref{table:Delta-hcxn}, and \figname\ref{fig:hcxn}.

Calculations for even members \ce{HC_{2k} N} show that, out of the two possible forms --- namely,
of polyyne type, \emph{e.g.},
\ce{H\bond{1}\dirad{C}\bond{1}C\bond{3}...\bond{1}C\bond{3}C\bond{1}C\bond{3}N} 
or of cumulene type, \emph{e.g.},
\ce{H\bond{1}C\bond{3}C\bond{1}\rad{C}\bond{2}C\bond{2}...\bond{2}C\bond{2}\rad{C}\bond{1}}\ce{C\bond{3}N} ---
it is the latter that occurs. To exemplify,
in \figname\ref{fig:bonds-hc8n-hc9n}b we depict the case of the \ce{HC8N} chain.
Our results for the even members \ce{HC_{2k} N} are presented in
Tables~\ref{table:H-hcxn} and \ref{table:Delta-hcxn}, 
and 
\figname{\ref{fig:hcxn}a.
They demonstrate an ``anomalous'' behavior; the triplet state is more stable than the singlet state.
The enthalpies of formation 
Table~\ref{table:H-hcxn} and 
\figname\ref{fig:hcxn}a)
for triplet are lower than for singlet,
the singlet-triplet separation energies 
(Table~\ref{table:Delta-hcxn}
and 
\figname\ref{fig:hcxn}b 
and c) are negative.

To sum up,
the electronic ground state for even members \ce{HC_{2k} N} is a triplet state,
which is in contrast with the singlet electronic ground state of the
odd members \ce{HC_{2k+1} N}.
\begin{figure*} % {hbtp}
  \centerline{\includegraphics[width=0.45\textwidth,angle=0]{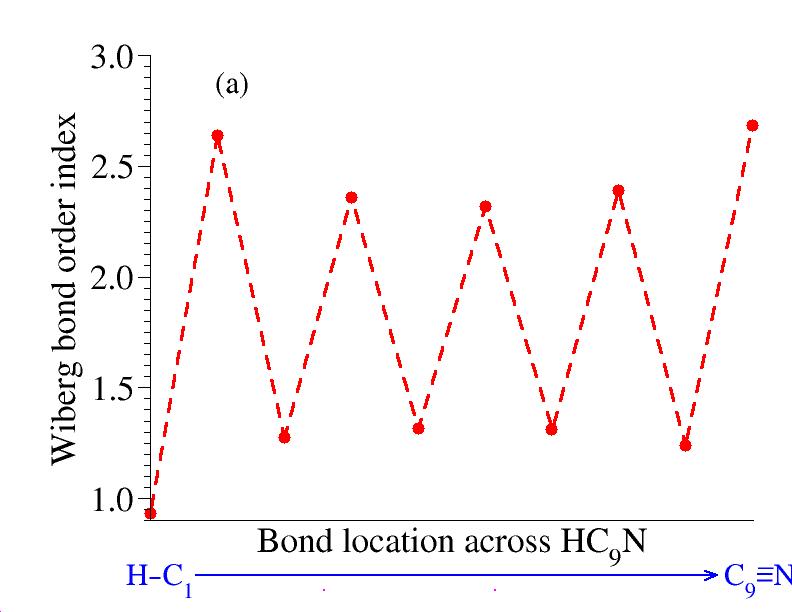}
  \includegraphics[width=0.45\textwidth,angle=0]{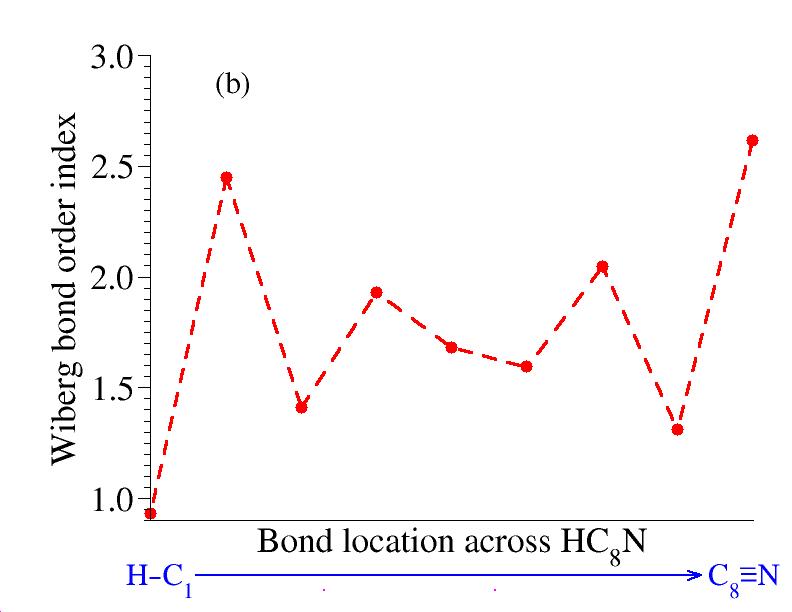}}
  \caption{Wiberg bond order indices for 
    % \replaced{\ce{HC_n N}}{\bad{\ce{HC_n H}}} 
    \ce{HC_n N}
    chains:
    (a) singlet \ce{HC9N} and (b) triplet \ce{HC8N}. The coordinates of these molecules
    at the corresponding energy minima as well as  the HOMO spatial distributions are presented in
    Tables~S11 and S10, % Tables~\ref{table:hc9n-xyz-singlet} and \ref{table:hc8n-xyz-triplet},
    and in
    {\figsname}S4 and S3, % \figname\ref{fig:homo-hc9n} and \ref{fig:homo-hc8n},
    respectively.
  }
  \label{fig:bonds-hc8n-hc9n}
\end{figure*}
\begin{table}[htbp] % [h!]
  \scriptsize % \small % \scriptsize % \footnotesize % \tiny
  \begin{center}
    \begin{threeparttable}
    %%%%%%%%%%%%%%%%%%%%%%%%%%%%
    \begin{tabular*}{0.47\textwidth}{@{\extracolsep{\fill}}lrrrr}
      \hline
      Molec.
      & $\Delta_{f} H^{0}_{0} $ & $\Delta_{f} H^{0}_{RT}\vert_{S}$
      & $\Delta_{f} H^{0}_{0} $ & $\Delta_{f} H^{0}_{RT}\vert_{T}$
      \\
      \hline
      % hcn_optimized_singlet_b3lyp_6311++g_3df3pd_n_nbo_ST_split_IP_EA_both_dft_and_ccsd_t.log, MA
      % hcn_optimized_triplet_b3lyp_6311++g_3df3pd_n_nbo_ST_split_IP_EA_both_dft_and_ccsd_t.log, MA
      \ce{HCN}   &  30.479 &  30.369 & 135.813 & 135.906 \\
      % hc2n_optimized_singlet_b3lyp_6311++g_3df3pd_n_IP_EA_ovgf_ST_split.log ulm
      % hc2n_optimized_triplet_b3lyp_6311++g_3df3pd_n_IP_EA_ovgf_ST_split.log ulm
      \ce{HC2N}  & 122.292 & 122.567 & 106.606 & 107.014 \\
      % hc3n_optimized_singlet_b3lyp_6311++g_3df3pd_n_nbo_ST_split_IP_EA_both_dft_and_ccsd_t.log MA
      % hc3n_triplet_b3lyp_6311++g_3df3pd_n_nbo_ST_split_IP_EA_both_dft_and_ccsd_t.log, MA
      \ce{HC3N}  &  88.613 &  88.933 & 174.805 & 175.320 \\ % triplet MA slightly different from ulm or another triplet ToDo!
% 
      % hc4n_optimized_singlet_b3lyp_6311++g_3df3pd_n_nbo_ST_split_IP_EA_both_dft_and_ccsd_t.log MA
      % hc4n_triplet_b3lyp_6311++g_3df3pd_n_nbo_ST_split_IP_EA_both_dft_and_ccsd_t.log MA
      \ce{HC4N}  & 164.202 & 165.109 & 146.902 & 147.703 \\
% 
% 
% 
      % hc5n_optimized_singlet_b3lyp_6311++g_3df3pd_n_nbo_ST_split_IP_EA_both_dft_and_ccsd_t.log, MA
      % hc5n_b3lyp_triplet_6311++g_3df3pd_zmat_n_nbo_ST_split_IP_EA_both_dft_and_ccsd_t.log, MA
      \ce{HC5N}  & 141.802 & 142.735 & 205.424 & 206.859 \\
% 
% 
% 
      % hc6n_optimized_singlet_b3lyp_6311++g_3df3pd_n_ST_split_IP_EA_both_dft_and_ccsd_t.log MA
      % hc6n_optimized_triplet_b3lyp_6311++g_3df3pd_n_nbo_ST_split_IP_EA_all_dft_triplet_ccsd_t_only.log MA
      % hc6n_optimized_triplet_b3lyp_6311++g_3df3pd_n_nbo_ST_split_IP_EA_both_dft_and_ccsd_t.log ulm forces ok
% 
% 
% 
% 
      %      \hline
      \ce{HC6N}  & 209.099 & 210.827 & 191.840 & 193.150 \\
% 
% 
% 
      % hc7n_optimized_singlet_b3lyp_6311++g_3df3pd_n_nbo_ST_split_IP_EA_both_dft_and_ccsd_t.log, MA
      % hc7n_optimized_triplet_b3lyp_6311++g_3df3pd_n_nbo_ST_split_IP_EA_both_dft_and_ccsd_t.log, MA
      \ce{HC7N}  & 193.670 & 195.288 & 246.645 & 248.688 \\
% 
% 
% 
% 
       %      \hline
      % hc8n_optimized_singlet_b3lyp_6311++g_3df3pd_n_nbo_ST_split_IP_EA_both_dft_and_ccsd_t.log MA
      % hc8n_optimized_triplet_b3lyp_6311++g_3df3pd_n_nbo_ST_split_IP_EA_both_dft_and_ccsd_t.log MA
      \ce{HC8N}  & 254.993 & 257.248 & 238.960 & 241.102 \\
% 
% 
% 
% 
% 
% 
% 
      % hc9n_optimized_singlet_b3lyp_6311++g_3df3pd_n_nbo_ST_split_IP_EA_both_dft_and_ccsd_t.log ulm
      % hc9n_reoptimized_triplet_b3lyp_6311++g_3df3pd_n_nbo_ST_split_IP_EA_both_dft_and_ccsd_t.log ulm
      \ce{HC9N}  & 245.201 & 247.476 & 290.266 & 292.834 \\
% 
% 
% 
% 
% 
      %      % hc10n_optimized_b3lyp_6311++g_3df3pd_nbo_ST_split.log, ulm
% 
% 
      % hc10n_optimized_singlet_b3lyp_6311++g_3df3pd_n_nbo_ST_split_IP_EA_both_dft_and_ccsd_t.log ulm
      % hc10n_optimized_triplet_b3lyp_6311++g_3df3pd_n_nbo_ST_split_IP_EA_both_dft_and_ccsd_t.log ulm
      \ce{HC10N} & 301.872 & 304.711 & 287.203 & 289.986 \\
% 
% 
% 
% 
% 
      % hc11n_optimized_singlet_b3lyp_6311++g_3df3pd_coord_n_nbo_ST_split.log, ulm ForcesNotChecked!
      % hc11n_optimized_triplet_b3lyp_6311++g_3df3pd_dft_ccsd_t_nbo_ST_split.log ulm  ForcesNotChecked!
      \ce{HC11N} \tnote{$\ast $} & 296.620 & 299.539 & 336.070 & 339.228 \\ % corrections OK!
% 
% 
% 
% 
% 
      % hc12n_optimized_singlet_b3lyp_6311++g_3df3pd_dft_ccsd_t_nbo_ST_split_IP_wo_EA.log ulm ForcesNotChecked!
      % hc12n_optimized_singlet_b3lyp_6311++g_3df3pd_n_nbo_ST_split_IP_EA_both_dft_and_ccsd_t.log ulm Forces Checked!
      % hc12n_b3lyp_triplet_6311++g_3df3pd_zmat_n_IP_EA.log, MA (corrections and forces ok)
      \ce{HC12N} & 349.939 & 353.385 & 336.219 & 339.636 \\
      \hline
    \end{tabular*}
    \begin{tablenotes}\footnotesize
    \item{$\ast $} Longest carbon chain ever claimed in astronomical observations \cite{Bell:97}
    \end{tablenotes}
    \end{threeparttable}
    %%%%%%%%%%%%%%%%%%%%%%%%%%%%%%%%%%%%%%%%%%%%%%%%%%%%%%%%%%%%%%%%%%%%%%%%%%%%%%%%%%%%%%
    \caption{Enthalpies of formation of linear \ce{HC_nN} chains at zero and room temperature (subscript $0$ and $RT$, respectively). 
      Notice that for the odd members 
      \ce{HC_{2k+1} N} the values for singlet (label $S$) are smaller than those for triplet (label $T$),
      while for the even members \ce{HC_{2k} N} the values for triplet are smaller than those for singlet.
    } 
    \label{table:H-hcxn}
\end{center}
\end{table}
\begin{figure*} % {hbtp}
  \centerline{\includegraphics[width=0.3\textwidth,angle=0]{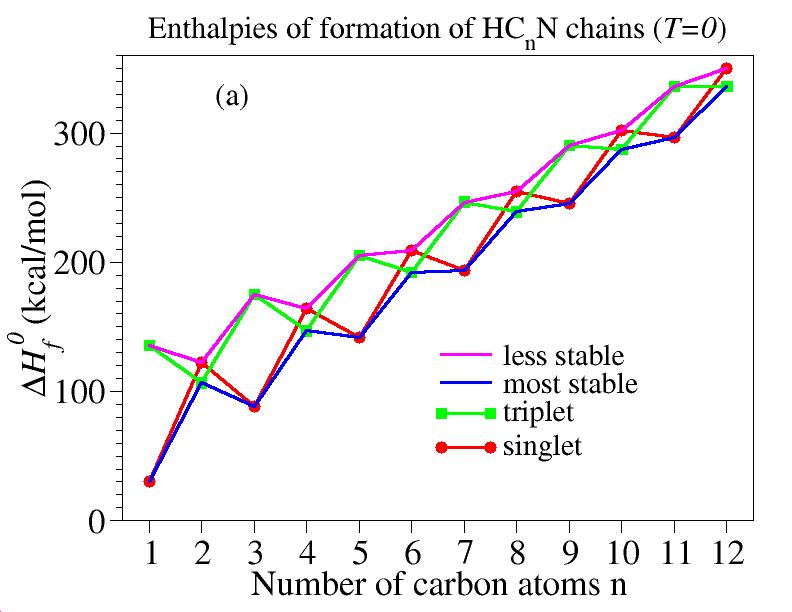}
          \includegraphics[width=0.3\textwidth,angle=0]{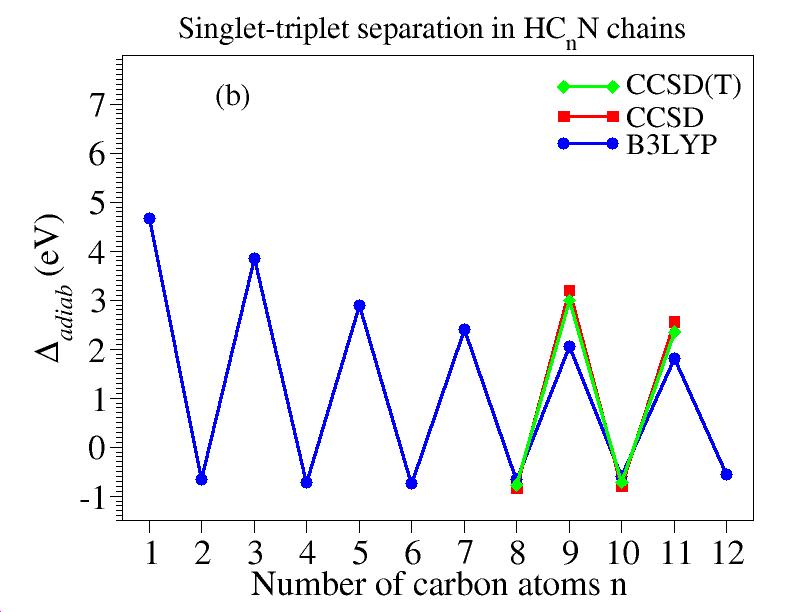}
              \includegraphics[width=0.3\textwidth,angle=0]{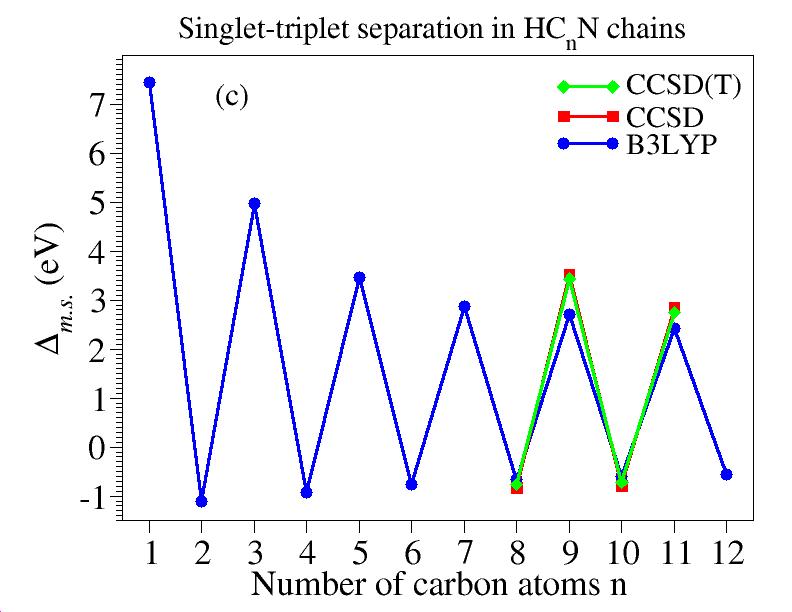}}
  \caption{Results for \ce{HC_n N} chains.
    (a) Enthalpies of formation for singlet and triplet chain isomers.
    (b) Adiabatic singlet-triplet separation energy $\Delta_{adiab}$.
    (c) Vertical singlet-triplet separation at the geometry of the most stable state $\Delta_{m.s.}$ (namely, singlet for odd members and triplet for even members,
    % \added[remark={\rem{New text added in response to the third comment of the second reviewer}}]
          {\emph{cf.}~Equation~(\ref{eq-Delta-m.s.})}).
    Lines are guide to the eye. The numerical values underlying this figure are presented in  Tables~\ref{table:H-hcxn} and \ref{table:Delta-hcxn}.
  }
  \label{fig:hcxn}
\end{figure*}
\begin{table}[htbp] % [h!]
  \scriptsize % \small % \footnotesize % \scriptsize % \footnotesize % \tiny
  \begin{center}
    \begin{tabular*}{0.47\textwidth}{@{\extracolsep{\fill}}rrrrr}
      \hline
      Molec. & Method 
      & $\Delta_{adiab}$     
      & $\Delta_{S}$     
      & $\Delta_{T}$     
      \\
      \hline
      % hcn_optimized_singlet_b3lyp_6311++g_3df3pd_n_nbo_ST_split_IP_EA_both_dft_and_ccsd_t.log, MA
      % hcn_optimized_triplet_b3lyp_6311++g_3df3pd_n_nbo_ST_split_IP_EA_both_dft_and_ccsd_t.log, MA
      \ce{HCN}   & B3LYP   &  4.670 &  7.438 &  3.141 \\
      \ce{HCN}   & B3LYP (corr)&  4.568 &  7.336 &  3.039 \\
      %%CCSD
% 
% 
      \hline
      % hc2n_optimized_singlet_b3lyp_6311++g_3df3pd_n_IP_EA_ovgf_ST_split.log ulm
      % hc2n_optimized_triplet_b3lyp_6311++g_3df3pd_n_IP_EA_ovgf_ST_split.log ulm
      \ce{HC2N}  & B3LYP   & -0.658 & -0.223 & -1.116 \\
      \ce{HC2N}  & B3LYP (corr)& -0.680 & -0.245 & -1.138 \\
      % hc2n_optimized_singlet_b3lyp_6311++g_3df3pd_n_nbo_ST_split_IP_EA_both_dft_and_ccsd_t.log MA
      % hc2n_optimized_triplet_b3lyp_6311++g_3df3pd_n_nbo_ST_split_IP_EA_both_dft_and_ccsd_t.log MA
      %%CCSD
% 
% 
      \hline
      \ce{HC3N}  & B3LYP   &  3.845 &  4.961 &  2.318 \\
      \ce{HC3N}  & B3LYP (corr)&  3.738 &  4.854 &  2.210 \\
      %%CCSD
% 
% 
      \hline
      % hc4n_optimized_singlet_b3lyp_6311++g_3df3pd_n_nbo_ST_split_IP_EA_both_dft_and_ccsd_t.log MA
      % hc4n_triplet_b3lyp_6311++g_3df3pd_n_nbo_ST_split_IP_EA_both_dft_and_ccsd_t.log MA
      \ce{HC4N}  & B3LYP   & -0.731 & -0.326 & -0.925 \\
      \ce{HC4N}  & B3LYP (corr)& -0.750 & -0.345 & -0.944 \\
      %%CCSD
% 
% 
      \hline
      % hc5n_optimized_singlet_b3lyp_6311++g_3df3pd_n_nbo_IP_EA_ST_split.log MA
      % hc5n_optimized_singlet_b3lyp_6311++g_3df3pd_n_nbo_IP_EA_ST_split.log MA
      \ce{HC5N}  & B3LYP   &  2.898 &  3.456 &  2.120 \\
      \ce{HC5N}  & B3LYP (corr)&  2.759 &  3.317 &  1.981 \\ 
      %%CCSD
% 
% 
      \hline
      % hc6n_optimized_singlet_b3lyp_6311++g_3df3pd_n_ST_split_IP_EA_both_dft_and_ccsd_t.log MA
      % hc6n_optimized_triplet_b3lyp_6311++g_3df3pd_n_nbo_ST_split_IP_EA_all_dft_triplet_ccsd_t_only.log MA
      % hc6n_optimized_triplet_b3lyp_6311++g_3df3pd_ST_split_IP_EA_ccsd_t.log MA
      % hc6n_optimized_triplet_b3lyp_6311++g_3df3pd_n_nbo_ST_split_IP_EA_both_dft_and_ccsd_t.log ulm forces ok
      \ce{HC6N}  & B3LYP   & -0.753 & -0.601 & -0.766 \\
      \ce{HC6N}  & B3LYP (corr)& -0.748 & -0.597 & -0.762 \\
      %%CCSD
% 
% 
      \hline
      % hc7n_optimized_singlet_b3lyp_6311++g_3df3pd_n_nbo_ST_split_IP_EA_both_dft_and_ccsd_t.log, MA
      % hc7n_optimized_triplet_b3lyp_6311++g_3df3pd_n_nbo_ST_split_IP_EA_both_dft_and_ccsd_t.log, MA
      \ce{HC7N}  & B3LYP   &  2.407 &  2.865 &  1.960 \\
      \ce{HC7N}  & B3LYP (corr)&  2.297 &  2.755 &  1.849 \\
      %%CCSD
% 
% 
      \hline
      % hc8n_optimized_singlet_b3lyp_6311++g_3df3pd_n_nbo_ST_split_IP_EA_both_dft_and_ccsd_t.log MA
      % hc8n_optimized_triplet_b3lyp_6311++g_3df3pd_n_nbo_ST_split_IP_EA_both_dft_and_ccsd_t.log MA
      \ce{HC8N}  & B3LYP   & -0.670 & -0.669 & -0.670 \\
      \ce{HC8N}  & B3LYP (corr)& -0.695 & -0.695 & -0.696 \\
      %%CCSD
% 
% 
      % hc8n_singlet_rob3lyp_6311++g_3df3pd_n_nbo_mayer_ST_split_IP_EA_dft_ccsd_t.log, ulm
      % hc8n_optimized_triplet_rob3lyp_6311++g_3df3pd_roh_n_nbo_mayer_ST_split_IP_EA_dft_ccsd_t.log, ulm
      \ce{HC8N} & CCSD    & -0.837 & -0.832 & -0.833 \\
      \ce{HC8N} & CCSD(T) & -0.774 & -0.766 & -0.768 \\
      % hc8n_singlet_rob3lyp_6311++g_3df3pd_n_nbo_mayer_ST_split_IP_EA_dft_ccsd_t.log, ulm
      % hc8n_triplet_rob3lyp_6311++g_3df3pd_n_nbo_mayer_ST_split_IP_EA_dft_ccsd_t.log, ulm
% 
% 

      \hline
      % hc9n_optimized_singlet_b3lyp_6311++g_3df3pd_n_nbo_ST_split_IP_EA_both_dft_and_ccsd_t.log ulm
      % hc9n_reoptimized_triplet_b3lyp_6311++g_3df3pd_n_nbo_ST_split_IP_EA_both_dft_and_ccsd_t.log ulm
      \ce{HC9N} & B3LYP   &  2.058  &  2.703 &  1.640 \\
      \ce{HC9N} & B3LYP (corr)&  1.954  &  2.599 &  1.536 \\
      %%CCSD
% 
% 
      \ce{HC9N} & CCSD    & 3.194 & 3.514 & 2.433 \\
      \ce{HC9N} & CCSD(T) & 2.999 & 3.430 & 2.494 \\
      \hline
% 
% 
% 
      % hc10n_optimized_singlet_b3lyp_6311++g_3df3pd_n_nbo_ST_split_IP_EA_both_dft_and_ccsd_t.log ulm
      % hc10n_optimized_triplet_b3lyp_6311++g_3df3pd_n_nbo_ST_split_IP_EA_both_dft_and_ccsd_t.log ulm
      \ce{HC10N} & B3LYP   & -0.607 & -0.606 & -0.607 \\
      \ce{HC10N} & B3LYP (corr)& -0.636 & -0.636 & -0.637 \\
      %%CCSD
% 
% 
      % hc10n_singlet_rob3lyp_6311++g_3df3pd_n_nbo_ST_split_IP_EA_both_dft_and_ccsd_t.log, ulm
      % hc10n_triplet_rob3lyp_6311++g_3df3pd_n_nbo_ST_split_IP_EA_both_dft_and_ccsd_t.log, ulm
      \ce{HC10N} & CCSD    & -0.807 & -0.787 & -0.789 \\
      \ce{HC10N} & CCSD(T) & -0.728 & -0.722 & -0.723 \\
      \hline
      % hc11n_optimized_singlet_b3lyp_6311++g_3df3pd_coord_n_nbo_ST_split.log, ulm
      % hc11n_optimized_triplet_b3lyp_6311++g_3df3pd_dft_ccsd_t_nbo_ST_split_IP_EA.log ulm
      \ce{HC11N} & B3LYP   &  1.815  & 2.413 &  1.411 \\
      \ce{HC11N} & B3LYP (corr)&  1.711  & 2.309 &  1.307 \\
      % hc11n_optimized_singlet_b3lyp_6311++g_3df3pd_n_nbo_ST_split_IP_EA_both_dft_and_ccsd_t.log ulm
      % hc11n_optimized_triplet_b3lyp_6311++g_3df3pd_dft_ccsd_t_nbo_ST_split.log ulm
      % hc11n_optimized_triplet_b3lyp_6311++g_3df3pd_ccsd_t_IP_EA.log ulm
      %%CCSD
% 
% 
      % hc11n_singlet_b3lyp_6311++g_3df3pd_n_nbo_mayer_ST_split_IP_EA_dft_ccsd_t.log, ulm, new (not labeled but it is roccsd!)
      % hc11n_optimized_triplet_rob3lyp_6311++g_3df3pd_roh_n_nbo_mayer_ST_split_IP_EA_dft_ccsd_t.log, ulm
      \ce{HC11N} & CCSD    & 2.551 & 2.839 & 1.721 \\
      \ce{HC11N} & CCSD(T) & 2.350 & 2.742 & 1.813 \\
      \hline
      % hc12n_optimized_singlet_b3lyp_6311++g_3df3pd_dft_ccsd_t_nbo_ST_split_IP_EA.log ulm
      % hc12n_b3lyp_triplet_6311++g_3df3pd_zmat_n_IP_EA.log, MA (corrections ok)
      \ce{HC12N} & B3LYP   & -0.563 & -0.562 & -0.563 \\
      \ce{HC12N} & B3LYP (corr)& -0.595 & -0.594 & -0.595 \\
      % hc12n_optimized_singlet_b3lyp_6311++g_3df3pd_dft_ccsd_t_nbo_ST_split_IP_wo_EA.log ulm ForcesNotChecked!
      % hc12n_optimized_triplet_b3lyp_6311++g_3df3pd_dft_ccsd_t_nbo_ST_split_IP_wo_EA.log ulm ForcesNotChecked!
      %%CCSD
% 
% 
      \hline
    \end{tabular*}      
    %%%%%%%%%%%%%%%%%%%%%%%%%%%%%%%%%%%%%%%%%%%%%%%%%%%%%%%%%%%%%%%%%%%%%%%%%%%%%%%%%%%%%%
    %%%%%%%%%%%%%%%%%%%%%%%%%%%%%%%%%%%%%%%%%%%%%%%%%%%%%%%%%%%%%%%%%%%%%%%%%%%%%%%%%%%%%%
    \caption{
      Adiabatic ($\Delta_{adiab}$) and vertical ($\Delta_{S,T}$) values of the singlet-triplet
      energy separation (in eV) for \ce{HC_{n} N} chains obtained within the methods indicated in the second column.
      The two vertical values shown here correspond to the optimized singlet ($\Delta_{S}$)
      and triplet ($\Delta_{T}$) geometries. Notice that the sign of $\Delta$
      indicates that the most stable isomers are singlets for odd members
      ($\Delta > 0$, \ce{HC_{2k+1} N}) and triplets for even members ($\Delta < 0$, \ce{HC_{2k} N}).
      Corrections due to zero-point motion (label \emph{corr}) were deduced within the DFT/B3LYP
      approach;
      % \added[remark={\rem{New text added in response to the first comment of the second reviewer}}]
            {see the last paragraph of \secname~\ref{sec:methods}.}
            % \added[remark={\rem{New text added in response to the second comment of the second reviewer}}]
                  {The almost equal values $\Delta_S \approx \Delta_T$ for larger even-members
                    (\ce{HC_{n} N}, $n=8,10,12$)
                    indicate that the singlet and triplet geometries are similar ($\mathbf{R}_S \approx \mathbf{R}_T$,
                    \emph{cf.}~Equation~(\ref{eq-Delta-S,T})).}
      %      \ib{HC$_{2n}$N triplet}, HC$_{2n+1}$N singlet. triplet \ce{HC3N} ulm and MA are different; their IR spectra are also different! IR spectra of s- and t-\ce{HC12N} exhibit some differences. Singlet and triplet \ce{HC4N} have significantly different (CCSD) IR spectra.
    }
    \label{table:Delta-hcxn}
\end{center}
\end{table}
\subsection{C$_{n}$S Homologous Series} % \subsection{\ce{C_{n} S} Homologous Series}
\label{sec:cxs}
In the examples presented in
{\secsname}\ref{sec:hcxh} and \ref{sec:hcxn},
we have seen that,
in contrast to (either even or odd) members of the same homologous series
consisting of ``normal'' (nonradical) species (\ce{HC_{2k} H} and \ce{HC_{2k+1} N}, respectively),
``non-normal'' (diradical) \cite{Abe:13} carbon chains
wherein the alternation of single and triple carbon-carbon bonds
(\ce{HC_{2k+1} H} and \ce{HC_{2k} N}, respectively)
is incompatible with standard
valence considerations possess triplet electronic ground states. Does this diradical character
necessarily make carbon-based chains adopting triplet ground states?

To answer this question, it is meaningful to investigate carbon-based chains wherein both even and odd members
of the molecular series possess two electrons that cannot be involved in covalent bonds.
Linear carbon chains terminated with a sulfur atom \ce{C_nS}
(\ce{S\bond{2}C\bond{2}...\bond{2}}\rdirad{C}), 
which we next consider and were also reported in astronomical observations 
(\emph{e.g.}, refs.~\citenum{Matthews:84,Cernicharo:87,Agundez:14}),
belong to this category.

As exemplified with the aid of the cases of the \ce{C6S} and \ce{C7S} chains
depicted in \figname\ref{fig:bonds-c6s-c7s}, calculations confirm the
cumulene-type structure both for even and odd members. This result is not surprising;
it can be expected based on chemical intuition. However, what is nontrivial is the fact
that 
the state corresponding to the lowest electronic energy is found to be a singlet state for odd members
\ce{C_{2k+1}S}, while for even members \ce{C_{2k}S} the lowest state is a triplet.
Indeed, the results of our calculations presented in
Table~\ref{table:H-cxs} and \figname\ref{fig:cxs}a show that the lower
enthalpies of formation correspond to singlet isomers for odd members \ce{C_{2k+1}S}
and to triplet isomers for even members \ce{C_{2k}S}. Alternatively rephrased,
the negative values of the singlet-triplet energy splitting $\Delta$ demonstrate
that the triplet state is the most stable for even members (\ce{C_{2k}S}),
in contrast to the odd members (\ce{C_{2k+1}S}) for which the most stable is a singlet,
as expressed by the positive $\Delta$-values (\emph{cf.}~Table~\ref{table:Delta-cxs}
and \figname\ref{fig:cxs}b and c).
\begin{figure*} % {hbtp}
  \centerline{\includegraphics[width=0.45\textwidth,angle=0]{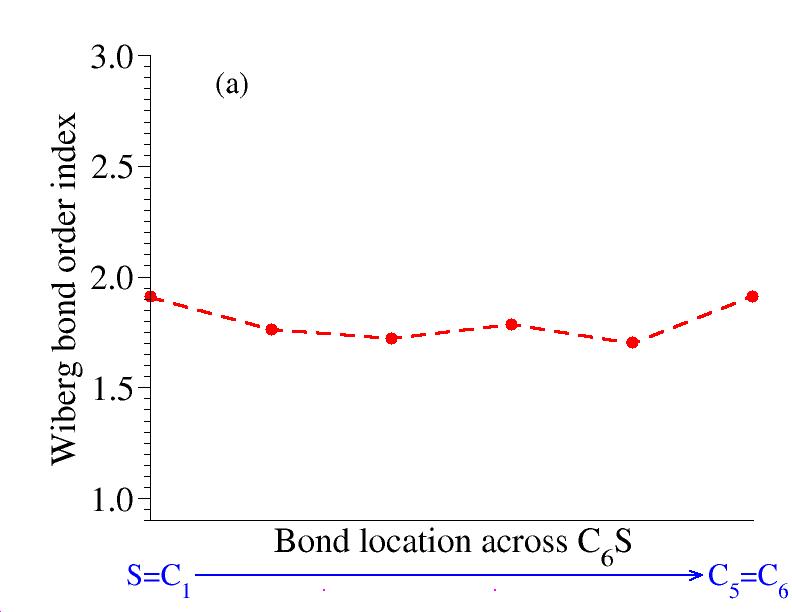}
    \includegraphics[width=0.45\textwidth,angle=0]{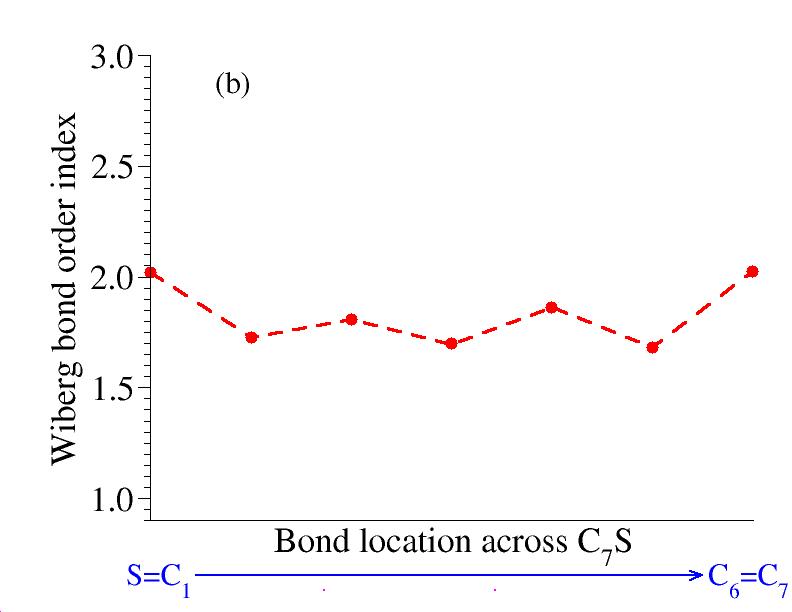}}
  \caption{Wiberg bond order indices for \ce{C_{n} S} chains:
    (a) triplet \ce{C6S} and (b) singlet \ce{C7S}.
    The coordinates of these molecules
    at the corresponding energy minima as well as  the HOMO spatial distributions are presented in
    Tables~S14 and S15  % Tables~\ref{table:c6s-xyz-triplet} and \ref{table:c7s-xyz-singlet},
    and in
    {\figsname}S5 and S6, % \figname\ref{fig:homo-c6s} and \figname\ref{fig:homo-c7s},
    respectively.
  }
  \label{fig:bonds-c6s-c7s}
\end{figure*}
\begin{table}[htbp] % [h!]
  \scriptsize % \small % \scriptsize % \footnotesize % \tiny
  \begin{center}
    \begin{threeparttable}
  %%%%%%%%%%%%%%%%%%%%%%%%%%%%
    \begin{tabular*}{0.47\textwidth}{@{\extracolsep{\fill}}lrrrr}
      \hline
      Molec.
      & $\Delta_{f} H^{0}_{0} $ & $\Delta_{f} H^{0}_{RT}\vert_{S}$
      & $\Delta_{f} H^{0}_{0} $ & $\Delta_{f} H^{0}_{RT}\vert_{T}$
      \\
      \hline
      % cs_singlet_b3lyp_6311++g_3df3pd_n_nbo_ST_split_IP_EA_both_dft_and_ccsd_t.log, ulm
      % cs_triplet_b3lyp_6311++g_3df3pd_n_nbo_ST_split_IP_EA_both_dft_and_ccsd_t.log, ulm
      \ce{CS}     &  70.934 &  69.634 & 146.830 & 147.617 \\
      \ce{C2S}    & 162.418 & 163.565 & 145.630 & 146.834 \\
      \ce{C3S}    & 135.505 & 136.805 & 190.378 & 191.944 \\
      \ce{C4S}    & 196.846 & 198.489 & 182.693 & 184.308 \\
      \ce{C5S} \tnote{$\ast $}   & 191.231 & 193.097 & 232.108 & 234.229 \\
      \ce{C6S}    & 243.319 & 245.553 & 232.214 & 234.417 \\
      \ce{C7S}    & 245.291 & 247.800 & 276.777 & 279.455 \\
      \hline
    \end{tabular*}      
    %%%%%%%%%%%%%%%%%%%%%%%%%%%%%%%%%%%%%%%%%%%%%%%%%%%%%%%%%%%%%%%%%%%%%%%%%%%%%%%%%%%%%%
    \begin{tablenotes}\footnotesize
    \item[$\ast $]
      Longest chain of this family astronomically observed \cite{Cernicharo:87,Agundez:14} % \ce{C5S} 
    \end{tablenotes}
    \end{threeparttable}
    %%%%%%%%%%%%%%%%%%%%%%%%%%%%%%%%%%%%%%%%%%%%%%%%%%%%%%%%%%%%%%%%%%%%%%%%%%%%%%%%%%%%%%
    \caption{Enthalpies of formation of linear \ce{C_nS} chains at zero and room temperature (subscript $0$ and $RT$, respectively). Notice that for the odd members 
      \ce{C_{2k+1} S} the values for singlet (label $S$) are smaller than those for triplet (label $T$),
      while for the even members \ce{C_{2k} S} the values for triplet are smaller than those for singlet.
    } 
    \label{table:H-cxs}
  \end{center}
\end{table}
\begin{figure*} % {hbtp}
  \centerline{\includegraphics[width=0.3\textwidth,angle=0]{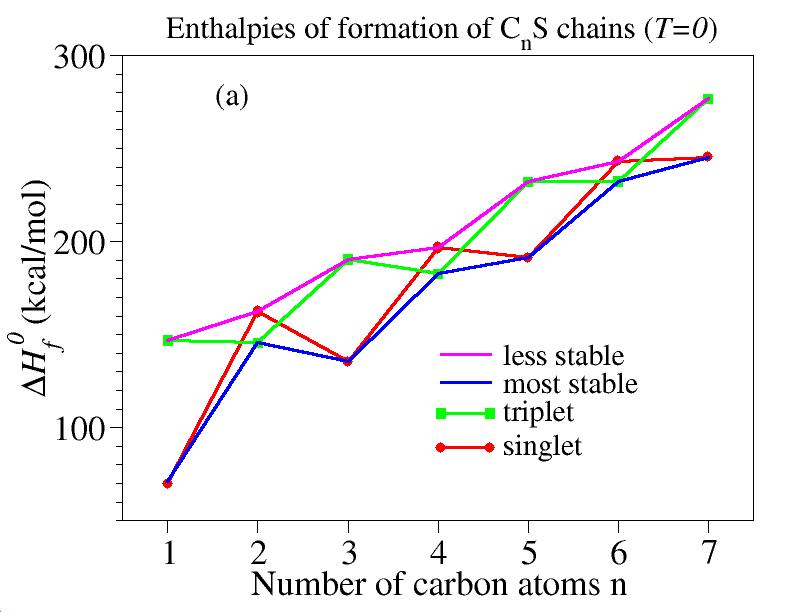}
    \includegraphics[width=0.3\textwidth,angle=0]{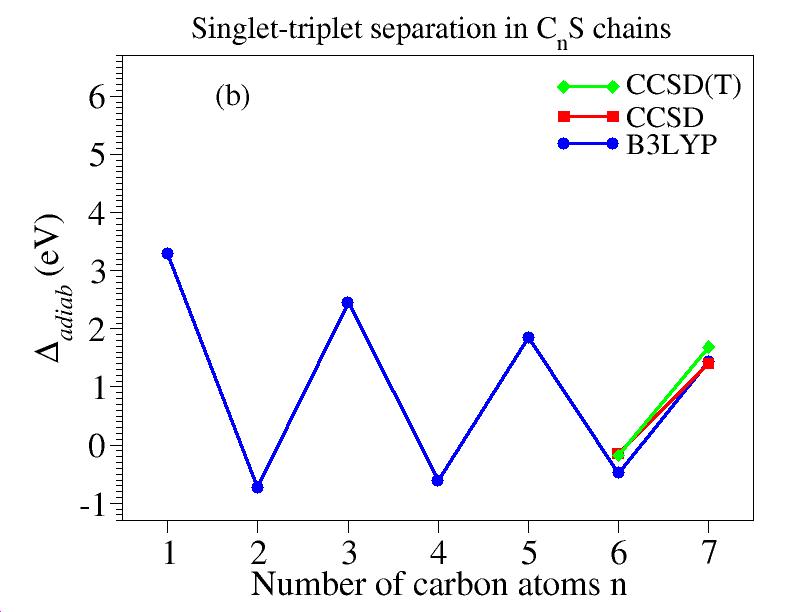}
              \includegraphics[width=0.3\textwidth,angle=0]{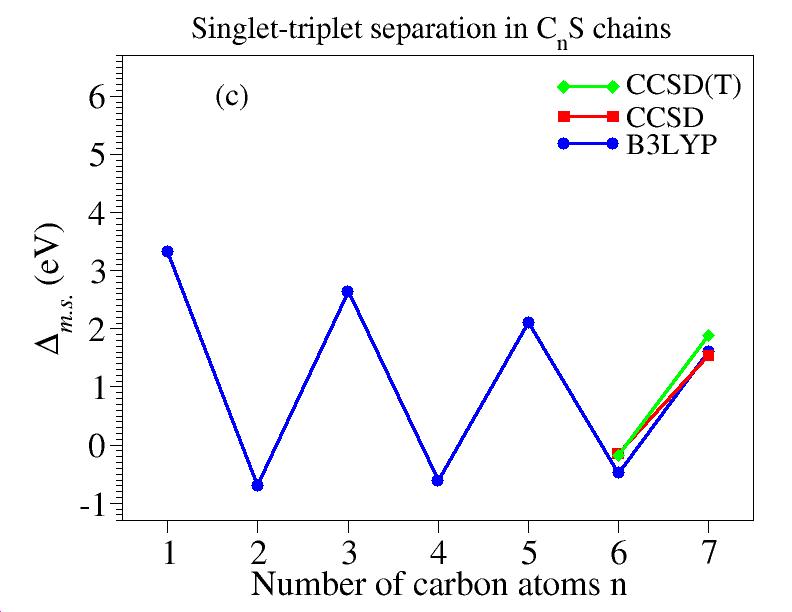}}
  \caption{Results for \ce{C_n S} chains. 
    (a) Enthalpies of formation for singlet and triplet chain isomers.
    (b) Adiabatic singlet-triplet separation energy $\Delta_{adiab}$.
    (c) Vertical singlet-triplet separation at the geometry of the most stable state $\Delta_{m.s.}$ (namely, singlet for odd members and triplet for even members,
    % \added[remark={\rem{New text added in response to the third comment of the second reviewer}}]
          {\emph{cf.}~Equation~(\ref{eq-Delta-m.s.})}).
    Lines are guide to the eye. The numerical values underlying this figure are presented in Tables~\ref{table:H-cxs} and \ref{table:Delta-cxs}.
  }
  \label{fig:cxs}
\end{figure*}
\begin{table}[htbp] % [h!]
  \scriptsize % \small % \footnotesize % \scriptsize % \footnotesize % \tiny
  \begin{center}
    \begin{tabular*}{0.47\textwidth}{@{\extracolsep{\fill}}rrrrr}
      \hline
      Molec. & Method 
      & $\Delta_{adiab}$     
      & $\Delta_{S}$     
      & $\Delta_{T}$     
      \\
      \hline
      \ce{CS}    & B3LYP   &  3.302 &  3.330 &  3.268 \\
      \ce{CS}    & B3LYP (corr)&  3.291 &  3.320 &  3.258 \\
      \hline
      \ce{C2S}   & B3LYP   & -0.723 & -0.691 & -0.700 \\  
      \ce{C2S}   & B3LYP (corr)& -0.728 & -0.696 & -0.705 \\
      \hline
      \ce{C3S}   & B3LYP   &  2.447 &  2.644 &  1.412 \\  
      \ce{C3S}   & B3LYP (corr)&  2.380 &  2.576 &  1.345 \\ 
      \hline
      \ce{C4S}   & B3LYP   & -0.613 & -0.613 & -0.614 \\  
      \ce{C4S}   & B3LYP (corr)& -0.614 & -0.613 & -0.614 \\  
      \hline
      \ce{C5S}   & B3LYP   &  1.848 &  2.107 &  1.787 \\  
      \ce{C5S}   & B3LYP (corr)&  1.773 &  2.031 &  1.711 \\  
      \hline
      \ce{C6S}   & B3LYP   & -0.480 & -0.480 & -0.481 \\  
      \ce{C6S}   & B3LYP (corr)& -0.482 & -0.481 & -0.482 \\ 
% 
% 
      % c6s_optimized_triplet_rob3lyp_6311++g_3df3pd_roh_n_nbo_mayer_ST_split_IP_EA_dft_ccsd_t_wo_IP_EA_ccsd_t.log, ulm
      % c6s_triplet_rob3lyp_6311++g_3df3pd_n_nbo_mayer_ST_split_IP_EA_dft_ccsd_t_wo_IP_EA_ccsd_t.log, ulm
      \ce{C6S} & CCSD    & -0.140 & -0.140 & -0.141 \\
      \ce{C6S} & CCSD(T) & -0.175 & -0.177 & -0.177 \\
      \hline
      \ce{C7S}   & B3LYP   &  1.442 &  1.615 &  1.407 \\
      \ce{C7S}   & B3LYP (corr)&  1.365 &  1.539 &  1.331 \\
      % 
% 
% 
      % c7s_singlet_rob3lyp_6311++g_3df3pd_n_nbo_mayer_ST_split_IP_EA_dft_ccsd_t.log, ulm
      % c7s_optimized_triplet_rob3lyp_6311++g_3df3pd_roh_n_nbo_mayer_ST_split_IP_EA_dft_ccsd_t.log, ulm
      \ce{C7S}  & CCSD    & 1.406 & 1.544 & 1.363 \\
      \ce{C7S}  & CCSD(T) & 1.687 & 1.882 & 1.691 \\
      \hline
    \end{tabular*}      
    %%%%%%%%%%%%%%%%%%%%%%%%%%%%%%%%%%%%%%%%%%%%%%%%%%%%%%%%%%%%%%%%%%%%%%%%%%%%%%%%%%%%%%
    %%%%%%%%%%%%%%%%%%%%%%%%%%%%%%%%%%%%%%%%%%%%%%%%%%%%%%%%%%%%%%%%%%%%%%%%%%%%%%%%%%%%%%
    \caption{
      Adiabatic ($\Delta_{adiab}$) and vertical ($\Delta_{S,T}$) values of the singlet-triplet
      energy separation (in eV) for \ce{C_{n} S} chains obtained within the methods indicated in the second column.
      The two vertical values shown here correspond to the optimized singlet ($\Delta_{S}$)
      and triplet ($\Delta_{T}$) geometries. Notice that the sign of $\Delta$
      indicates that the most stable isomers are singlets for odd members
      ($\Delta > 0$, \ce{C_{2k+1} S}) and triplets for even members ($\Delta < 0$, \ce{C_{2k} S}).
      Corrections due to zero-point motion (label \emph{corr}) were deduced within the DFT/B3LYP
      approach;
      % \added[remark={\rem{New text added in response to the first comment of the second reviewer}}]
            {see the last paragraph of \secname~\ref{sec:methods}.}
    }
    \label{table:Delta-cxs}
  \end{center}
\end{table}
\subsection{C$_{n}$O Homologous Series} % \subsection{\ce{C_{n} O} Homologous Series}
\label{sec:cxo}
Let us next consider linear carbon chains terminated with an oxygen atom (\ce{C_nO}).
Similar to \ce{C_{n} S}, \ce{C_{n} O} chains also possess two electrons that cannot be involved in covalent bonds.
Such chains were also reported in astronomical observations.\cite{Matthews:84}

Calculations confirm again the cumulene-type structure
of the ground state \ce{O\bond{2}C\bond{2}...\bond{2}}\rdirad{C}
irrespective whether the number of carbon atoms is even or odd.
\figname\ref{fig:bonds-c6o-c7o} illustrates this fact for the specific
case of \ce{C6O} and \ce{C7O} chains. Inspection reveals only slight differences
between this figure and \figname\ref{fig:bonds-c6s-c7s}, which refers to the
isoelectronic \ce{C6S} and \ce{C7S} chains.
The slightly broader range of the bond order index variation in \figname\ref{fig:bonds-c6o-c7o}
as compared to \figname\ref{fig:bonds-c6s-c7s}
can be attributed to the oxygen electronegativity ($\chi_{O}^{Pauling}=3.44$, $\chi_{O}^{Allen}=3.610$),
which is larger than the sulfur electronegativity ($\chi_{S}^{Pauling}=2.58$, $\chi_{S}^{Allen}=2.589$).
\begin{figure*} % {hbtp}
  \centerline{\includegraphics[width=0.45\textwidth,angle=0]{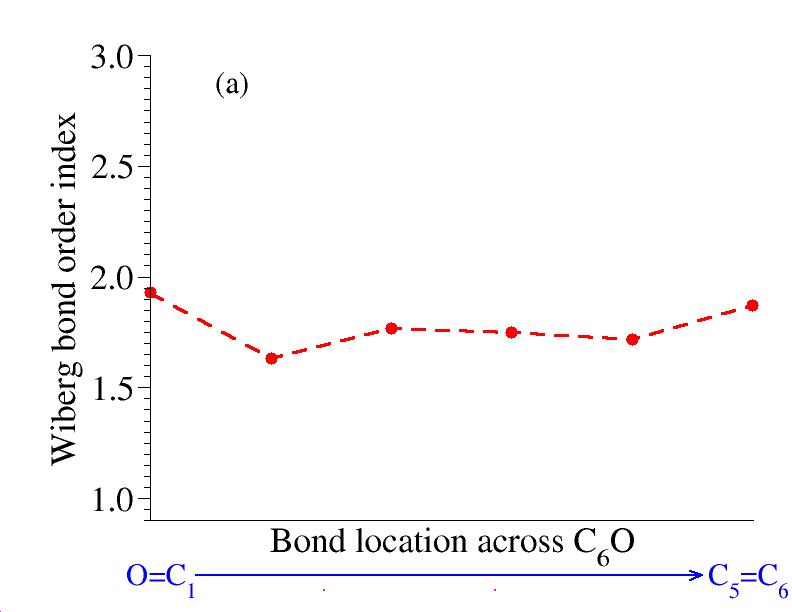}
    \includegraphics[width=0.45\textwidth,angle=0]{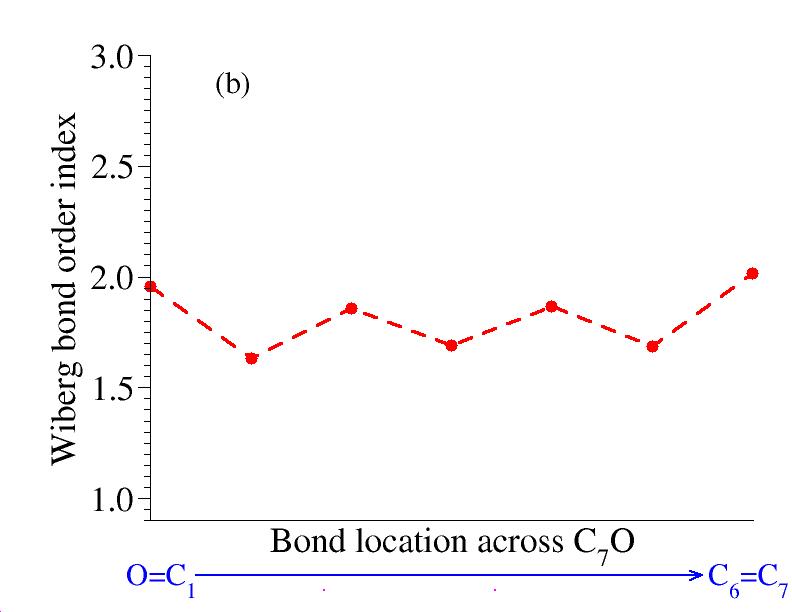}}
  \caption{Wiberg bond order indices for \ce{C_{n} O} chains:
    (a) triplet \ce{C6O} and (b) singlet \ce{C7O}.
    The coordinates of these molecules
    at the corresponding energy minima as well as  the HOMO spatial distributions are presented in
    Tables~S18 and S19 % Tables~\ref{table:c6o-xyz-triplet} and \ref{table:c7o-xyz-singlet},
    and in
    {\figsname}S7 and S8, % \figname\ref{fig:homo-c6o} and \figname\ref{fig:homo-c7o},
    respectively.
  }
  \label{fig:bonds-c6o-c7o}
\end{figure*}
Calculations also confirm the alternation of the singlet and triplet states 
of the \ce{C_nO} chains with increasing number of carbon atoms found for the
isoelectronic \ce{C_nS} chains reported in
{\secname}\ref{sec:cxs}.
The values of the enthalpies of formations
(Table~\ref{table:H-cxo} and \figname\ref{fig:cxo}a) and the singlet-triplet separations
(Table~\ref{table:Delta-cxo} and \figsname\ref{fig:cxo}b and \ref{fig:cxo}c) demonstrate
that, for even members (\ce{C_{2k} O}), the triplet state is more stable than the singlet state
while the opposite holds true for the odd members (\ce{C_{2k+1} O}).
\begin{table}[htbp] % [h!]
  \scriptsize % \small % \scriptsize % \footnotesize % \tiny
  \begin{center}
    \begin{threeparttable}
  %%%%%%%%%%%%%%%%%%%%%%%%%%%%
    \begin{tabular*}{0.47\textwidth}{@{\extracolsep{\fill}}lrrrr}
      \hline
      Molec.
      & $\Delta_{f} H^{0}_{0} $ & $\Delta_{f} H^{0}_{RT}\vert_{S}$
      & $\Delta_{f} H^{0}_{0} $ & $\Delta_{f} H^{0}_{RT}\vert_{T}$
      \\
      \hline
      % co_singlet_b3lyp_6311++g_3df3pd_n_nbo_ST_split_IP_EA_both_dft_and_ccsd_t_ovgf.log MA
      % co_triplet_b3lyp_6311++g_3df3pd_n_nbo_ST_split_IP_EA_both_dft_and_ccsd_t_ovgf.log MA
      \ce{CO}     &  -24.140 & -23.356  & 110.602 & 111.387 \\
      % oc2_singlet_b3lyp_6311++g_3df3pd_n_nbo_ST_split_IP_EA_both_dft_and_ccsd_t.log MA, forces ok
      % oc2_triplet_b3lyp_6311++g_3df3pd_n_nbo_ST_split_IP_EA_both_dft_and_ccsd_t.log MA, forces ok
      \ce{C2O}    &  110.227 &  111.332 &  85.497 &  86.550 \\
% 
      % oc3_singlet_b3lyp_6311++g_3df3pd_n_nbo_ST_split_IP_EA_both_dft_and_ccsd_t.log MA
      % oc3_triplet_b3lyp_6311++g_3df3pd_n_nbo_ST_split_IP_EA_both_dft_and_ccsd_t.log MA
      \ce{C3O} \tnote{$\ast $}   &   76.170 &   77.389 & 143.236 & 144.734 \\  
      % oc4_singlet_b3lyp_6311++g_3df3pd_n_nbo_ST_split_IP_EA_both_dft_and_ccsd_t.log MA
      % oc4_triplet_b3lyp_6311++g_3df3pd_n_nbo_ST_split_IP_EA_both_dft_and_ccsd_t.log MA
% 
      \ce{C4O}    &  149.118 &  150.669 & 133.039 & 134.532 \\ 
% 
      % c4o_singlet_cbs-qb3.log, ulm: s=-227.459189391; zmps=0.018336; ens=0.024427; hms=-227.440854; zpm=zpms; en=ens; hm=hms;
      % c4o_triplet_cbs-qb3.log, ulm: t=-227.484916268; zpmt=0.019002; ent=0.024745; hmt=-227.465914; zpm=zpmt; en=ent; hm=hmt;
% 
% 
% 
      % oc5_singlet_b3lyp_6311++g_3df3pd_n_nbo_ST_split_IP_EA_both_dft_and_ccsd_t.log MA
      % oc5_triplet_b3lyp_6311++g_3df3pd_n_nbo_ST_split_IP_EA_both_dft_and_ccsd_t.log MA
      \ce{C5O}    &  135.765 &  137.516 & 186.814 & 188.994 \\
      % c5o_singlet_cbs-qb3.log, ulm: s=-265.611953557; zpms=0.025391; ens=0.031964; hms=-265.155963; zpm=zpms; en=ens; hm=hms;
      % c5o_triplet_cbs-qb3.log, ulm: t=-265.526867975; zmpt=0.021040; ent=0.028646; hmt=-265.505828; zpm=zpmt; en=ent; hm=hmt;
% 
% 
% 
% 
      % oc6_singlet_b3lyp_6311++g_3df3pd_n_nbo_ST_split_IP_EA_both_dft_and_ccsd_t.log MA
      % oc6_triplet_b3lyp_6311++g_3df3pd_n_nbo_ST_split_IP_EA_both_dft_and_ccsd_t.log MA
      \ce{C6O}    &  193.343 &  195.458 & 181.261 & 183.327 \\
% 
% 
% 
% 
% 
      % oc7_singlet_b3lyp_6311++g_3df3pd_n_nbo_ST_split_IP_EA_both_dft_and_ccsd_t.log MA
      % oc7_triplet_b3lyp_6311++g_3df3pd_n_nbo_ST_split_IP_EA_both_dft_and_ccsd_t.log MA
      \ce{C7O}    &  190.940 &  193.324 & 228.914 & 231.450 \\
      % oc8_singlet_b3lyp_6311++g_3df3pd_n_nbo_ST_split_IP_EA_both_dft_and_ccsd_t.log MA
      % oc8_triplet_b3lyp_6311++g_3df3pd_n_nbo_ST_split_IP_EA_both_dft_and_ccsd_t.log MA
% 
      % oc9_singlet_b3lyp_6311++g_3df3pd_n_nbo_ST_split_IP_EA_both_dft_and_ccsd_t.log ulm
      % oc9_triplet_b3lyp_6311++g_3df3pd_n_nbo_ST_split_IP_EA_both_dft_and_ccsd_t.log ulm
% 
      % oc10_singlet_b3lyp_6311++g_3df3pd_n_nbo_ST_split_IP_EA_both_dft_and_ccsd_t.log ulm
      % oc10_triplet_b3lyp_6311++g_3df3pd_n_nbo_ST_split_IP_EA_both_dft_and_ccsd_t.log ulm
% 
      % oc11_singlet_b3lyp_6311++g_3df3pd_n_nbo_ST_split_IP_EA_both_dft_and_ccsd_t.log ulm
      % oc11_triplet_b3lyp_6311++g_3df3pd_n_nbo_ST_split_IP_EA_both_dft_and_ccsd_t.log ulm
% 
      % oc12_singlet_b3lyp_6311++g_3df3pd_n_nbo_ST_split_IP_EA_both_dft_and_ccsd_t.log ulm
      % oc12_triplet_b3lyp_6311++g_3df3pd_IP_EA_ccsd_t.log ulm
% 
      \hline
    \end{tabular*}      
    %%%%%%%%%%%%%%%%%%%%%%%%%%%%%%%%%%%%%%%%%%%%%%%%%%%%%%%%%%%%%%%%%%%%%%%%%%%%%%%%%%%%%%
    \begin{tablenotes}\footnotesize
    \item[$\ast $] Longest chain of this family astronomically observed \cite{Matthews:84} % \ce{C3O} 
    \end{tablenotes}
    \end{threeparttable}
    %%%%%%%%%%%%%%%%%%%%%%%%%%%%%%%%%%%%%%%%%%%%%%%%%%%%%%%%%%%%%%%%%%%%%%%%%%%%%%%%%%%%%%
    \caption{Enthalpies of formation of linear \ce{C_nO} chains at zero and room temperature (subscript $0$ and $RT$, respectively). Notice that for the odd members 
      \ce{C_{2k+1} O} the values for singlet (label $S$) are smaller than those for triplet (label $T$),
      while for the even members \ce{C_{2k} O} the values for triplet are smaller
      than those for singlet.
    }
    \label{table:H-cxo}
  \end{center}
\end{table}
\begin{figure*} % {hbtp}
  \centerline{\includegraphics[width=0.3\textwidth,angle=0]{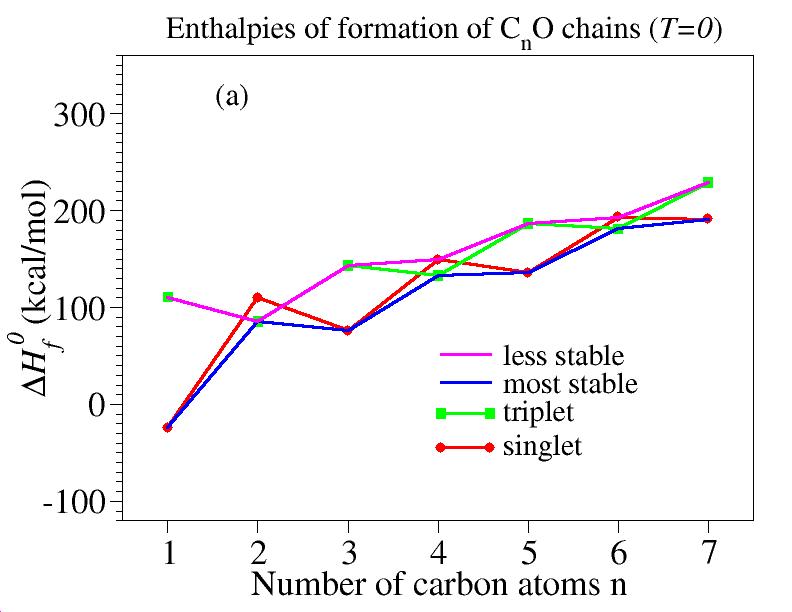}
      \includegraphics[width=0.3\textwidth,angle=0]{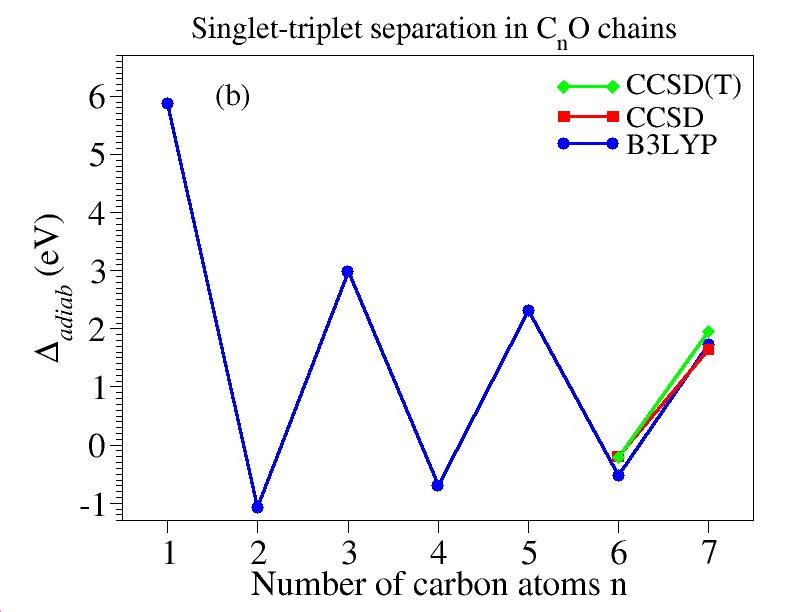}
  \includegraphics[width=0.3\textwidth,angle=0]{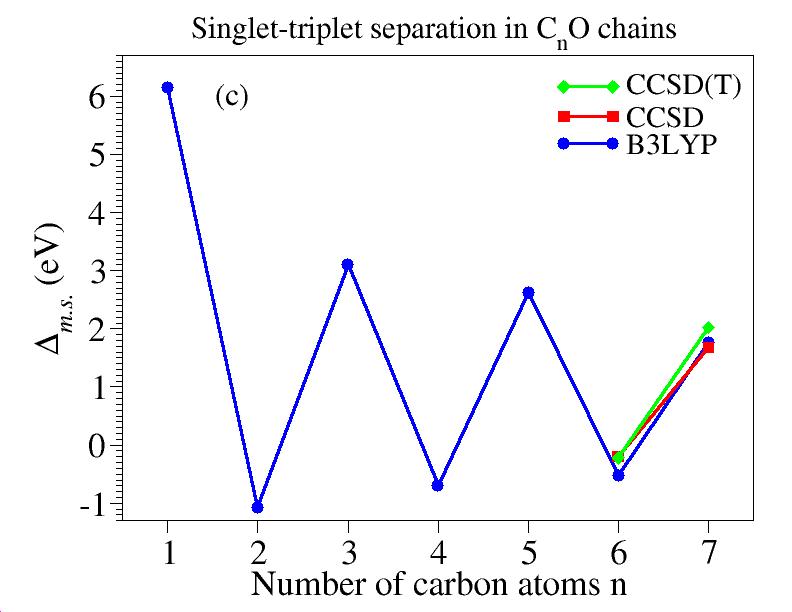}}
  \caption{Results for \ce{C_n O} chains.
    (a) Enthalpies of formation for singlet and triplet chain isomers.
    (b) Adiabatic singlet-triplet separation energy $\Delta_{adiab}$.
    (c) Vertical singlet-triplet separation at the geometry of the most stable state $\Delta_{m.s.}$ (namely, singlet for odd members and triplet for even members,
    % \added[remark={\rem{New text added in response to the third comment of the second reviewer}}]
          {\emph{cf.}~Equation~(\ref{eq-Delta-m.s.})}).
     Lines are guide to the eye. The numerical values underlying this figure are presented in  Tables~\ref{table:H-cxo} and \ref{table:Delta-cxo}.
  }
  \label{fig:cxo}
\end{figure*}
\begin{table}[htbp] % [h!]
  \scriptsize % \small % \footnotesize % \scriptsize % \footnotesize % \tiny
  \begin{center}
    \begin{tabular*}{0.47\textwidth}{@{\extracolsep{\fill}}rrrrr}
      \hline
      Molec. & Method 
      & $\Delta_{adiab}$     
      & $\Delta_{S}$     
      & $\Delta_{T}$     
      \\
      \hline
      % co_singlet_b3lyp_6311++g_3df3pd_n_nbo_ST_split_IP_EA_both_dft_and_ccsd_t_ovgf.log MA
      % co_triplet_b3lyp_6311++g_3df3pd_n_nbo_ST_split_IP_EA_both_dft_and_ccsd_t_ovgf.log MA
      \ce{CO}    & B3LYP   &  5.869  &   6.148 &  5.572 \\
      \ce{CO}    & B3LYP (corr)&  5.843  &   6.122 &  5.545 \\
      %%CCSD
% 
% 
      \hline
      % oc2_singlet_b3lyp_6311++g_3df3pd_n_nbo_ST_split_IP_EA_both_dft_and_ccsd_t.log MA, forces ok
      % oc2_triplet_b3lyp_6311++g_3df3pd_n_nbo_ST_split_IP_EA_both_dft_and_ccsd_t.log MA, forces ok
      \ce{C2O}   & B3LYP   & -1.078  &  -1.063 & -1.072 \\
      \ce{C2O}   & B3LYP (corr)& -1.072  &  -1.058 & -1.067 \\
      %%CCSD
% 
% 
      \hline
      % oc3_singlet_b3lyp_6311++g_3df3pd_n_nbo_ST_split_IP_EA_both_dft_and_ccsd_t.log MA
      % oc3_triplet_b3lyp_6311++g_3df3pd_n_nbo_ST_split_IP_EA_both_dft_and_ccsd_t.log MA
      \ce{C3O}   & B3LYP   &  2.988  &   3.104 &  2.143 \\
      \ce{C3O}   & B3LYP (corr)&  2.908  &   3.024 &  2.063 \\
      %%CCSD
% 
% 
      \hline
      % oc4_singlet_b3lyp_6311++g_3df3pd_n_nbo_ST_split_IP_EA_both_dft_and_ccsd_t.log MA
      % oc4_triplet_b3lyp_6311++g_3df3pd_n_nbo_ST_split_IP_EA_both_dft_and_ccsd_t.log MA
      \ce{C4O}   & B3LYP   &  -0.698 &  -0.698 & -0.699 \\
      \ce{C4O}   & B3LYP (corr)&  -0.697 &  -0.697 & -0.698 \\
      %%CCSD
% 
% 
      \hline
      % oc5_singlet_b3lyp_6311++g_3df3pd_n_nbo_ST_split_IP_EA_both_dft_and_ccsd_t.log MA
      % oc5_triplet_b3lyp_6311++g_3df3pd_n_nbo_ST_split_IP_EA_both_dft_and_ccsd_t.log MA
      \ce{C5O}   & B3LYP   &   2.314 &   2.626 &  2.221 \\
      \ce{C5O}   & B3LYP (corr)&   2.214 &   2.525 &  2.121 \\
      %%CCSD
% 
% 
      \hline
      % oc6_singlet_b3lyp_6311++g_3df3pd_n_nbo_ST_split_IP_EA_both_dft_and_ccsd_t.log MA
      % oc6_triplet_b3lyp_6311++g_3df3pd_n_nbo_ST_split_IP_EA_both_dft_and_ccsd_t.log MA
      \ce{C6O}   & B3LYP   &  -0.523 &  -0.523 & -0.524 \\     
      \ce{C6O}   & B3LYP (corr)&  -0.524 &  -0.523 & -0.524 \\
      % oc6_singlet_b3lyp_6311++g_3df3pd_n_nbo_ST_split_IP_EA_both_dft_and_ccsd_t.log MA
      % oc6_triplet_b3lyp_6311++g_3df3pd_n_nbo_ST_split_IP_EA_both_dft_and_ccsd_t.log MA
      %%CCSD
% 
% 
      % oc6_singlet_rob3lyp_6311++g_3df3pd_n_nbo_mayer_ST_split_IP_EA_dft_ccsd_t_wo_IP_and_EA_ccsd_t.log, ulm
      % oc6_triplet_rob3lyp_6311++g_3df3pd_n_nbo_mayer_ST_split_IP_EA_dft_ccsd_t_wo_IP_and_EA_ccsd_t.log, ulm
      \ce{C6O} & CCSD    & -0.194 & -0.193 & -0.194 \\
      \ce{C6O} & CCSD(T) & -0.217 & -0.219 & -0.220 \\
      \hline
      % oc7_singlet_b3lyp_6311++g_3df3pd_n_nbo_ST_split_IP_EA_both_dft_and_ccsd_t.log MA
      % oc7_triplet_b3lyp_6311++g_3df3pd_n_nbo_ST_split_IP_EA_both_dft_and_ccsd_t.log MA
      \ce{C7O}   & B3LYP   &   1.725 &   1.761 &  1.684 \\
      \ce{C7O}   & B3LYP (corr)&   1.647 &   1.683 &  1.606 \\
      %%CCSD
      % oc7_singlet_b3lyp_6311++g_3df3pd_n_nbo_ST_split_IP_EA_both_dft_and_ccsd_t.log MA
      % oc7_triplet_b3lyp_6311++g_3df3pd_n_nbo_ST_split_IP_EA_both_dft_and_ccsd_t.log MA
% 
% 
      % oc7_singlet_rob3lyp_6311++g_3df3pd_n_nbo_mayer_ST_split_IP_EA_dft_ccsd_t.log, ulm
      % oc7_optimized_triplet_rob3lyp_6311++g_3df3pd_roh_n_nbo_mayer_ST_split_IP_EA_dft_ccsd_t.log, ulm
      \ce{C7O} &  CCSD    & 1.645 & 1.681 & 1.595 \\
      \ce{C7O} &  CCSD(T) & 1.956 & 2.025 & 1.962 \\
% 
      % oc8_singlet_b3lyp_6311++g_3df3pd_n_nbo_ST_split_IP_EA_both_dft_and_ccsd_t.log MA
      % oc8_triplet_b3lyp_6311++g_3df3pd_n_nbo_ST_split_IP_EA_both_dft_and_ccsd_t.log MA
% 
% 
% 
% 
% 
% 
% 
% 
% 
% 
% 
% 
% 
% 
% 
% 
% 
% 
% 
% 
% 
% 
% 
% 
% 
% 
% 
% 
% 
% 
% 
% 
      \hline
    \end{tabular*}      
    %%%%%%%%%%%%%%%%%%%%%%%%%%%%%%%%%%%%%%%%%%%%%%%%%%%%%%%%%%%%%%%%%%%%%%%%%%%%%%%%%%%%%%
    %%%%%%%%%%%%%%%%%%%%%%%%%%%%%%%%%%%%%%%%%%%%%%%%%%%%%%%%%%%%%%%%%%%%%%%%%%%%%%%%%%%%%%
    \caption{
      Adiabatic ($\Delta_{adiab}$) and vertical ($\Delta_{S,T}$) values of the singlet-triplet
      energy separation (in eV) for \ce{C_{n} O} chains obtained within the methods indicated in the second column.
      The two vertical values shown here correspond to the optimized singlet ($\Delta_{S}$)
      and triplet ($\Delta_{T}$) geometries. Notice that the sign of $\Delta$
      indicates that the most stable isomers are singlets for odd members
      ($\Delta > 0$, \ce{C_{2k+1} O}) and triplets for even members ($\Delta < 0$, \ce{C_{2k} O}).
      Corrections due to zero-point motion (label \emph{corr}) were deduced within the DFT/B3LYP
      approach;
      % \added[remark={\rem{New text added in response to the first comment of the second reviewer}}]
            {see the last paragraph of \secname~\ref{sec:methods}.}
      }
    \label{table:Delta-cxo}
  \end{center}
\end{table}
\subsection{OC$_{n}$O Homologous Series} % \subsection{\ce{OC_{n} O} Homologous Series}
\label{sec:ocxo}
Out of the chains investigated in
{\secsname}\ref{sec:hcxh} and \ref{sec:hcxn},
all diradical species
\ce{HC_{2k+1}H} and \ce{HC_{2k}N} turned out to possess triplet ground states.
The results of
{\secsname}\ref{sec:cxs} and \ref{sec:cxo}
indicated that
the diradical character of a carbon chain does not automatically imply that the most stable state
is a triplet. Although all \ce{C_{n}S} and \ce{C_{n}O} chains are diradical species,
only the \ce{C_{2k}S} and \ce{C_{2k}O} chains considered have triplet ground states; 
the \ce{C_{2k+1}S} and \ce{C_{2k+1}O} chains possess singlet ground states.

Conversely, because the most stable isomers of the
``normal'' (nonradical) chains \ce{HC_{2k}H} and \ce{HC_{2k+1}N}
examined in
{\secsname}\ref{sec:hcxh} and \ref{sec:hcxn}
were found to be a singlet state,
one may next ask whether all nonradical carbon-based chains possess singlet ground states.
In order to inquire this possibility, we will next consider carbon chains having oxygen atoms
attached at the both ends (\ce{OC_n O}).

Irrespective whether the number of carbon atoms is even or odd, the cumulene-type structure
\ce{O\bond{2}C\bond{2}...\bond{2}C\bond{2}O} characterized by successive double bonds along the chain
ensures that all valence electrons are involved in chemical (double) bonds. Results of calculations
confirming this behavior are depicted in \figname\ref{fig:bonds-oc6o-oc7o}.

Still, in spite of the fact that both even and odd members of this family are ``normal'' molecules
with a similar (cumulene-type) structure, their most stable form is not necessarily
a singlet state. As seen in Table~\ref{table:H-ocxo} and \figname\ref{fig:ocxo}a, 
only odd members (\ce{OC_{2k+1}O}) have enthalpies of formation for singlet lower than for triplet.
However, for even members (\ce{OC_{2k}O}), triplet isomers possess enthalpies of formation lower
than for singlets. The fact that the most stable form is a singlet state for odd members but a triplet
for even members can alternatively be seen by inspecting the single-triplet energy separations
$\Delta$ (\emph{cf.}~Table~\ref{table:Delta-ocxo}, and \figname\ref{fig:ocxo}b and c),
which are positive for \ce{OC_{2k+1}O} but negative for \ce{OC_{2k}O}.

To conclude this subsection, carbon-based chains can have a triplet ground state
notwithstanding the fact that the corresponding species have always``normal'' (nonradical) character.
Parenthetically, one can still note that, in carbon chains, a triplet state yields an overall enforcement
of the cumulenic character even in species (\emph{e.g.},~\ce{HC12H}, \figname S11 of the {\si})
where the most stable isomer is of singlet polyyne type compatible with
the standard rules of valence.
\begin{figure*} % {hbtp}
  \centerline{\includegraphics[width=0.45\textwidth,angle=0]{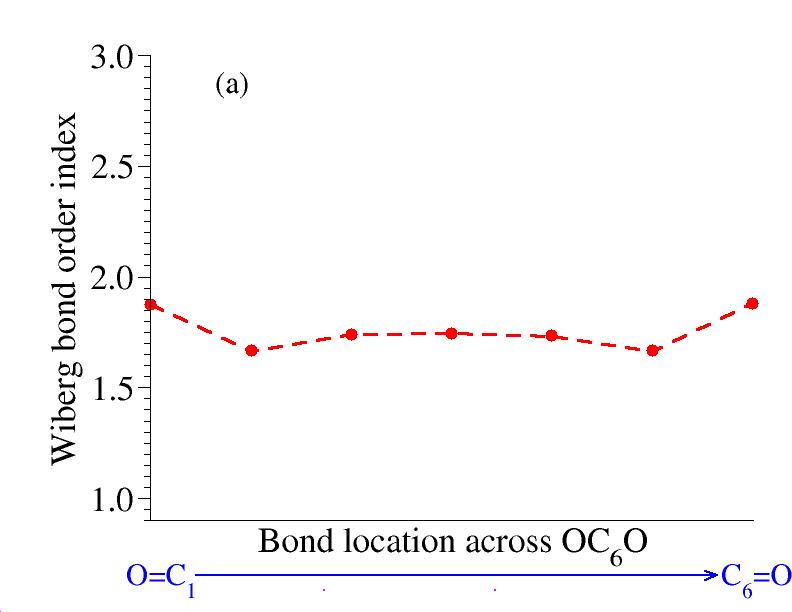}
  \includegraphics[width=0.45\textwidth,angle=0]{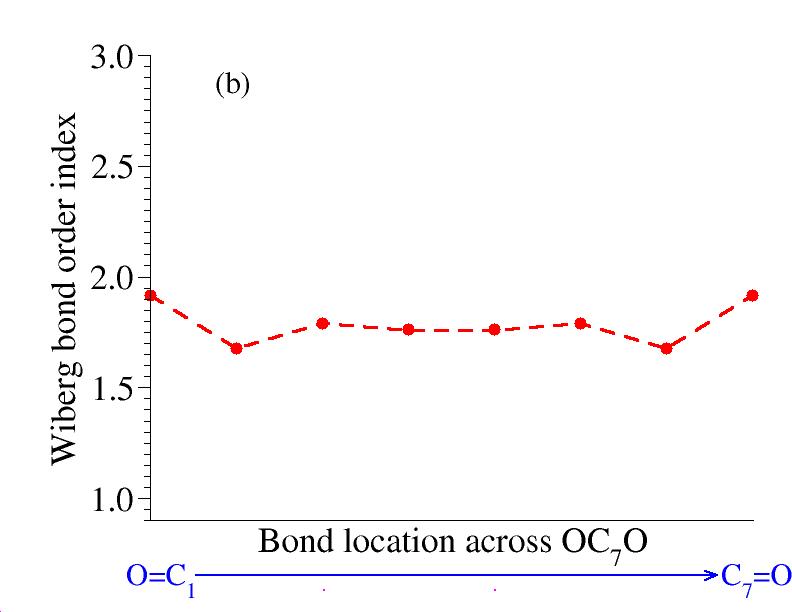}}
  \caption{Wiberg bond order indices for \ce{OC_n O} chains:
    (a) triplet \ce{OC6O} and (b) singlet \ce{OC7O}.
    The coordinates of these molecules
    at the corresponding energy minima as well as the HOMO spatial distributions are presented in
    Tables~S22 and S23 % Tables~\ref{table:oc6o-xyz-triplet} and \ref{table:oc7o-xyz-singlet}, 
    and in
    {\figsname}S9 and S10 % \figname\ref{fig:homo-oc6o} and \figname\ref{fig:homo-oc7o},
    respectively.
  }
  \label{fig:bonds-oc6o-oc7o}
\end{figure*}
\begin{table}[htbp] % [h!]
  \scriptsize % \small % \scriptsize % \footnotesize % \tiny
  \begin{center}
    \begin{threeparttable}
  %%%%%%%%%%%%%%%%%%%%%%%%%%%%
    \begin{tabular*}{0.47\textwidth}{@{\extracolsep{\fill}}lrrrr}
      \hline
      Molec.
      & $\Delta_{f} H^{0}_{0} $ & $\Delta_{f} H^{0}_{RT}\vert_{S}$
      & $\Delta_{f} H^{0}_{0} $ & $\Delta_{f} H^{0}_{RT}\vert_{T}$
      \\
      \hline
      % oco_optimized_singlet_b3lyp_6311++g_3df3pd_n_nbo_ST_split_IP_EA_both_dft_and_ccsd_t.log MA
      % oco_optimized_triplet_b3lyp_6311++g_3df3pd_n_nbo_ST_split_IP_EA_both_dft_and_ccsd_t.log MA
      \ce{OCO}   & -94.430 & -94.529  & 16.393 & 16.655 \\
      % OC2O singlet is unstable; seems to dissociate into 2 x CO
      % oc2o_optimized_triplet_b3lyp_6311++g_3df3pd_n_nbo_ST_split_IP_EA_both_dft_and_ccsd_t.log MA
      \ce{OC2O}  & ---     &          & -7.416  & -6.796 \\
      % oc3o_optimized_singlet_b3lyp_6311++g_3df3pd_n_nbo_ST_split_IP_EA_both_dft_and_ccsd_t.log MA
      % oc3o_optimized_triplet_b3lyp_6311++g_3df3pd_n_nbo_ST_split_IP_EA_both_dft_and_ccsd_t.log MA
      \ce{OC3O}  & -29.465 & -26.888 & 43.260 & 44.194 \\
      % oc4o_optimized_singlet_b3lyp_6311++g_3df3pd_n_nbo_ST_split_IP_EA_both_dft_and_ccsd_t.log MA
      % oc4o_optimized_triplet_b3lyp_6311++g_3df3pd_n_nbo_ST_split_IP_EA_both_dft_and_ccsd_t.log MA
      \ce{OC4O}  & 44.030 & 45.323  & 31.070 & 32.001 \\
      % oc5o_optimized_singlet_b3lyp_6311++g_3df3pd_n_nbo_ST_split_IP_EA_both_dft_and_ccsd_t.log MA
      % oc5o_optimized_triplet_b3lyp_6311++g_3df3pd_n_nbo_ST_split_IP_EA_both_dft_and_ccsd_t.log MA
      \ce{OC5O}  & 29.782 & 30.933  & 84.659 & 86.351 \\
      % oc6o_optimized_singlet_b3lyp_6311++g_3df3pd_n_nbo_ST_split_IP_EA_both_dft_and_ccsd_t.log MA
      % oc6o_optimized_triplet_b3lyp_6311++g_3df3pd_n_nbo_ST_split_IP_EA_both_dft_and_ccsd_t.log MA
      \ce{OC6O}  & 88.330 & 90.169  & 78.024 & 79.532 \\
      % oc7o_optimized_singlet_b3lyp_6311++g_3df3pd_n_nbo_ST_split_IP_EA_both_dft_and_ccsd_t.log MA
      % oc7o_optimized_triplet_b3lyp_6311++g_3df3pd_n_nbo_ST_split_IP_EA_both_dft_and_ccsd_t.log MA
      \ce{OC7O}  & 85.409 & 87.231  & 126.150 & 128.203 \\
      % oc8o_optimized_singlet_b3lyp_6311++g_3df3pd_n_nbo_ST_split_IP_EA_both_dft_and_ccsd_t.log ulm
      % oc8o_optimized_triplet_b3lyp_6311++g_3df3pd_n_nbo_ST_split_IP_EA_both_dft_and_ccsd_t.log ulm
% 
      % oc8o_singlet_cbs-qb3.log, ulm:
      % oc8o_triplet_cbs-qb3.log, ulm: t=-455.199099615; zpmt=0.045287; ent=0.045287; hmt=-455.153812; hm=hmt; zpm=zpmt; en=ent;
% 
      % oc9o_optimized_singlet_b3lyp_6311++g_3df3pd_n_nbo_ST_split_IP_EA_both_dft_and_ccsd_t.log ulm
      % oc9o_optimized_triplet_b3lyp_6311++g_3df3pd_n_nbo_ST_split_IP_EA_both_dft_and_ccsd_t.log ulm
% 
      % oc10o_optimized_singlet_b3lyp_6311++g_3df3pd_n_nbo_ST_split_IP_EA_both_dft_and_ccsd_t.log ulm
      % oc10o_optimized_triplet_b3lyp_6311++g_3df3pd_n_nbo_ST_split_IP_EA_both_dft_and_ccsd_t.log ulm ->
% 
      % oc11o_optimized_singlet_b3lyp_6311++g_3df3pd_n_nbo_ST_split_IP_EA_both_dft_and_ccsd_t.log ulm ->
      % oc11o_optimized_triplet_b3lyp_6311++g_3df3pd_n_nbo_ST_split_IP_EA_both_dft_and_ccsd_t.log ulm ->
% 
      % oc12o_optimized_singlet_b3lyp_6311++g_3df3pd_n_nbo_ST_split_IP_EA_both_dft_and_ccsd_t.log ulm ->
      % oc12o_optimized_triplet_b3lyp_6311++g_3df3pd_n_nbo_ST_split_IP_EA_both_dft_and_ccsd_t.log ulm ->
% 
      \hline
    \end{tabular*}      
    %%%%%%%%%%%%%%%%%%%%%%%%%%%%%%%%%%%%%%%%%%%%%%%%%%%%%%%%%%%%%%%%%%%%%%%%%%%%%%%%%%%%%%
% 
% 
% 
    \end{threeparttable}
    %%%%%%%%%%%%%%%%%%%%%%%%%%%%%%%%%%%%%%%%%%%%%%%%%%%%%%%%%%%%%%%%%%%%%%%%%%%%%%%%%%%%%%
    \caption{Enthalpies of formation of linear \ce{OC_nO} chains at zero and room temperature (subscript $0$ and $RT$, respectively). 
      Notice that for the odd members 
      \ce{OC_{2k+1} O} the values for singlet (label $S$) are smaller than those for triplet (label $T$),
      while for the even members \ce{OC_{2k} O} the values for triplet are smaller than those for singlet.
      (Values for singlet \ce{OC_2 O} are missing; calculations for
      geometry optimization invariably yielded two spatially separated \ce{CO} dimers.) 
      }
    \label{table:H-ocxo}
  \end{center}
\end{table}
\begin{figure*} % {hbtp}
  \centerline{\includegraphics[width=0.3\textwidth,angle=0]{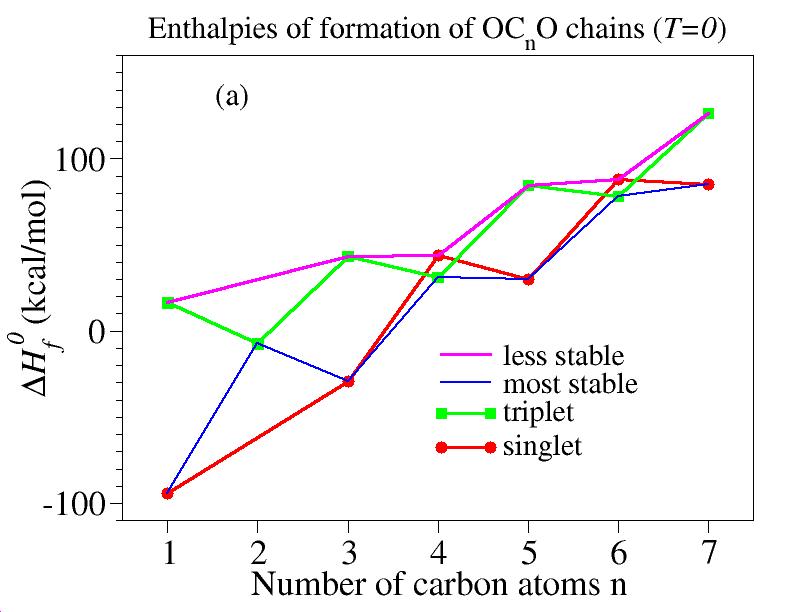}
    \includegraphics[width=0.3\textwidth,angle=0]{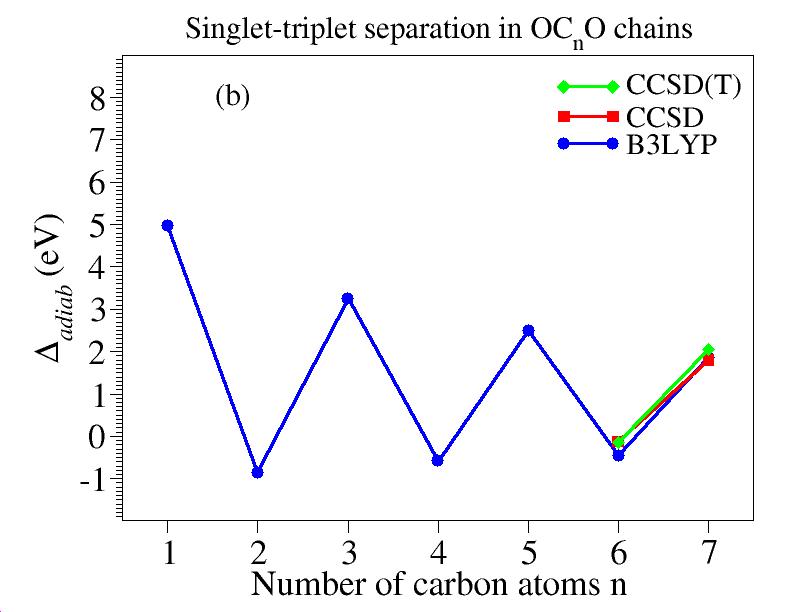}
      \includegraphics[width=0.3\textwidth,angle=0]{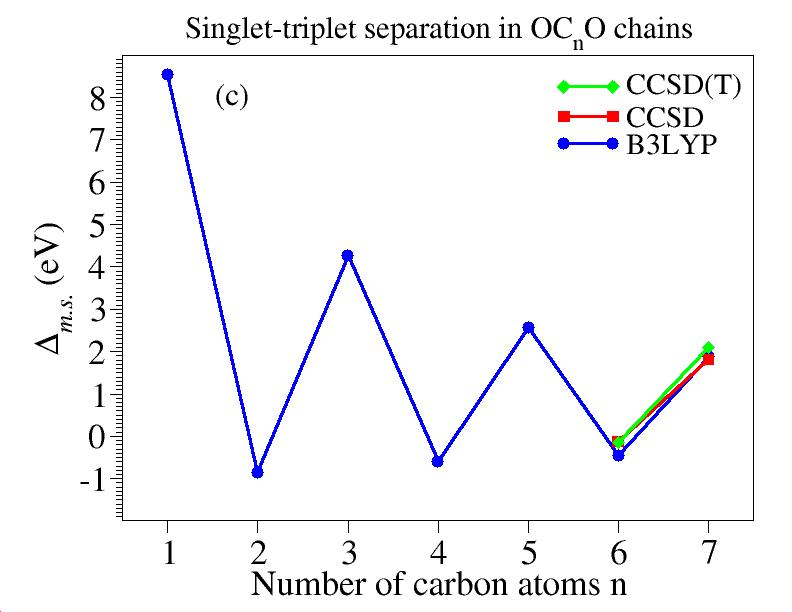}}
  \caption{Results for \ce{OC_n O} chains.
    (a) Enthalpies of formation for singlet and triplet chain isomers.
    (b) Adiabatic singlet-triplet separation energy $\Delta_{adiab}$.
    (c) Vertical singlet-triplet separation at the geometry of the most stable state $\Delta_{m.s.}$ (namely, singlet for odd members and triplet for even members,
    % \added[remark={\rem{New text added in response to the third comment of the second reviewer}}]
          {\emph{cf.}~Equation~(\ref{eq-Delta-m.s.})}).
    Lines are guide to the eye. The numerical values underlying this figure are presented in
    Tables~\ref{table:H-ocxo} and \ref{table:Delta-ocxo}. (Values for singlet \ce{OC_2 O} are missing; calculations for
      geometry optimization invariably yielded two spatially separated \ce{CO} dimers.)
  }
  \label{fig:ocxo}
\end{figure*}
% 
% 
    % this rotates with 90 deg but does bring nothing: \begin{landscape}
\begin{table}[htbp] % [h!]
  \scriptsize % \small % \footnotesize % \scriptsize % \footnotesize % \tiny
  \begin{center}
    \begin{threeparttable}
    \begin{tabular*}{0.47\textwidth}{@{\extracolsep{\fill}}rrrrr}
      \hline
      Molec. & Method 
      & $\Delta_{adiab}$     
      & $\Delta_{S}$     
      & $\Delta_{T}$     
      \\
      \hline
      % oco_optimized_singlet_b3lyp_6311++g_3df3pd_n_nbo_ST_split_IP_EA_both_dft_and_ccsd_t.log MA
      % oco_optimized_triplet_b3lyp_6311++g_3df3pd_n_nbo_ST_split_IP_EA_both_dft_and_ccsd_t.log MA
      \ce{OCO}   & B3LYP   &   4.977   &  8.549 &  3.001  \\
      \ce{OCO}   & B3LYP (corr)&   4.806   &  8.378 &  2.830  \\
      %%CCSD
% 
% 
      \hline
      % OC2O singlet is unstable; seems to dissociate into 2 x CO
      % oc2o_optimized_triplet_b3lyp_6311++g_3df3pd_n_nbo_ST_split_IP_EA_both_dft_and_ccsd_t.log MA
      \ce{OC2O}  & B3LYP   &   ---     & ---   & -0.856 \\
      \ce{OC2O}  & B3LYP (corr)&   ---     & ---   &   ---  \\
      %%CCSD
% 
% 
      \hline
      % oc3o_optimized_singlet_b3lyp_6311++g_3df3pd_n_nbo_ST_split_IP_EA_both_dft_and_ccsd_t.log MA
      % oc3o_optimized_triplet_b3lyp_6311++g_3df3pd_n_nbo_ST_split_IP_EA_both_dft_and_ccsd_t.log MA
      \ce{OC3O}  & B3LYP   &   3.247 &  4.273  &  1.130 \\
      \ce{OC3O}  & B3LYP (corr)&   3.154 &  4.179  &  1.037 \\
      %%CCSD
% 
% 
      \hline
      % oc4o_optimized_singlet_b3lyp_6311++g_3df3pd_n_nbo_ST_split_IP_EA_both_dft_and_ccsd_t.log MA
      % oc4o_optimized_triplet_b3lyp_6311++g_3df3pd_n_nbo_ST_split_IP_EA_both_dft_and_ccsd_t.log MA
      \ce{OC4O}  & B3LYP   &  -0.583 & -0.583  & -0.584 \\
      \ce{OC4O}  & B3LYP (corr)&  -0.562 & -0.562  & -0.562 \\
      %%CCSD
% 
% 
      \hline
      % oc5o_optimized_singlet_b3lyp_6311++g_3df3pd_n_nbo_ST_split_IP_EA_both_dft_and_ccsd_t.log MA
      % oc5o_optimized_triplet_b3lyp_6311++g_3df3pd_n_nbo_ST_split_IP_EA_both_dft_and_ccsd_t.log MA
      \ce{OC5O}  & B3LYP   &   2.491 &  2.570  &  1.724 \\
      \ce{OC5O}  & B3LYP (corr)&   2.380 &  2.459  &  1.612 \\
      %%CCSD
% 
% 
      \hline
      % oc6o_optimized_singlet_b3lyp_6311++g_3df3pd_n_nbo_ST_split_IP_EA_both_dft_and_ccsd_t.log MA
      % oc6o_optimized_triplet_b3lyp_6311++g_3df3pd_n_nbo_ST_split_IP_EA_both_dft_and_ccsd_t.log MA
      \ce{OC6O}  & B3LYP   & -0.447 &  -0.447  & -0.447 \\
      \ce{OC6O}  & B3LYP (corr)& -0.447 &  -0.447  & -0.448 \\
      %%CCSD
      % oc6o_optimized_singlet_b3lyp_6311++g_3df3pd_n_nbo_ST_split_IP_EA_both_dft_and_ccsd_t.log MA
      % oc6o_optimized_triplet_b3lyp_6311++g_3df3pd_n_nbo_ST_split_IP_EA_both_dft_and_ccsd_t.log MA
% 
% 
      % oc6o_singlet_rob3lyp_6311++g_3df3pd_n_nbo_mayer_ST_split_IP_EA_dft_ccsd_t_wo_IP_EA_ccsd_t.log
      % oc6o_optimized_triplet_rob3lyp_6311++g_3df3pd_roh_n_nbo_mayer_ST_split_IP_EA_dft_ccsd_t_wo_IP_EA_roccsd_t.log, ulm
      \ce{OC6O} &  CCSD    & -0.124 &  -0.126 & -0.126 \\
      \ce{OC6O} &  CCSD(T) & -0.142 &  -0.145 & -0.145 \\
      \hline
      % oc7o_optimized_singlet_b3lyp_6311++g_3df3pd_n_nbo_ST_split_IP_EA_both_dft_and_ccsd_t.log MA
      % oc7o_optimized_triplet_b3lyp_6311++g_3df3pd_n_nbo_ST_split_IP_EA_both_dft_and_ccsd_t.log MA
      \ce{OC7O}  & B3LYP   &  1.855 &  1.873  &  1.836 \\
      \ce{OC7O}  & B3LYP (corr)&  1.767 &  1.786  &  1.748 \\
      %%CCSD
      % oc7o_optimized_singlet_b3lyp_6311++g_3df3pd_n_nbo_ST_split_IP_EA_both_dft_and_ccsd_t.log MA
      % oc7o_optimized_triplet_b3lyp_6311++g_3df3pd_n_nbo_ST_split_IP_EA_both_dft_and_ccsd_t.log MA      
% 
% 
      % oc7o_singlet_rob3lyp_6311++g_3df3pd_n_nbo_mayer_ST_split_IP_EA_dft_ccsd_t.log, ulm
      % oc7o_optimized_triplet_rob3lyp_6311++g_3df3pd_roh_n_nbo_mayer_ST_split_IP_EA_dft_ccsd_t.log, ulm
      \ce{OC7O}  &   CCSD     & 1.802 & 1.822 & 1.778 \\
      \ce{OC7O}  &   CCSD(T)  & 2.057 & 2.107 & 2.072 \\
% 
      % oc8o_optimized_singlet_b3lyp_6311++g_3df3pd_n_nbo_ST_split_IP_EA_both_dft_and_ccsd_t.log ulm
      % oc8o_optimized_triplet_b3lyp_6311++g_3df3pd_n_nbo_ST_split_IP_EA_both_dft_and_ccsd_t.log ulm
% 
% 
% 
% 
% 
      % oc9o_optimized_singlet_b3lyp_6311++g_3df3pd_n_nbo_ST_split_IP_EA_both_dft_and_ccsd_t.log ulm
      % oc9o_optimized_triplet_b3lyp_6311++g_3df3pd_n_nbo_ST_split_IP_EA_both_dft_and_ccsd_t.log ulm
% 
% 
% 
% 
% 
      % oc10o_optimized_singlet_b3lyp_6311++g_3df3pd_n_nbo_ST_split_IP_EA_both_dft_and_ccsd_t.log ulm 
      % oc10o_optimized_triplet_b3lyp_6311++g_3df3pd_n_nbo_ST_split_IP_EA_both_dft_and_ccsd_t.log ulm 
% 
% 
% 
% 
% 
      % oc11o_optimized_singlet_b3lyp_6311++g_3df3pd_n_nbo_ST_split_IP_EA_both_dft_and_ccsd_t.log ulm 
      % oc11o_optimized_triplet_b3lyp_6311++g_3df3pd_n_nbo_ST_split_IP_EA_both_dft_and_ccsd_t.log ulm 
% 
% 
% 
% 
% 
      % oc12o_optimized_singlet_b3lyp_6311++g_3df3pd_n_nbo_ST_split_IP_EA_both_dft_and_ccsd_t.log ulm ->
      % oc12o_optimized_triplet_b3lyp_6311++g_3df3pd_n_nbo_ST_split_IP_EA_both_dft_and_ccsd_t.log ulm ->
% 
% 
% 
% 
      \hline
    \end{tabular*}      
    \end{threeparttable}
    %%%%%%%%%%%%%%%%%%%%%%%%%%%%%%%%%%%%%%%%%%%%%%%%%%%%%%%%%%%%%%%%%%%%%%%%%%%%%%%%%%%%%%
    \caption{
      Adiabatic ($\Delta_{adiab}$) and vertical ($\Delta_{S,T}$) values of the singlet-triplet
      energy separation (in eV) for \ce{O C_{n} O} chains obtained within the methods indicated in the second column.
      The two vertical values shown here correspond to the optimized singlet ($\Delta_{S}$)
      and triplet ($\Delta_{T}$) geometries. Notice that the sign of $\Delta$
      indicates that the most stable isomers are singlets for odd members
      ($\Delta > 0$, \ce{OC_{2k+1} O}) and triplets for even members ($\Delta < 0$, \ce{OC_{2k} O}).
      Corrections due to zero-point motion (label \emph{corr}) were deduced within the DFT/B3LYP
      approach;
      % \added[remark={\rem{New text added in response to the first comment of the second reviewer}}]
            {see the last paragraph of \secname~\ref{sec:methods}.}
      %      Diffferent IR spectra of \ce{OC3O} because singlet is ĺinear while triplet is nonlinear. \ib{OC$_{2n}$O triplet}, OC$_{2n+1}$O singlet. Different IR spectra of singlet and \ce{OC11O}/\ce{OC9O} although both have linear conformation. IR spectra for s- and t-OC7O, OC6O are similar.
    }
    \label{table:Delta-ocxo}
  \end{center}
\end{table}
\subsection{Infrared and UV-Visible Absorption of Carbon Chains in Singlet and Triplet States}
\label{sec:ir_uv-vis}
One of the main aims of this paper is to emphasize that
the lowest electronic state of carbon-based chains of interest for astrochemistry systematically
switches back and forth between singlet and triplet as the carbon chain length increases.
A detailed analysis of the differences between properties of the lowest
singlet and triplet states for each carbon-based chain of astrochemical interest is certainly important,
\emph{e.g.}, for adequately processing data acquired (or to be acquired) in astronomical observations.
This analysis is out of the scope of the present paper and will be deferred to a forthcoming publication.

Certainly, the nature of the ground state (\emph{i.e.}, singlet \emph{vs} triplet)
can have a pronounced impact on various observable molecular properties.
While important differences between properties of singlet and triplet states of
various molecular species would not be surprising in general, the
specific example presented below reveals that the impact can be significant even in situations less expected.

For illustration, let us consider \ce{HC11H} chains. By inspection the HOMO spatial distributions of 
\ce{HC11H} chains (\emph{cf.}~{\figname}S1) % (\emph{cf.}~\figname\ref{fig:homo-hc11h})
one can conclude that the differences between the singlet and triplet states are rather minor.
Nevertheless,
the infrared spectra presented in \figname\ref{fig:ir-hc11h} (calculated within the harmonic approximation)
reveal significant differences between the singlet and triplet \ce{HC11H} isomers.

Less surprisingly, calculations showed that
the spin multitplicity has little impact on the high frequency strong peak depicted in
\figname\ref{fig:ir-hc11h} due to the stretching mode of the \ce{C\bond{1}H} bonds with pronounced
s-character; the corresponding frequency values
($\nu_{\ce{C\bond{1}H}}^{S}=3462.46\,\mbox{cm}^{-1}$ and $\nu_{\ce{C\bond{1}H}}^{T}=3461.93\,\mbox{cm}^{-1}$)
are identical within the numerical accuracy. However, as visible in \figname\ref{fig:ir-hc11h},
differences between the singlet and 
triplet \ce{HC11H} infrared spectra significant especially in the spectral range $\sim 400 - 800\,\mbox{cm}^{-1}$
characteristic for \ce{CCH} bending modes.
We referred above to vibrational properties
because infrared (vibrational) spectroscopy is an important tool for detecting
symmetric molecules like \ce{HC11H}; given the fact that their dipole moment vanishes, rotational
spectroscopy cannot be utilized in such cases.
\begin{figure*} % {hbtp}
  \centerline{
    \includegraphics[width=0.45\textwidth,angle=0]{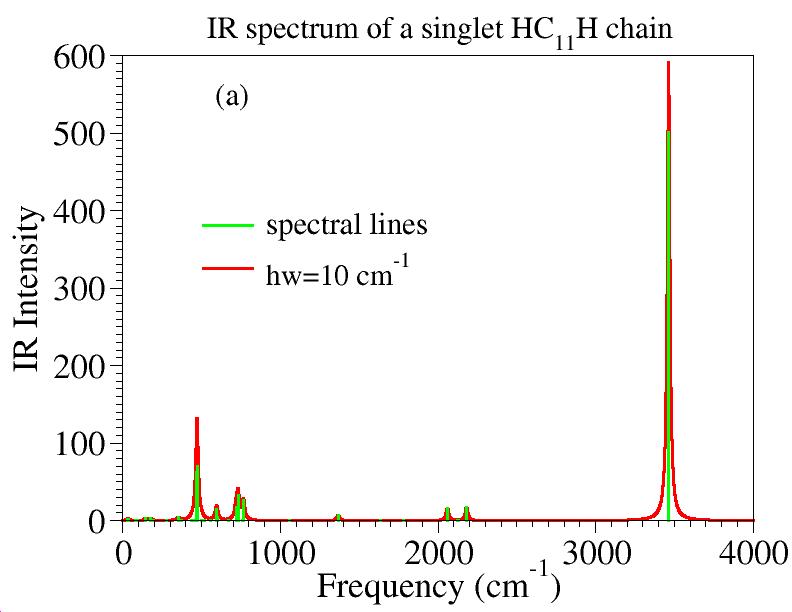}
  \includegraphics[width=0.45\textwidth,angle=0]{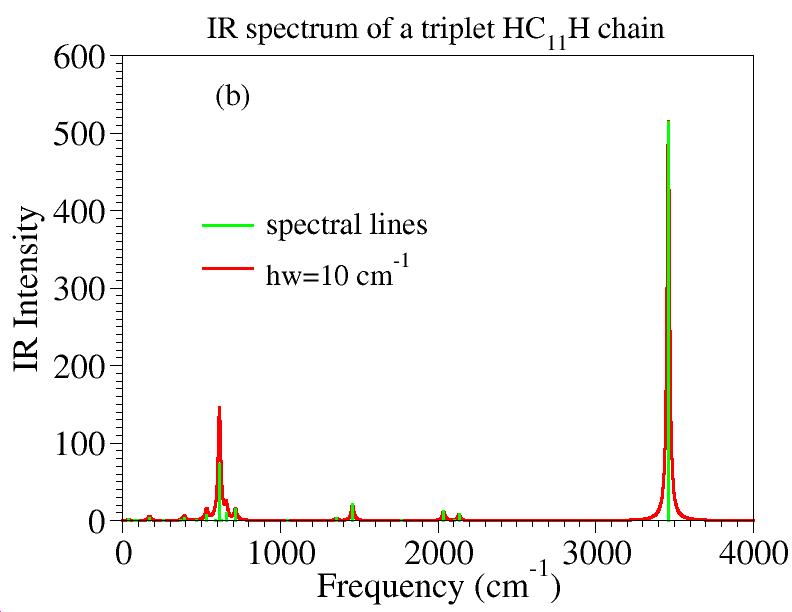}}
  \caption{Infrared spectra of (a) singlet and (b) triplet \ce{HC11H} chains.
     The solid red lines were deduced by convoluting the spectral lines (depicted in green)
    computed within a DFT/B3LYP/6-311++g(3df, 3pd) approach by using Lorentzian distributions
    whose halfwidth (hw) is indicated in the legend.
  }
  \label{fig:ir-hc11h}
\end{figure*}
Another example revealing the impact of the nature of the ground state on the
infrared spectra is presented in the {\si}; see
{\figsname}S12a and S12b. % \figname\ref{fig:ir-hc12h}a and \figname\ref{fig:ir-hc12h}b,

In addition to the significant impact on the vibrational transitions underlying
the differences between the singlet and triplet
infrared spectra of \figsname\ref{fig:ir-hc11h}a and \ref{fig:ir-hc11h}b,
we can also mention a similar effect on the UV-visible absorption.
To exemplify, our TD-DFT/CAM-B3LYP/6-311++g(3df, 3pd)
calculations yielded a value $\varepsilon_{T} = 2.614$\,eV ($474$\,nm)
for the lowest $\mbox{A}\,^{3}\Sigma^{-} - \mbox{X}\,^{3}\Sigma^{-}$ electronic transition.
Similar calculations done for the lowest electronic transition
of the singlet \ce{HC11H} found a significantly different
value of $\varepsilon_{S} = 2.303$\,eV ($538$\,nm).

One should still note here that estimates based on the methods utilized above
may not be sufficiently accurate to be directly compared with measured data.

Unfortunately, \emph{ab initio} (\emph{e.g.},~CC-)methods
to compute vibrational properties for such large 
and open shell molecular species,
which  can be employed to smaller closed-shell chains,\cite{Doney:18}
are currently prohibitive.
Still, we believe that that the significant differences between
\figname\ref{fig:ir-hc11h}a and b (and between \figname S12a and b of the {\si}) % \figname\ref{fig:ir-hc11h}
represent a convincing argument that experiments --- which the present theoretical study hopes
to motivate ---
can discriminate between the singlet and triplet \ce{HC11H} (and \ce{HC12H}) chains.

Likewise, the above TD-DFT-based $\varepsilon_{S,T}$-values may not be
accurate enough for quantitative purposes. 
Still, given the fact that \emph{ab initio} calculations
for excited states of open-shell systems at such large molecular sizes
represents a formidable challenge for theory, the magnitude of the
difference $\varepsilon_{T} - \varepsilon_{S}$ found above should be sufficiently
significant to encourage companion experimental efforts.
\subsection{A \emph{Sui Generis} Even-Odd Effect that Confounds Straightforward Chemical Intuition}
\label{sec:sui-generis}
It is known that successive addition of repeat
units to homologous classes of systems may result in properties periodically switching
back and forth in a manner similar to that depicted in our
\figsname\ref{fig:hcxh}, \ref{fig:hcxn}, \ref{fig:cxs}, 
\ref{fig:cxo}, and \ref{fig:ocxo}.
Examples of this kind include, \emph{e.g.}, the case of aromatic $4n+2$ (H\"uckel) and $4n$ (anti-H\"uckel)
cyclic polyenes (annulenes) dating back to the early days of quantum mechanics
\cite{Hueckel:31a,Hueckel:31b,Hueckel:32,London:37,Pople:66} or their more recent
counterpart in mesoscopic and quantum dot rings
\cite{Baldea:99a,Baldea:99b,Baldea:2001a,Baldea:2002}
as well as a variety of even-odd effects in multilayers,\cite{Wu:16} self-assembled monolayers,\cite{Tao:07}
molecular electronic devices,\cite{CuevasScheer:17,Baldea:2015e} or atomic nuclei.

Compared to those and all other cases of which we are aware,
the even-odd effect reported previously \cite{Fan:89} and extensively
analyzed in this paper for
carbon-based chains of astrochemical interest is unique.
This even-odd effect does not merely consist of an alternation of a
certain (ground state) property upon size increasing. Rather, it is the very nature of the electronic ground state
that switches back and forth between a singlet and triplet state. Specifically,
we showed here that, depending on the chemical nature of the terminal heteroatom(s),  
the chain possesses a singlet ground state
for a certain parity (\emph{i.e.},~even or odd) of the number of carbon atoms,
which switches to a triplet ground state for the other parity (odd or even, respectively)
of this number.

It is worth emphasizing that the singlet-triplet alternation discussed above is a subtle effect
that confounds straightforward intuition. 
We first considered
({\secsname}\ref{sec:hcxh} and \ref{sec:hcxn})
carbon chain families, wherein members of a certain parity are normal
(closed-shell, nonradical) molecules having all 
valence electrons paired in chemical bonds between adjacent atoms
while members of the opposite parity are diradicals.
Chains of \ce{HC_n H} and \ce{HC_n N}
(as well as \ce{NC_n N}, a case not explicitly discussed because of its similarity with the first two)
belonging to this category seemed to fit an intuitive rule of thumb
(ground state of normal chains is a singlet, diradical chains have triplet ground states).
In disagreement to that, subsequent examples examined revealed that not all diradical chains 
(\ce{C_n S},
{\secname}\ref{sec:cxs}
and
\ce{C_n O},
{\secname}\ref{sec:cxo})
have a triplet ground state, while not all normal (nonradical) chains
(\ce{O C_n O},
{\secname}\ref{sec:ocxo})
possess singlet ground states.
\subsection{Remarks on Some Recent Studies on Carbon Chains Done in the Astrophysics/Astronomical Community}
\label{sec:etim}
Without any intention to target the entire astrophysical/astrochemical community,
we have to emphasize that our present findings disagree with some recent studies
\cite{Etim:16a,Etim:16b,Etim16aImplicitACS} claiming that,
for all carbon-based chains with an even number of electrons like those examined above
singlet isomers are the most stable. 
Refuting that claim and drawing attention on the fact that triplet states deserve consideration
in astrochemical context
(\emph{cf.}~{\secname}\ref{sec:astronomy}
below) was an important aim of conducting the extensive theoretical investigation presented in this paper.

To demonstrate that, even if not explicitly stated, ref.~\citenum{Etim:16a} tacitly
admitted singlet electronic ground states,
in Tables~S1--S4 of the {\si} we present results reported in that work
along with $\Delta_{f} H^{0}$-estimates for both singlets and triplets based on
the DFT/B3LYP/6-311++g(3df, 3pd) approach 
underlying Tables~\ref{table:H-hcxh}, \ref{table:H-hcxn}, \ref{table:H-cxo}, \ref{table:H-cxs}, and \ref{table:H-ocxo}.
As visible from the inspection of Tables~S1--S4,
the $\Delta_{f} H^{0}$-values of ref.~\citenum{Etim:16a} agree with our $\Delta_{f} H^{0}$ values
obtained for the lowest singlet states. So, although not explicitly stated in ref.~\citenum{Etim:16a},
the results presented there do refer to the singlet states.
The small differences between $\Delta_{f} H^{0}$-values of
Tables~\ref{table:H-hcxh}, \ref{table:H-hcxn}, \ref{table:H-cxo}, \ref{table:H-cxs}, and \ref{table:H-ocxo} and those of ref.~\citenum{Etim:16a}
may reflect the slightly different methods
utilized. These differences due to the different approaches
are comparable to those between the B3LYP/6-311++g(3df, 3pd)-based values and the  $\Delta_{f} H^{0}$-estimates
obtained by using the CBS-QB3 protocol as implemented in GAUSSIAN 16, which are also shown in 
Tables~S1--S4 % Tables~\ref{table:suppl-H-hcxh}, \ref{table:suppl-H-hcxn}, \ref{table:suppl-H-cxs}, and \ref{table:suppl-H-cxo}
for comparison purposes.
\subsection{Computational Issues}
\label{sec:ccsd_t}
Important insight from a computational perspective can be gained from the comparison between the various
estimates presented in Tables~\ref{table:Delta-hcxh}, \ref{table:Delta-hcxn}, \ref{table:Delta-cxs},
\ref{table:Delta-cxo}, and \ref{table:Delta-ocxo}.

First, the results presented there emphasize limitations of popular Delta-SCF 
\cite{Gunnarson:89} or Delta-DFT \cite{Baldea:2012i,Baldea:2013b,Baldea:2014c}
methods to estimate the singlet-triplet energy separation. 
Properties like the singlet-triplet separation $\Delta$ examined above
as well as other quantities, \emph{e.g.},~ionization or electron attachment energies
can be expressed as differences between pertaining total molecular energies.
Similar to other cases,\cite{Baldea:2014c,Baldea:2014e,Baldea:2017e}
the comparison with state-of-the-art approaches based on coupled-cluster expansions
(\emph{cf.}~Tables~\ref{table:Delta-hcxh}, \ref{table:Delta-hcxn}, \ref{table:Delta-cxs},
\ref{table:Delta-cxo}, and \ref{table:Delta-ocxo}) reveals
differences of several tenth of electronvolt. Such differences are significantly
larger than the accuracy needed for reliable astrochemical modeling
or achieved in experiments.\cite{Hansen:08,Li:09} % \cite{Hansen:08,Li:09,Kaiser:10}

More importantly, the differences between CCSD and CCSD(T) estimates appear
to be unusually large.
In ``normal'' cases, where electron correlations are ``moderately'' strong,
quantities obtained within the CCSD(T) approach
do not substantially differ from those based on the CCSD approach.
The reason is that, within the CCSD(T) approach,
triple excitations are only \emph{perturbatively} treated
on top of the single and double excitations embodied within the CCSD approach.
Such a perturbative approach can be hardly justified in cases like those shown above for
longer chains, where the differences between CCSD(T) and CCSD estimates
amount to $\sim 0.3 - 0.4$ \,eV (\emph{cf.}~Tables~\ref{table:Delta-cxs}, \ref{table:Delta-cxo}, and
\ref{table:Delta-ocxo}). This state of affairs calls for further clarification.
Possible further developments in this direction
include other theoretical treatments (\emph{e.g.}, SAC-CI \cite{Ehara:05}),
CC approaches including higher order electron correlations (\emph{e.g.},~CCSDT or CC3 \cite{Christiansen:06})
or based on the (particularly suited for one-dimensional systems) density matrix renormalization group
(DMRG).\cite{White:99,Chan:02,Chan:09}
\subsection{Relevance for Astronomical Detectability}
\label{sec:astronomy}
From an astrochemical perspective, 
the inspection of the results presented above unravels a striking correlation.
Across the various families of linear carbon-chains presently considered,
the members that were not identified in astronomical data
are preponderantly those for which our present calculations predict a triplet ground state.
Examples will be presented below in support of this assertion.

\ce{C5S} is the longest chain detected in space \cite{Cernicharo:87} from the \ce{C_n S} family.
However, the shorter \ce{C4S} chain was not reported so far. Still, the enthalpy of formation for the 
most stable \ce{C4S} chain (which is a \emph{triplet}) is lower than that for the most stable
(singlet) \ce{C5S} chain observed (\emph{cf.}~Table~\ref{table:H-cxs}). On the contrary, 
the less stable singlet \ce{C4S} isomer has an enthalpy of formation higher than that of the
singlet \ce{C5S} chain. So, it is not unexpected that it is more difficult to detect \ce{C4S} chains
implicitly assuming that they are singlets than to detect (singlet)  \ce{C5S} chains.

A similar situation is encountered in the \ce{HC_n H} series.
Out of the \ce{HC_n H} family, \ce{HC6H} is the longest chain observed.\cite{Cernicharo:01}
Nevertheless, the shorter \ce{HC5H} chain was not detected in space so far. Its most stable
(\emph{triplet}) isomer is more stable than the most stable (singlet) \ce{HC6H} chain
(\emph{cf.}~Table~\ref{table:H-hcxh}) although the singlet \ce{HC5H} chain is thermodynamically
less stable than the singlet \ce{HC6H} chain.

Most notorious is the case of the \ce{HC_n N} family. The even members \ce{HC6N}, \ce{HC8N}, and \ce{HC10N}
were not observed in space; nevertheless, the next larger odd members
\ce{HC7N},\cite{Kroto:78}
\ce{HC9N},\cite{Broten:78}
and \ce{HC11N} \cite{Bell:97} were detected in astronomical data.
The inspection of Table~\ref{table:H-hcxn} reveals that, again, it is the most stable (\emph{triplet})
isomer of the even members that is more stable than the most stable (singlet) isomer of the next larger
isomer of the odd members.
A \emph{triplet} \ce{HC6N} chain is more stable thermodynamically than a singlet \ce{HC7N} chain,
a \emph{triplet} \ce{HC8N} chain is more stable than a singlet \ce{HC9N} chain, and
a \emph{triplet} \ce{HC10N} chain is more stable than a singlet \ce{HC11N} chain,\cite{hc10n,Baldea:2019e} 
although a \emph{singlet} \ce{HC6N} chain is less stable thermodynamically than a singlet \ce{HC7N} chain,
a \emph{singlet} \ce{HC8N} chain is less stable than a singlet \ce{HC9N} chain, and
a \emph{singlet} \ce{HC10N} chain is less stable than a singlet \ce{HC11N} chain.

In all cases of the aforementioned type, ref.~\citenum{Etim:16a} arrived at an opposite
conclusion on the thermodynamical in/stability across various chain families
because of the incorrect (implicit\cite{Etim16aImplicitACS})
assumption that the most stable chains of species with an even number of electrons are always singlets.

In Introduction we mentioned that information on the presently considered carbon-based chains
are very scarce. Given this data scarcity, it may be tempting to resort to
empirical interpolation/extrapolation schemes;\cite{Etim:16b}
namely, to utilize available properties of molecular species of a certain homologous family 
to deduce unavailable properties of other molecular species, and use the latter
in attempting to reveal their presence by processing the astronomical data.
However, such a procedure will inherently fail. Properties of, \emph{e.g.}, \emph{triplet} \ce{HC6N}, \ce{HC8N} or
\ce{HC10N} chains can by no means be obtained by interpolating/extrapolating properties
of \emph{singlet} \ce{HC7N}, \ce{HC9N} and \ce{HC11N} chains. The reason of this failure should be clear;
we have seen that
differences between the singlet and triplet isomers of a given carbon chain
may be important even in situations where one can expect that they are less significant
(\emph{cf.}~{\secname}\ref{sec:ir_uv-vis}).
\section{Summary}
\label{sec:summary}
In this paper, we demonstrated that, in families of carbon-based chains,
most stable members of one parity (even or odd)
are singlets while most stable
members of the opposite parity (odd or even) are triplets.
This is a \emph{sui generis} effect qualitatively different from other even-odd effects
known from studies over decades in other areas.
\cite{Hueckel:31a,Hueckel:31b,Hueckel:32,London:37,Baldea:99a,Baldea:99b,Baldea:2001a,Baldea:2002,Tao:07,Baldea:2015e,Wu:16,CuevasScheer:17}

From a more general perspective, the present paper aimed at bridging astronomy/astrophysics
and physical/computational chemistry.
The results reported here are most significant in the context of the observability in space.
As elaborated in
{\secname}\ref{sec:astronomy},
we strongly believe that it is not coincidental that
absent in the list of various homologous series detected in space are
just carbon-based chains for which our calculations predict a triplet ground state.
Emphasizing that most stable isomers of carbon-based chains of the type considered in our paper
are not invariably singlets is important because this was incorrectly claimed 
in some recent studies done in the astronomical/astrophysical community.\cite{Etim:16b,Etim:16a}
Prior to the present extensive study, a series of authors drew attention on the fact that triplet isomers
of carbon-based chains are more stable than singlet isomers: 
\ce{HC_{2k+1}H},\cite{Fan:89,Ball:00} \ce{C_{2k}},\cite{Fan:89,Maier:14}
\ce{HC4N},\cite{Aoki:93,Aoki:94,Gutowski:96,Kim:97}
\ce{NC5N},\cite{Smith:94b}
\ce{C_{2k}O},\cite{Ohshima:95} % k=1,2,3,4
\ce{C_{2,4}S},\cite{McGuire:18b}, and \ce{HC6N}.\cite{Maier:01}

Last but not least, from a computational perspective, the present results are important because they emphasize
that electron correlations in carbon-based chains are unusually strong. The analysis of
{\secname}\ref{sec:ccsd_t}
indicated that the ``standard gold'' CCSD(T) quantum chemical
approach may not be sufficiently accurate to quantify these strong electron correlations.
This is an important challenge for the community of theoretical chemistry
that deserves further consideration.
\section*{Acknowledgment}
I thank Jochen Schirmer for stimulating discussions.
Financial support for this research
provided by the Deu\-tsche For\-schungs\-ge\-mein\-schaft (DFG grant BA 1799/3-1,2)
and partial computational support by the State of Baden-W\"urttemberg through bwHPC
and the DFG through grant no INST 40/467-1 FUGG
are gratefully acknowledged.
{
\providecommand{\latin}[1]{#1}
\makeatletter
\providecommand{\doi}
  {\begingroup\let\do\@makeother\dospecials
  \catcode`\{=1 \catcode`\}=2 \doi@aux}
\providecommand{\doi@aux}[1]{\endgroup\texttt{#1}}
\makeatother
\providecommand*\mcitethebibliography{\thebibliography}
\csname @ifundefined\endcsname{endmcitethebibliography}
  {\let\endmcitethebibliography\endthebibliography}{}

}
\renewcommand{\theequation}{S\arabic{equation}}
\setcounter{equation}{0}
\renewcommand{\thefigure}{S\arabic{figure}}
\setcounter{figure}{0}
\renewcommand{\thetable}{S\arabic{table}}
\setcounter{table}{0}
\renewcommand{\thesection}{S\arabic{section}}
\setcounter{section}{0}
\renewcommand{\thepage}{S\arabic{page}}
\setcounter{page}{1}
\renewcommand{\thefootnote}{\alph{footnote}}

\newpage

\begin{center}
         Supporting Information \\ for \\
  Alternation of Singlet and Triplet % Ground
  States % upon Adding Repeat Units
  in Carbon-Based Chain Molecules.
  Results of an Extensive Theoretical Study
\end{center}
\vspace{1ex}

\section{Quantum Chemical Calculations for Triplet States: Unrestricted \emph{versus} Restricted Open Shell Approaches}
\label{sec:ub3lyp-roccsd_t}
Applying the unrestricted
formalism to open-shell triplet states may be problematic because of the related spin contamination.
The unrestricted DFT calculations done in this study (\emph{cf.}~{\secname}2 in the main text)
posed no special problem. In all cases, we found values $\left\langle \mathbf{S^2}\right\rangle \approx 2.15$
before annihilation of the first spin contaminant and $\left\langle \mathbf{S^2}\right\rangle \approx 2.01$
after annihilation. As expected in view of the insignificant departure from the exact value
$\left\langle \mathbf{S^2}\right\rangle = 2$,
in spot checks, we explicitly found that differences between results obtained within 
unrestricted (UB3LYP) and restricted open shell (ROB3LYP) calculations are altoghether negligible.

Things completely change in the case of CC-calculations. For the triplet \ce{HC9N}, unrestricted CC calculations
yield values (which are typical) $\left\langle \mathbf{S^2}\right\rangle \approx 3.63$
before annihilation of the first spin contaminant and $\left\langle \mathbf{S^2}\right\rangle \approx 4.66$
after annihilation.
In view of these unacceptable values, it is not at all surprising that properties
obtained within unrestricted CC calculations are inacceptable. To illustrate,
(unrestricted) UCCSD calculations yielded a dipole $D=5.586$\,debye.
The value obtained from (restricted open shell) ROCCSD calculations
is $D \approx 7.2$\,debye; % $D=7.192$\,debye;
it reasonably agrees with the UB3LYP/ROB3LYP-based values of 
$D\approx 6.8$\,debye.
\section{Remarks on Previous Work Reporting Single-Triplet Alternation}
\label{sec:To-Fan89}
As mentioned in the main text, a singlet-triplet alternation was previously 
\cite{Fan:89} reported to occur in linear carbon (\ce{C_n}, $6 \leq n \leq 10$)
and hydrogen-terminated carbon (\ce{HC_n H}, $2 \leq n \leq 10$) chains.
Indicating the possibility that such linear molecules can have triplet ground states is certainly a remarkable
finding of ref.~\citenum{Fan:89}. Without any polemical intention, we have to note, however,
that the results obtained there were inherently limited by the level of theory
(RHF with DZ or DZP basis sets) and
the computational facilities available at the time it was completed.

To illustrate the \emph{quantitative} inaccuracies of the approach of ref.~\citenum{Fan:89},
let us refer to the case the 
\ce{HC11H} chain. At the triplet optimum, the singlet state is predicted to lie at
$-\Delta_{T}^{RHF/DZ}=4.348$\,eV above the triplet at the optimized geometry of the latter.
According to Table~2  % \ref{table:Delta-hcxh})
of the main text, with the largest Pople 6-311++g(3df, 3pd)
basis sets the B3LYP, CCSD, and CCSD(T) yield $\Delta_{T} = -0.592; -0.783; -0.717$\,eV, respectively.

Concerning the results of ref.~\citenum{Fan:89}, we should still note that the authors themselves were aware
of the fact that their RHF-based approach may be incorrect. They even pointed out that electron
correlations, which escape the RHF framework, may play an essential role
to correctly describe the singlet-triplet separation.
\section{Additional Data for Enthalpies of Formation}
\label{sec:suppl-H}
In addition to the results presented in the main text
(\emph{cf.}~Tables~1, 3, 5, 7, and 9), % (\ref{table:H-hcxh}, \ref{table:H-hcxn}, \ref{table:H-cxs}, \ref{table:H-cxo}, and \ref{table:H-ocxo}),
Tables~\ref{table:suppl-H-hcxh}, \ref{table:suppl-H-hcxn}, \ref{table:suppl-H-cxs}, and \ref{table:suppl-H-cxo}
collect results for enthalpies of formation obtained
\ib{at the B3LYP/CBSB7 level of theory, CBSB7 being a standard basis set 6-311G(2d,d,p)\,(5D,\,7F) utilized
to enforce convergence to the complete basis (CBS) limit in conjunction with}
the CBS-QB3 protocol as implemented in GAUSSIAN 16 \cite{g16}, the enthalpies of formation obtained within the genuine CBS-QB3 protocol,
as well as values reported in ref.~\citenum{Etim:16a}.
\emph{We note that due to misunderstandings related to the GAUSSIAN output files
these B3LYP/CBSB7-based results for enthalpies of formation were inadequately referred to as CBS-QB3-based values for $\Delta_f H_{0,RT}^{0}$ in Tables S1 to S4 in ref.~\citenum{Baldea:2019g}.}

To demonstrate that ref.~\citenum{Etim:16a} implicitly (and incorrectly) took for granted that all \ce{HC}$_{n}$\ce{H}, \ce{HC}$_{n}$\ce{N}, \ce{C}$_{n}$\ce{O}, and \ce{C}$_{n}$\ce{S}
possess singlet ground states, in Tables~\ref{table:suppl-H-hcxh}, \ref{table:suppl-H-hcxn}, \ref{table:suppl-H-cxs}, and \ref{table:suppl-H-cxo} we have also included values of those authors. The small differences between their results and our results for singlet states are due to the different methods employed to compute the enthalpies of formation. These differences are comparable to the differences from the values deduced obtained within the CBS-Q3B protocol \cite{g16}, as visible in the same Tables~\ref{table:suppl-H-hcxh}, \ref{table:suppl-H-hcxn}, \ref{table:suppl-H-cxs}, and \ref{table:suppl-H-cxo}.

\begin{table}[htbp] % [h!]
  \scriptsize % \footnotesize % \small \footnotesize % \scriptsize % \footnotesize % \tiny
  \begin{center}
    \begin{threeparttable}
    %%%%%%%%%%%%%%%%%%%%%%%%%%%%
    \begin{tabular*}{0.54\textwidth}{@{\extracolsep{\fill}}lrrrr}
      \hline
      Molec.
      & $\Delta_{f} H^{0}_{0}$ & $\Delta_{f} H^{0}_{RT}\vert_{S}$
      & $\Delta_{f} H^{0}_{0}$ & $\Delta_{f} H^{0}_{RT}\vert_{T}$
      \\
      \hline
      % hc2h_optimized_singlet_b3lyp_6311++g_3df3pd_dft_ccsd_ST_split.log MA; 
      % hc2h_optimized_triplet_b3lyp_6311++g_3df3pd_dft_ccsd_ST_split.log MA
      % hc2h_optimized_singlet_b3lyp_6311++g_3df3pd_n_IP_EA_ovgf_ST_split.log, MA; forces ok
      % hc2h_optimized_singlet_b3lyp_6311++g_3df3pd_n_nbo_ST_split_IP_EA_both_dft_and_ccsd_t.log, MA; forces ok
% 
      % hc3h_optimized_singlet_b3lyp_6311++g_3df3pd_n_IP_EA_ovgf_ST_split.log, MA; forces ok
      % hc3h_optimized_triplet_b3lyp_6311++g_3df3pd_n_IP_EA_ovgf_ST_split.log, ulm; forces ok
% 
      % hc4h_optimized_singlet_b3lyp_6311++g_3df3pd_n_IP_EA_ovgf_ST_split.log, ulm; forces ok
      % hc4h_optimized_triplet_b3lyp_6311++g_3df3pd_n_IP_EA_ovgf_ST_split.log, MA; forces ok
% 
% 
      % hc5h_b3lyp_singlet_6311++g_3df3pd_zmat_nbo_IP_EA_ST_split_ovgf.log, MA; forces ok
      % hc5h_b3lyp_triplet_6311++g_3df3pd_zmat_nbo_IP_EA_ST_split_ovgf.log, MA; forces ok
      \ce{HC5H} \tnote{$\dagger$}  & 183.280 & 184.472 & 166.461 & 167.299 \\
      % hc6h_optimized_singlet_b3lyp_6311++g_3df3pd_n_IP_EA_ovgf_ST_split.log, MA; forces ok
      % hc6h_triplet_b3lyp_6311++g_3df3pd_n_IP_EA_ovgf_ST_split.log, MA; forces ok
% 
      % hc5h_triplet_cbs-qb3.log, ulm: t=-191.566534340; zmpt=0.037715; ent=0.044500; hmt=-191.528820; hm=hmt; en=ent; zpm=zpmt
      \ce{HC5H} \tnote{*} % B3LYP/CBSB7 
      & 189.935 & 191.144 & 172.182 & 173.169 \\
      \ce{HC5H} \tnote{$\aleph$} % CBS-QB3 
      & 193.822 & 198.320 & 175.078 & 179.358 \\     
      \ce{HC5H} \tnote{$\ddagger $} % ref.~\citenum{Etim:16a}
      & 185.283 & --- & --- & --- \\
      \hline
      \ce{HC6H} \tnote{$\dagger $}  & 163.773 & 164.660 & 227.837 & 229.260 \\
      \ce{HC6H} \tnote{*} % B3LYP/CBSB7
      & 170.909 & 171.863 & 235.42 & 236.907 \\
      \ce{HC6H} \tnote{$\aleph$} % CBS-QB3
      & 166.273 & 170.773 & 242.988 & 248.019 \\
      \ce{HC6H} \tnote{$\ddagger $} % ref.~\citenum{Etim:16a}
      & 161.678 & --- & --- & --- \\
      \hline
      % hc7h_optimized_singlet_b3lyp_6311++g_3df3pd_n_IP_EA_ovgf_ST_split.log, ulm; forces ok
      % hc7h_optimized_triplet_b3lyp_6311++g_3df3pd_zmat_n_IP_EA_ST_split_ovgf.log, ulm; forces ok
      \ce{HC7H} \tnote{$\dagger$}  & 229.287 & 231.208 & 212.236 & 213.732 \\
      % hc8h_optimized_singlet_b3lyp_6311++g_3df3pd_n_IP_EA_ovgf_ST_split.log, ulm; forces ok
      % hc8h_optimized_triplet_b3lyp_6311++g_3df3pd_n_IP_EA_ovgf_ST_split.log, ulm; forces ok
% 
% 
      \ce{HC7H} \tnote{*} % B3LYP/CBSB7
      & 237.178 & 239.330 & 220.553 & 222.089 \\
      \ce{HC7H} \tnote{$\aleph$} % CBS-QB3
      & 243.944 & 249.893 & 224.401 & 229.74 \\
      \ce{HC7H} \tnote{$\ddagger $} % ref.~\citenum{Etim:16a}
      & 226.691 & --- & --- & --- \\
      \hline
      \ce{HC8H} \tnote{$\dagger$}  & 215.390 & 216.935 & 270.313 & 272.668 \\
      \ce{HC8H} \tnote{*} % B3LYP/CBSB7 
      & 224.660 & 226.344 & 280.042 & 282.585 \\
      \ce{HC8H} \tnote{$\aleph$} % CBS-QB3
      & 220.208 & 225.947 & 290.236 & 296.832 \\
      \ce{HC8H} \tnote{$\ddagger $}
      & 212.684 & --- & --- & --- \\
      \hline
      % hc9h_optimized_singlet_b3lyp_6311++g_3df3pd_n_IP_EA_ovgf_ST_split.log, ulm; forces ok
      % hc9h_triplet_b3lyp_6311++g_3df3pd_n_IP_EA_ovgf_ST_split.log, MA; forces ok
      \ce{HC9H}  \tnote{$\dagger$} & 275.106 & 277.356 & 259.606 & 261.482 \\
      % hc9h_singlet_cbs-qb3.log, ulm: s=-343.915194016; zpms=0.059596; ens=0.070394; hms=-343.855598; hm=hms; en=ens; zpm=zpms;
      % hc9h_triplet_cbs-qb3.log, ulm: t=-343.939360120; zpmt=0.058627; ent=0.068736; hmt=-343.880733; hm=hmt; en=ent; zpm=zpmt;
      \ce{HC9H} \tnote{*} % B3LYP/CBSB7               
      & 285.570 & 288.076 & 269.797 & 271.871 \\
      \ce{HC9H} \tnote{$\aleph$} % CBS-QB3
      & 293.872 & 300.685 & 273.161 & 279.545 \\
      \hline
      % hc10h_optimized_singlet_b3lyp_6311++g_3df3pd_n_IP_EA_ovgf_ST_split.log, ulm; forces ok
      % hc10h_triplet_b3lyp_6311++g_3df3pd_n_IP_EA_ovgf_ST_split.log, ulm; forces ok
      \ce{HC10H} \tnote{$\dagger$} & 266.706 & 268.921 & 314.956 & 317.451 \\
      \ce{HC10H} \tnote{*} % B3LYP/CBSB7
      & 278.310 & 280.664 & 326.43 & 329.279 \\  
      \ce{HC10H} \tnote{$\aleph$} % CBS-QB3
      & 274.294 & 281.211 & 334.920 & 342.332 \\
      \hline
      % hc11h_optimized_singlet_b3lyp_6311++g_3df3pd_n_ST_split_IP_EA.log, ulm, 
      % hc11h_optimized_triplet_b3lyp_6311++g_3df3pd_n_IP_EA_ovgf_ST_split.log, ulm; forces ok
      \ce{HC11H} \tnote{$\dagger$} & 322.336 & 325.149 & 307.943 & 310.455 \\
      \ce{HC11H} \tnote{*} % B3LYP/CBSB7
      & 334.832 & 338.981 & 320.16 & 322.979 \\
      \ce{HC11H} \tnote{$\aleph$} % CBS-QB3
      & 344.598 & 352.556 & 324.389 & 332.026 \\
 %     \hline
      % hc12h_optimized_singlet_b3lyp_6311++g_3df3pd_n_IP_EA_ovgf_ST_split.log, ulm; forces ok
      % hc12h_triplet_b3lyp_6311++g_3df3pd_n_IP_EA_ovgf_ST_split.log, ulm; forces ok
 %     \ce{HC12H} & 317.923 & 320.797 & 360.526 & 363.516 \\
      \hline
    \end{tabular*}
    \begin{tablenotes}\footnotesize
    \item[$\dagger $] Based on the DFT/B3LYP/6-311++(3df, 3pd) approach, same as in the main text
    \item[*] Based on the DFT/B3LYP/CBSB7
    \item[$\aleph$] CBS-QB3 protocol
    \item[$\ddagger $] From ref.~\citenum{Etim:16a}
    \end{tablenotes}
    \end{threeparttable}
     %%%%%%%%%%%%%%%%%%%%%%%%%%%%%%%%%%%%%%%%%%%%%%%%%%%%%%%%%%%%%%%%%%%%%%%%%%%%%%%%%%%%%%
    \caption{Enthalpies of formation of \ce{HC_{n} H} chains. Notice that the values of ref.~\citenum{Etim:16a}
      correspond to singlet states. (The small differences between the values obtained singlet states within
      the DFT/B3LYP/6-311++g(3df, 3pd) approach and those of ref.~\citenum{Etim:16a} are due to the different
      method utilized there; they are comparable to the differences from the values obtained by means of the
      CBS-QB3 protocol \cite{g16}.)
    }
    % \ib{Compare with Table~9 and Fig.~10 of ref.~\citenum{Etim:16a}. IR spectrum of the singlet (nonlinear) is significantly different from that of the triplet (triplet is linear).}
    \label{table:suppl-H-hcxh}
  \end{center}
\end{table}
\begin{table}[htbp] % [h!]
  \scriptsize % \small % \scriptsize % \footnotesize % \tiny
  \begin{center}
    \begin{threeparttable}
    %%%%%%%%%%%%%%%%%%%%%%%%%%%%
    \begin{tabular*}{0.54\textwidth}{@{\extracolsep{\fill}}lrrrr}
      \hline
      Molec.
      & $\Delta_{f} H^{0}_{0} $ & $\Delta_{f} H^{0}_{RT}\vert_{S}$
      & $\Delta_{f} H^{0}_{0} $ & $\Delta_{f} H^{0}_{RT}\vert_{T}$
      \\
      \hline
      % hcn_optimized_singlet_b3lyp_6311++g_3df3pd_n_nbo_ST_split_IP_EA_both_dft_and_ccsd_t.log, MA
      % hcn_optimized_triplet_b3lyp_6311++g_3df3pd_n_nbo_ST_split_IP_EA_both_dft_and_ccsd_t.log, MA
% 
      % hc2n_optimized_singlet_b3lyp_6311++g_3df3pd_n_IP_EA_ovgf_ST_split.log ulm
      % hc2n_optimized_triplet_b3lyp_6311++g_3df3pd_n_IP_EA_ovgf_ST_split.log ulm
% 
      % hc3n_optimized_singlet_b3lyp_6311++g_3df3pd_n_nbo_ST_split_IP_EA_both_dft_and_ccsd_t.log MA
      % hc3n_triplet_b3lyp_6311++g_3df3pd_n_nbo_ST_split_IP_EA_both_dft_and_ccsd_t.log, MA
% 
% 
      % hc4n_optimized_singlet_b3lyp_6311++g_3df3pd_n_nbo_ST_split_IP_EA_both_dft_and_ccsd_t.log MA
      % hc4n_triplet_b3lyp_6311++g_3df3pd_n_nbo_ST_split_IP_EA_both_dft_and_ccsd_t.log MA
      \ce{HC4N} \tnote{$\dagger $} & 164.202 & 165.109 & 146.902 & 147.703 \\
      \ce{HC4N} \tnote{$\ddagger $} % ref.~\citenum{Etim:16a}
      & 161.602    & --- & --- & --- \\
      \hline
      % hc5n_optimized_singlet_b3lyp_6311++g_3df3pd_n_nbo_ST_split_IP_EA_both_dft_and_ccsd_t.log, MA
      % hc5n_b3lyp_triplet_6311++g_3df3pd_zmat_n_nbo_ST_split_IP_EA_both_dft_and_ccsd_t.log, MA
      \ce{HC5N} \tnote{$\dagger $} & 141.802 & 142.735 & 205.424 & 206.859 \\
% 
% 
% 
      % hc6n_optimized_singlet_b3lyp_6311++g_3df3pd_n_ST_split_IP_EA_both_dft_and_ccsd_t.log MA
      % hc6n_optimized_triplet_b3lyp_6311++g_3df3pd_n_nbo_ST_split_IP_EA_all_dft_triplet_ccsd_t_only.log MA
      % hc6n_optimized_triplet_b3lyp_6311++g_3df3pd_n_nbo_ST_split_IP_EA_both_dft_and_ccsd_t.log ulm forces ok
      \ce{HC5N}  \tnote{*} % B3LYP/CBSB7
      & 149.434 & 150.348 & failed to & converge \\
      \ce{HC5N}  \tnote{$\aleph$} % CBS-QB3
      & 145.889 & 150.124 & failed to & converge \\
      \ce{HC5N} \tnote{$\ddagger $} % ref.~\citenum{Etim:16a}
      & 140.566 & --- & --- & --- \\
      \hline
      \ce{HC6N} \tnote{$\dagger $} & 209.099 & 210.827 & 191.840 & 193.150 \\
      \ce{HC6N} \tnote{$\ddagger $} % ref.~\citenum{Etim:16a}
      & 211.115    & --- & --- & --- \\
      \hline
      % hc7n_optimized_singlet_b3lyp_6311++g_3df3pd_n_nbo_ST_split_IP_EA_both_dft_and_ccsd_t.log, MA
      % hc7n_optimized_triplet_b3lyp_6311++g_3df3pd_n_nbo_ST_split_IP_EA_both_dft_and_ccsd_t.log, MA
      \ce{HC7N} \tnote{$\dagger $} & 193.670 & 195.288 & 246.645 & 248.688 \\
      \ce{HC7N}  \tnote{*} % B3LYP/CBSB7
      & 203.421 & 205.106 & 262.585 & 264.528 \\
      \ce{HC7N}  \tnote{$\aleph $} % CBS-QB3
      & 200.090 & 205.605 & 272.288 & 278.058 \\
      \ce{HC7N} \tnote{$\ddagger $} % ref.~\citenum{Etim:16a}
      & 191.824 & --- & --- & --- \\
      \hline
      % hc8n_optimized_singlet_b3lyp_6311++g_3df3pd_n_nbo_ST_split_IP_EA_both_dft_and_ccsd_t.log MA
      % hc8n_optimized_triplet_b3lyp_6311++g_3df3pd_n_nbo_ST_split_IP_EA_both_dft_and_ccsd_t.log MA
      \ce{HC8N} \tnote{$\dagger $} & 254.993 & 257.248 & 238.960 & 241.102 \\
      \ce{HC8N} \tnote{*} % B3LYP/CBSB7
      & 265.992 & 268.358 & 249.702 & 251.938 \\
      \ce{HC8N} \tnote{$\aleph$} % CBS-QB3
      & 274.618 & 280.994 & 272.288 & 278.058 \\
      \ce{HC8N} \tnote{$\ddagger $} % ref.~\citenum{Etim:16a}
      & 263.903    & --- & --- & --- \\
      \hline
      % hc9n_optimized_singlet_b3lyp_6311++g_3df3pd_n_nbo_ST_split_IP_EA_both_dft_and_ccsd_t.log ulm
      % hc9n_reoptimized_triplet_b3lyp_6311++g_3df3pd_n_nbo_ST_split_IP_EA_both_dft_and_ccsd_t.log ulm
      \ce{HC9N} \tnote{$\dagger $} & 245.201 & 247.476 & 290.266 & 292.834 \\
      \ce{HC9N} \tnote{*} % B3LYP/CBSB7
      & 257.342 & 259.680 & 302.446 & 305.095 \\
      \ce{HC9N} \tnote{$\aleph $} % CBS-QB3
      & 254.474 & 261.151 & 309.626 & 316.615 \\
      \ce{HC9N} \tnote{$\ddagger $} % ref.~\citenum{Etim:16a}
      & 242.928 & --- & --- & --- \\
      \hline
      %      % hc10n_optimized_b3lyp_6311++g_3df3pd_nbo_ST_split.log, ulm
% 
% 
      % hc10n_optimized_singlet_b3lyp_6311++g_3df3pd_n_nbo_ST_split_IP_EA_both_dft_and_ccsd_t.log ulm
      % hc10n_optimized_triplet_b3lyp_6311++g_3df3pd_n_nbo_ST_split_IP_EA_both_dft_and_ccsd_t.log ulm
      \ce{HC10N} \tnote{$\dagger $} & 301.872 & 304.711 & 287.203 & 289.986 \\
      \ce{HC10N} \tnote{*} % B3LYP/CBSB7
      & 315.453  & 318.318 & 300.514 & 303.336 \\
      \ce{HC10N} \tnote{$\aleph $} % CBS-QB3
      & 326.120 & 333.577 & 305.986 & 313.402 \\
      \ce{HC10N} \tnote{$\ddagger $} % ref.~\citenum{Etim:16a}
      & 317.217    & --- & --- & --- \\
      \hline
      % hc11n_optimized_singlet_b3lyp_6311++g_3df3pd_coord_n_nbo_ST_split.log, ulm ForcesNotChecked!
      % hc11n_optimized_triplet_b3lyp_6311++g_3df3pd_dft_ccsd_t_nbo_ST_split.log ulm  ForcesNotChecked!
      \ce{HC11N} \tnote{$\dagger $} & 296.620 & 299.539 & 336.070 & 339.228 \\ % corrections OK!
      \ce{HC11N} \tnote{*} % B3LYP/CBSB7
      & 310.585 & 313.719 & 350.093 & 353.494 \\
      \ce{HC11N} \tnote{$\aleph $} % CBS-QB3
      & 308.204 & 316.186 & 366.987 & 375.329 \\
      \ce{HC11N} \tnote{$\ddagger $} % ref.~\citenum{Etim:16a}
      & 292.191  & --- & --- & --- \\
      \hline
      % hc12n_optimized_singlet_b3lyp_6311++g_3df3pd_dft_ccsd_t_nbo_ST_split_IP_wo_EA.log ulm ForcesNotChecked!
      % hc12n_optimized_singlet_b3lyp_6311++g_3df3pd_n_nbo_ST_split_IP_EA_both_dft_and_ccsd_t.log ulm Forces Checked!
      % hc12n_b3lyp_triplet_6311++g_3df3pd_zmat_n_IP_EA.log, MA (corrections and forces ok)
      \ce{HC12N} \tnote{$\dagger $} & 349.939 & 353.385 & 336.219 & 339.636 \\
      \ce{HC12N} \tnote{$\ddagger $} % ref.~\citenum{Etim:16a}
      & 372.551    & --- & --- & --- \\
      \hline
    \end{tabular*}
    \begin{tablenotes}\footnotesize
    \item{$\dagger $} Based on the DFT/B3LYP/6-311++(3df, 3pd) approach, same as in the main text
      % Longest chain ever claimed \ce{HC11N} \cite{Bell:97}.
    \item[*] CBS-QB3 protocol
    \item[$\ddagger $] From ref.~\citenum{Etim:16a}
    \end{tablenotes}
    \end{threeparttable}
    %%%%%%%%%%%%%%%%%%%%%%%%%%%%%%%%%%%%%%%%%%%%%%%%%%%%%%%%%%%%%%%%%%%%%%%%%%%%%%%%%%%%%%
    \caption{Enthalpies of formation of HC$_{n}$N chains.
      Notice that the values of ref.~\citenum{Etim:16a}
      correspond to singlet states. (The small differences between the values obtained singlet states within
      the DFT/B3LYP/6-311++g(3df, 3pd) approach and those of ref.~\citenum{Etim:16a} are due to the different
      method utilized there; they are comparable to the differences from the values obtained by means of the
      CBS-QB3 protocol \cite{g16}.)
      %      \ib{HC$_{2n}$N triplet}, HC$_{2n+1}$N singlet. triplet \ce{HC3N} ulm and MA are different; their IR spectra are also different! IR spectra of s- and t-\ce{HC12N} exhibit some differences. Singlet and triplet \ce{HC4N} have significantly different (CCSD) IR spectra.
    }
    \label{table:suppl-H-hcxn}
\end{center}
\end{table}
\begin{table}[htbp] % [h!]
  \scriptsize % \small % \scriptsize % \footnotesize % \tiny
  \begin{center}
    \begin{threeparttable}
  %%%%%%%%%%%%%%%%%%%%%%%%%%%%
    \begin{tabular*}{0.54\textwidth}{@{\extracolsep{\fill}}lrrrr}
      \hline
      Molec.
      & $\Delta_{f} H^{0}_{0} $ & $\Delta_{f} H^{0}_{RT}\vert_{S}$
      & $\Delta_{f} H^{0}_{0} $ & $\Delta_{f} H^{0}_{RT}\vert_{T}$
      \\
      \hline
      % cs_singlet_b3lyp_6311++g_3df3pd_n_nbo_ST_split_IP_EA_both_dft_and_ccsd_t.log, ulm
      % cs_triplet_b3lyp_6311++g_3df3pd_n_nbo_ST_split_IP_EA_both_dft_and_ccsd_t.log, ulm
      \ce{CS} \tnote{$\dagger $} &  70.934 &  69.634 & 146.830 & 147.617 \\
      \ce{CS} \tnote{$\ast $} % B3LYP/CBSB7
      & 73.630 & 74.411 & 149.652 & 150.440 \\
      \ce{CS} \tnote{$\aleph $} % CBS-QB3
      & 65.464 & 67.546 & 144.920 & 147.008 \\
      \ce{CS} \tnote{$\ddagger $} % ref.~\citenum{Etim:16a}
      & 72.404 & --- & --- & --- \\
      \hline
      \ce{C2S} \tnote{$\dagger $} & 162.418 & 163.565 & 145.630 & 146.834 \\
      \ce{C2S} \tnote{$\ast $} % B3LYP/CBSB7
      & 167.080 & 168.219 & 150.033 & 151.226 \\
      \ce{C2S} \tnote{$\aleph $} % CBS-QB3
      & 159.061 & 161.754 & 149.895 & 152.641 \\
      \ce{C2S} \tnote{$\ddagger $} % ref.~\citenum{Etim:16a}
      & 141.307 & --- & --- & --- \\
      \hline
      \ce{C3S} \tnote{$\dagger $} & 135.505 & 136.805 & 190.378 & 191.944 \\
      \ce{C3S} \tnote{$\ast $} % B3LYP/CBSB7
      & 141.215 & 142.514 & 195.902 & 197.456 \\
      \ce{C3S} \tnote{$\aleph $} % CBS-QB3
      & 137.088 & 140.194 & 197.569 & 200.931 \\
      \ce{C3S} \tnote{$\ddagger $} % ref.~\citenum{Etim:16a}
      & 136.292 & --- & --- & --- \\
      \hline
      \ce{C4S} \tnote{$\dagger $} & 196.846 & 198.489 & 182.693 & 184.308 \\
      \ce{C4S} \tnote{$\ast $} % B3LYP/CBSB7
      & 203.886 & 205.563 & 189.699 & 191.346 \\
      \ce{C4S} \tnote{$\aleph $} % CBS-QB3
      & 199.634 & 203.374 & 192.348 & 196.058 \\
      \ce{C4S} \tnote{$\ddagger $} % ref.~\citenum{Etim:16a}
      & 196.958 & --- & --- & --- \\
      \hline
      \ce{C5S} \tnote{$\dagger $}   & 191.231 & 193.097 & 232.108 & 234.229 \\
      \ce{C5S} \tnote{$\ast $} % B3LYP/CBSB7
      & 199.412 & 201.276 & 240.452 & 242.625 \\
      \ce{C5S} \tnote{$\aleph $} % CBS-QB3
      & 199.486 & 203.668 & 246.506 & 250.997 \\
      \ce{C5S} \tnote{$\ddagger $} % ref.~\citenum{Etim:16a}
      & 190.684 & --- & --- & --- \\
      \hline
      \ce{C6S} \tnote{$\dagger $} & 243.319 & 245.553 & 232.214 & 234.417 \\
      \ce{C6S} \tnote{$\ddagger $}
      & 253.179 & 255.336 & 241.955 & 244.109 \\
      \ce{C6S} \tnote{$\ddagger $} % ref.~\citenum{Etim:16a}
      & 254.207 & --- & --- & --- \\
      \hline
      \ce{C7S} \tnote{$\dagger $} & 245.291 & 247.800 & 276.777 & 279.455 \\
      \ce{C7S} \tnote{*} % B3LYP/CBSB7   
      & 255.720 & 258.298 & 290.27 & 293.178 \\
      \ce{C7S} \tnote{$\aleph $} % CBS-QB3
      & 259.573 & 264.977 & 291.448 & 297.183 \\
      \ce{C7S} \tnote{$\ddagger $} % ref.~\citenum{Etim:16a}
      & 243.788 & --- & --- & --- \\
      \hline
    \end{tabular*}      
    %%%%%%%%%%%%%%%%%%%%%%%%%%%%%%%%%%%%%%%%%%%%%%%%%%%%%%%%%%%%%%%%%%%%%%%%%%%%%%%%%%%%%%
    \begin{tablenotes}\footnotesize
    \item[$\dagger $] Based on the DFT/B3LYP/6-311++(3df, 3pd) approach, same as in the main text
    \item[$\ast $] B3LYP/CBSB7
    \item[$\aleph $] CBS-QB3 protocol
    \item[$\ddagger $] From ref.~\citenum{Etim:16a}
    \end{tablenotes}
    \end{threeparttable}
    %%%%%%%%%%%%%%%%%%%%%%%%%%%%%%%%%%%%%%%%%%%%%%%%%%%%%%%%%%%%%%%%%%%%%%%%%%%%%%%%%%%%%%
    \caption{Enthalpies of formation of \ce{C_{n} S} chains.
      Notice that the values of ref.~\citenum{Etim:16a}
      correspond to singlet states. (The small differences between the values obtained singlet states within
      the DFT/B3LYP/6-311++g(3df, 3pd) approach and those of ref.~\citenum{Etim:16a} are due to the different
      method utilized there; they are comparable to the differences from the values obtained by means of the
      CBS-QB3 protocol \cite{g16}.)
    }
    \label{table:suppl-H-cxs}
  \end{center}
\end{table}
\begin{table}[htbp] % [h!]
  \scriptsize % \small % \scriptsize % \footnotesize % \tiny
  \begin{center}
    \begin{threeparttable}
  %%%%%%%%%%%%%%%%%%%%%%%%%%%%
    \begin{tabular*}{0.54\textwidth}{@{\extracolsep{\fill}}lrrrr}
      \hline
      Molec.
      & $\Delta_{f} H^{0}_{0} $ & $\Delta_{f} H^{0}_{RT}\vert_{S}$
      & $\Delta_{f} H^{0}_{0} $ & $\Delta_{f} H^{0}_{RT}\vert_{T}$
      \\
      \hline
      % co_singlet_b3lyp_6311++g_3df3pd_n_nbo_ST_split_IP_EA_both_dft_and_ccsd_t_ovgf.log MA
      % co_triplet_b3lyp_6311++g_3df3pd_n_nbo_ST_split_IP_EA_both_dft_and_ccsd_t_ovgf.log MA
      \ce{CO} \tnote{$\dagger $}    &  -24.140 & -23.356  & 110.602 & 111.387 \\
      % oc2_singlet_b3lyp_6311++g_3df3pd_n_nbo_ST_split_IP_EA_both_dft_and_ccsd_t.log MA, forces ok
      % oc2_triplet_b3lyp_6311++g_3df3pd_n_nbo_ST_split_IP_EA_both_dft_and_ccsd_t.log MA, forces ok
      \ce{CO} \tnote{$\ddagger $}
      & -23.127 & --- & --- & --- \\
      \hline
      \ce{C2O} \tnote{$\dagger $}   &  110.227 &  111.332 &  85.497 &  86.550 \\
      \ce{C2O} \tnote{$\ddagger $} 
      & 91.369 & --- & --- & --- \\
      \hline
      % oc3_singlet_b3lyp_6311++g_3df3pd_n_nbo_ST_split_IP_EA_both_dft_and_ccsd_t.log MA
      % oc3_triplet_b3lyp_6311++g_3df3pd_n_nbo_ST_split_IP_EA_both_dft_and_ccsd_t.log MA
      \ce{C3O} \tnote{$\dagger $}   &   76.170 &   77.389 & 143.236 & 144.734 \\  
      % oc4_singlet_b3lyp_6311++g_3df3pd_n_nbo_ST_split_IP_EA_both_dft_and_ccsd_t.log MA
      % oc4_triplet_b3lyp_6311++g_3df3pd_n_nbo_ST_split_IP_EA_both_dft_and_ccsd_t.log MA
      \ce{C3O} \tnote{$\ddagger $} & 75.328 & --- & --- & --- \\
      \hline
      \ce{C4O} \tnote{$\dagger $}   &  149.118 &  150.669 & 133.039 & 134.532 \\
      \ce{C4O} \tnote{$\ast $} % B3LYP/CBSB7
      & 154.374 & 149.427 & 138.648 & 140.212 \\
      % c4o_singlet_cbs-qb3.log, ulm: s=-227.459189391; zmps=0.018336; ens=0.024427; hms=-227.440854; zpm=zpms; en=ens; hm=hms;
      % c4o_triplet_cbs-qb3.log, ulm: t=-227.484916268; zpmt=0.019002; ent=0.024745; hmt=-227.465914; zpm=zpmt; en=ent; hm=hmt;
      \ce{C4O} \tnote{$\aleph $} % CBS-QB3
      & 153.646 & 157.480 & 145.245 & 148.862 \\
      \ce{C4O} \tnote{$\ddagger $}
      & 148.640 & --- & --- & --- \\
      \hline
      % oc5_singlet_b3lyp_6311++g_3df3pd_n_nbo_ST_split_IP_EA_both_dft_and_ccsd_t.log MA
      % oc5_triplet_b3lyp_6311++g_3df3pd_n_nbo_ST_split_IP_EA_both_dft_and_ccsd_t.log MA
      \ce{C5O} \tnote{$\dagger $}  &  135.765 &  137.516 & 186.814 & 188.994 \\
      % c5o_singlet_cbs-qb3.log, ulm: s=-265.611953557; zpms=0.025391; ens=0.031964; hms=-265.155963; zpm=zpms; en=ens; hm=hms;
      % c5o_triplet_cbs-qb3.log, ulm: t=-265.526867975; zmpt=0.021040; ent=0.028646; hmt=-265.505828; zpm=zpmt; en=ent; hm=hmt;
      \ce{C5O} \tnote{$\ast $} % B3LYP/CBSB7
      & 142.551 & 144.386 & 193.213 & 196.975 \\
      \ce{C5O} \tnote{$\aleph $} % CBS-QB3
      & 145.935 & 150.076 & 202.711 & 207.499 \\
      \ce{C5O} \tnote{$\ddagger $} 
      & 133.782 & --- & --- & --- \\
      \hline
      % oc6_singlet_b3lyp_6311++g_3df3pd_n_nbo_ST_split_IP_EA_both_dft_and_ccsd_t.log MA
      % oc6_triplet_b3lyp_6311++g_3df3pd_n_nbo_ST_split_IP_EA_both_dft_and_ccsd_t.log MA
      \ce{C6O} \tnote{$\dagger $} &  193.343 &  195.458 & 181.261 & 183.327 \\
      \ce{C6O} \tnote{$\ast $} % B3LYP/CBSB7
      & 201.115 & 203.454 & 189.021 & 191.305 \\
      \ce{C6O} \tnote{$\aleph $} % CBS-QB3
      & 203.944 & 208.844 & 198.524 & 203.370 \\
      \ce{C6O} \tnote{$\ddagger $}
      & 204.526 & --- & --- & --- \\
% 
      % oc7_singlet_b3lyp_6311++g_3df3pd_n_nbo_ST_split_IP_EA_both_dft_and_ccsd_t.log MA
      % oc7_triplet_b3lyp_6311++g_3df3pd_n_nbo_ST_split_IP_EA_both_dft_and_ccsd_t.log MA
% 
      % oc8_singlet_b3lyp_6311++g_3df3pd_n_nbo_ST_split_IP_EA_both_dft_and_ccsd_t.log MA
      % oc8_triplet_b3lyp_6311++g_3df3pd_n_nbo_ST_split_IP_EA_both_dft_and_ccsd_t.log MA
% 
      % oc9_singlet_b3lyp_6311++g_3df3pd_n_nbo_ST_split_IP_EA_both_dft_and_ccsd_t.log ulm
      % oc9_triplet_b3lyp_6311++g_3df3pd_n_nbo_ST_split_IP_EA_both_dft_and_ccsd_t.log ulm
% 
      % oc10_singlet_b3lyp_6311++g_3df3pd_n_nbo_ST_split_IP_EA_both_dft_and_ccsd_t.log ulm
      % oc10_triplet_b3lyp_6311++g_3df3pd_n_nbo_ST_split_IP_EA_both_dft_and_ccsd_t.log ulm
% 
      % oc11_singlet_b3lyp_6311++g_3df3pd_n_nbo_ST_split_IP_EA_both_dft_and_ccsd_t.log ulm
      % oc11_triplet_b3lyp_6311++g_3df3pd_n_nbo_ST_split_IP_EA_both_dft_and_ccsd_t.log ulm
% 
      % oc12_singlet_b3lyp_6311++g_3df3pd_n_nbo_ST_split_IP_EA_both_dft_and_ccsd_t.log ulm
      % oc12_triplet_b3lyp_6311++g_3df3pd_IP_EA_ccsd_t.log ulm
% 
      \hline
    \end{tabular*}      
    %%%%%%%%%%%%%%%%%%%%%%%%%%%%%%%%%%%%%%%%%%%%%%%%%%%%%%%%%%%%%%%%%%%%%%%%%%%%%%%%%%%%%%
    \begin{tablenotes}\footnotesize
    \item[$\dagger $] Based on the DFT/B3LYP/6-311++(3df, 3pd) approach, same as in the main text
      % Longest chain of this family astronomically observed \cite{Matthews:84}. % \ce{C3O} 
    \item[$\ast $] B3LYP/CBSB7
    \item[$\aleph $] CBS-QB3 protocol
    \item[$\ddagger $] From ref.~\citenum{Etim:16a}
    \end{tablenotes}
    \end{threeparttable}
    %%%%%%%%%%%%%%%%%%%%%%%%%%%%%%%%%%%%%%%%%%%%%%%%%%%%%%%%%%%%%%%%%%%%%%%%%%%%%%%%%%%%%%
    \caption{Enthalpies of formation of C$_{n}$O chains.
      Notice that the values of ref.~\citenum{Etim:16a}
      correspond to singlet states. (The small differences between the values obtained singlet states within
      the DFT/B3LYP/6-311++g(3df, 3pd) approach and those of ref.~\citenum{Etim:16a} are due to the different
      method utilized there; they are comparable to the differences from the values obtained by means of the
      CBS-QB3 protocol \cite{g16}.)
      }
    \label{table:suppl-H-cxo}
  \end{center}
\end{table}
\section{HOMO Spatial Distributions and Cartesian Coordinates of Representative Carbon-Based Chains}
\label{sec:homos}
\figsname\ref{fig:homo-hc11h}, \ref{fig:homo-hc12h}, \ref{fig:homo-hc8n}, \ref{fig:homo-hc9n},
\ref{fig:homo-c6s}, \ref{fig:homo-c7s}, \ref{fig:homo-c6o}, \ref{fig:homo-c7o},
\ref{fig:homo-oc6o}, and \ref{fig:homo-oc7o} depict HOMO spatial distributions of singlet and triplet
isomers of representative carbon chains analyzed in the main text.
\begin{figure*} % {hbtp}
  \centerline{
    \includegraphics[width=0.3\textwidth,angle=0]{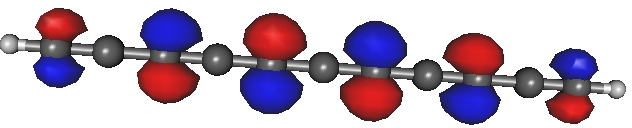}
    \includegraphics[width=0.3\textwidth,angle=0]{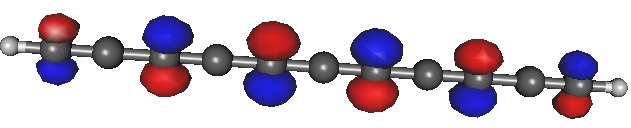}
    \includegraphics[width=0.3\textwidth,angle=0]{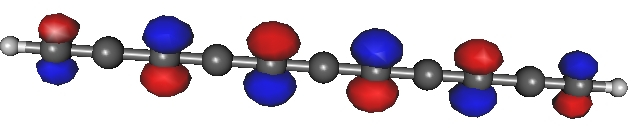}
  }
  \caption{HOMO distributions of \ce{HC11H} chains. Left to right: singlet; triplet alpha; triplet beta.}
  \label{fig:homo-hc11h}
\end{figure*}
\begin{figure*} % {hbtp}
  \centerline{
    \includegraphics[width=0.3\textwidth,angle=0]{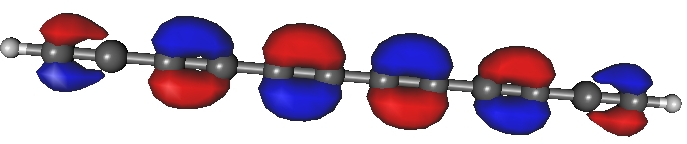}
    \includegraphics[width=0.3\textwidth,angle=0]{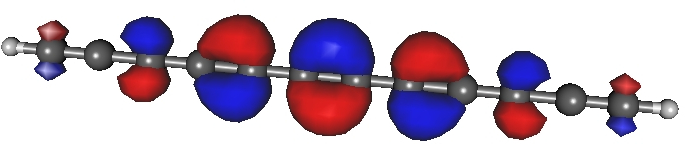}
    \includegraphics[width=0.3\textwidth,angle=0]{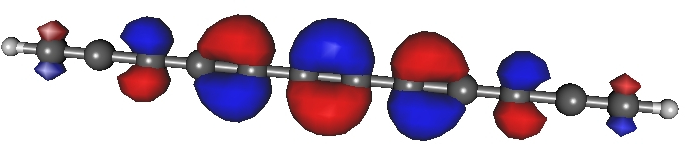}
  }
  \caption{HOMO distributions of \ce{HC12H} chains. Left to right: singlet; triplet alpha; triplet beta.}
  \label{fig:homo-hc12h}
\end{figure*}
\begin{figure*} % {hbtp}
  \centerline{
    \includegraphics[width=0.3\textwidth,angle=0]{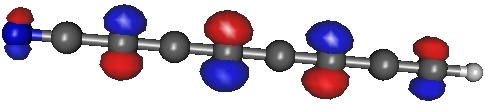}
    \includegraphics[width=0.3\textwidth,angle=0]{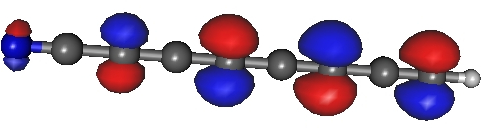}
    \includegraphics[width=0.3\textwidth,angle=0]{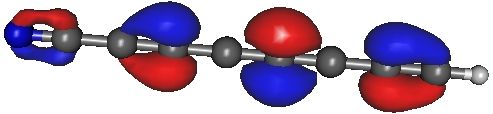}
  }
  \caption{HOMO distributions of \ce{HC8N} chains. Left to right: singlet; triplet alpha; triplet beta.}
  \label{fig:homo-hc8n}
\end{figure*}
\begin{figure*} % {hbtp}
  \centerline{
    \includegraphics[width=0.3\textwidth,angle=0]{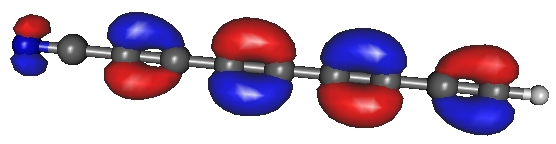}
    \includegraphics[width=0.3\textwidth,angle=0]{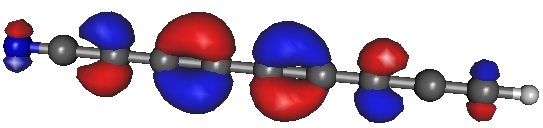}
    \includegraphics[width=0.3\textwidth,angle=0]{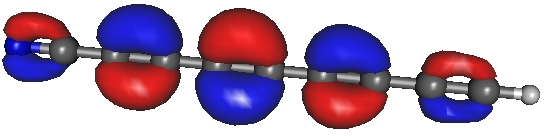}
  }
  \caption{HOMO distributions of \ce{HC9N} chains. Left to right: singlet; triplet alpha; triplet beta.}
  \label{fig:homo-hc9n}
\end{figure*}
\begin{figure*} % {hbtp}
  \centerline{
    \includegraphics[width=0.3\textwidth,angle=0]{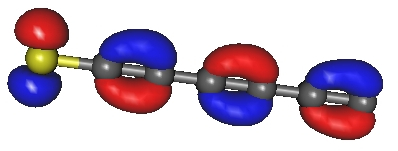}
    \includegraphics[width=0.3\textwidth,angle=0]{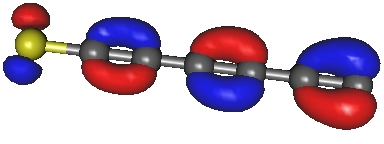}
    \includegraphics[width=0.3\textwidth,angle=0]{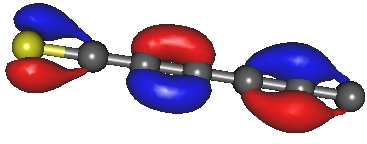}
  }
  \caption{HOMO distributions of \ce{C6S} chains. Left to right: singlet; triplet alpha; triplet beta.}
  \label{fig:homo-c6s}
\end{figure*}
\begin{figure*} % {hbtp}
  \centerline{
    \includegraphics[width=0.3\textwidth,angle=0]{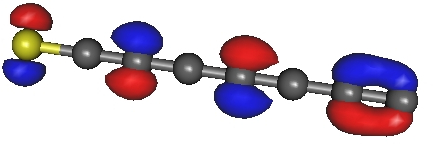}
    \includegraphics[width=0.3\textwidth,angle=0]{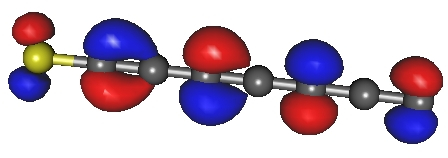}
    \includegraphics[width=0.3\textwidth,angle=0]{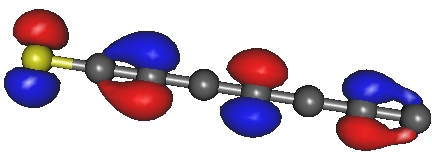}
  }
  \caption{HOMO distributions of \ce{C7S} chains. Left to right: singlet; triplet alpha; triplet beta.}
  \label{fig:homo-c7s}
\end{figure*}
\begin{figure*} % {hbtp}
  \centerline{
    \includegraphics[width=0.3\textwidth,angle=0]{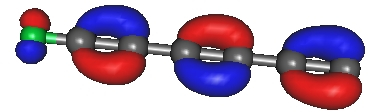}
    \includegraphics[width=0.3\textwidth,angle=0]{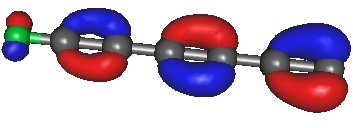}
    \includegraphics[width=0.3\textwidth,angle=0]{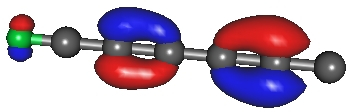}
  }
  \caption{HOMO distributions of \ce{C6O} chains. Left to right: singlet; triplet alpha; triplet beta.}
  \label{fig:homo-c6o}
\end{figure*}
\begin{figure*} % {hbtp}
  \centerline{
    \includegraphics[width=0.3\textwidth,angle=0]{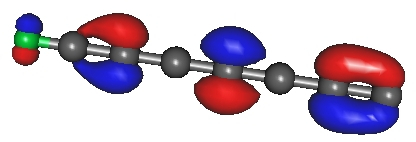}
    \includegraphics[width=0.3\textwidth,angle=0]{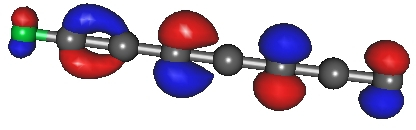}
    \includegraphics[width=0.3\textwidth,angle=0]{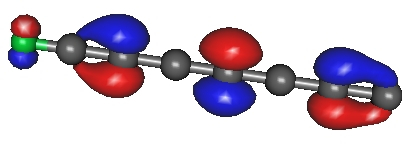}
  }
  \caption{HOMO distributions of \ce{C7O} chains. Left to right: singlet; triplet alpha; triplet beta.}
  \label{fig:homo-c7o}
\end{figure*}
\begin{figure*} % {hbtp}
  \centerline{
    \includegraphics[width=0.3\textwidth,angle=0]{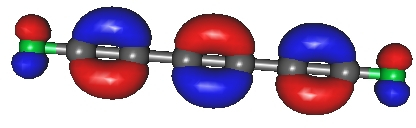}
    \includegraphics[width=0.3\textwidth,angle=0]{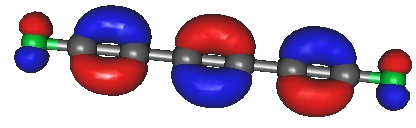}
    \includegraphics[width=0.3\textwidth,angle=0]{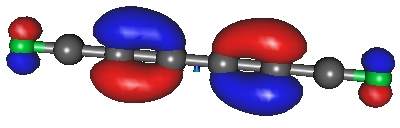}
  }
  \caption{HOMO distributions of \ce{OC6O} chains. Left to right: singlet; triplet alpha; triplet beta.}
  \label{fig:homo-oc6o}
\end{figure*}
\begin{figure*} % {hbtp}
  \centerline{
    \includegraphics[width=0.3\textwidth,angle=0]{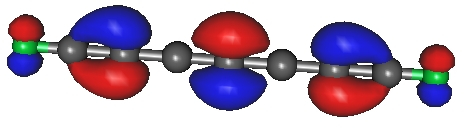}
    \includegraphics[width=0.3\textwidth,angle=0]{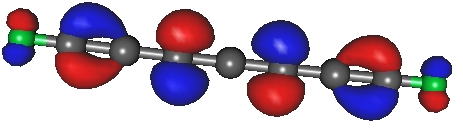}
    \includegraphics[width=0.3\textwidth,angle=0]{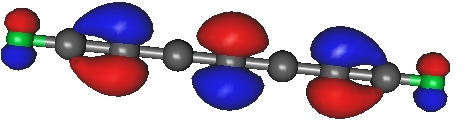}
  }
  \caption{HOMO distributions of \ce{OC7O} chains. Left to right: singlet; triplet alpha; triplet beta.}
  \label{fig:homo-oc7o}
\end{figure*}

The corresponding Cartesian coordinates are presented in
Tables~\ref{table:hc11h-xyz-singlet}, \ref{table:hc11h-xyz-triplet},
\ref{table:hc12h-xyz-singlet}, \ref{table:hc12h-xyz-triplet},
\ref{table:hc8n-xyz-singlet}, \ref{table:hc8n-xyz-triplet},
\ref{table:hc9n-xyz-singlet}, \ref{table:hc9n-xyz-triplet},
\ref{table:c6s-xyz-singlet}, \ref{table:c6s-xyz-triplet},
\ref{table:c7s-xyz-singlet}, \ref{table:c7s-xyz-triplet},
\ref{table:c6o-xyz-singlet}, \ref{table:c6o-xyz-triplet},
\ref{table:c7o-xyz-singlet}, \ref{table:c7o-xyz-triplet},
\ref{table:oc6o-xyz-singlet}, \ref{table:oc6o-xyz-triplet},
\ref{table:oc7o-xyz-singlet}, and \ref{table:oc7o-xyz-triplet}, respectively.
\begin{table}[htbp] % [h!]
  \begin{center}
    %%%%%%%%%%%%%%%%%%%%%%%%%%%%
    \begin{tabular*}{0.49\textwidth}{@{\extracolsep{\fill}}rrrr}
      \hline
Atom &                    X        &   Y           &     Z         \\
\hline
 H   &               0.00000000    &  0.00000000   &   7.42219600  \\ 
 C   &               0.00000000    &  0.00000000   &   6.36073600  \\ 
 C   &               0.00000000    &  0.00000000   &   5.14476600  \\ 
 C   &               0.00000000    &  0.00000000   &   3.81191600  \\ 
 C   &               0.00000000    &  0.00000000   &   2.56889600  \\ 
 C   &               0.00000000    &  0.00000000   &   1.26974600  \\ 
 C   &               0.00000000    &  0.00000000   &   0.00001600  \\ 
 C   &               0.00000000    &  0.00000000   &  -1.26971400  \\ 
 C   &               0.00000000    &  0.00000000   &  -2.56886400  \\ 
 C   &               0.00000000    &  0.00000000   &  -3.81187400  \\ 
 C   &               0.00000000    &  0.00000000   &  -5.14473400  \\ 
 C   &               0.00000000    &  0.00000000   &  -6.36085400  \\ 
 H   &               0.00000000    &  0.00000000   &  -7.42243400  \\ 
\hline
    \end{tabular*}      
    %%%%%%%%%%%%%%%%%%%%%%%%%%%%%%%%%%%%%%%%%%%%%%%%%%%%%%%%%%%%%%%%%%%%%%%%%%%%%%%%%%%%%%
    \caption{Geometry of a singlet \ce{HC11H} chain optimized at DFT/B3LYP/6-311++g(3df, 3pd) level. All coordinates in angstrom.}
    \label{table:hc11h-xyz-singlet}
  \end{center}
\end{table}
\begin{table}[htbp] % [h!]
  \begin{center}
    %%%%%%%%%%%%%%%%%%%%%%%%%%%%
    \begin{tabular*}{0.49\textwidth}{@{\extracolsep{\fill}}rrrr}
      \hline
Atom &                    X        &   Y           &     Z         \\
\hline
 H   &                 7.42392900    &  0.21171400    &  0.06333600 \\ 
 C   &                 6.36288700    &  0.18177100    &  0.05413900 \\ 
 C   &                 5.14726900    &  0.14738000    &  0.04354300 \\ 
 C   &                 3.81387300    &  0.10951800    &  0.03186600 \\ 
 C   &                 2.57049600    &  0.07395900    &  0.02112100 \\ 
 C   &                 1.27053900    &  0.03655400    &  0.01010800 \\ 
 C   &                 0.00007100    & -0.00006400    & -0.00053200 \\ 
 C   &                -1.27067200    & -0.03670000    & -0.01098800 \\ 
 C   &                -2.57038800    & -0.07391900    & -0.02168300 \\ 
 C   &                -3.81395000    & -0.10939700    & -0.03191100 \\ 
 C   &                -5.14723200    & -0.14726200    & -0.04290600 \\ 
 C   &                -6.36289500    & -0.18177400    & -0.05294200 \\ 
 H   &                -7.42391600    & -0.21210500    & -0.06223100 \\ 
\hline
    \end{tabular*}      
    %%%%%%%%%%%%%%%%%%%%%%%%%%%%%%%%%%%%%%%%%%%%%%%%%%%%%%%%%%%%%%%%%%%%%%%%%%%%%%%%%%%%%%
    \caption{Geometry of a triplet \ce{HC11H} chain optimized at DFT/B3LYP/6-311++g(3df, 3pd) level. All coordinates in angstrom.}
    \label{table:hc11h-xyz-triplet}
  \end{center}
\end{table}
\begin{table}[htbp] % [h!]
  \begin{center}
    %%%%%%%%%%%%%%%%%%%%%%%%%%%%
    \begin{tabular*}{0.49\textwidth}{@{\extracolsep{\fill}}rrrr}
      \hline
Atom &                    X        &   Y           &     Z         \\
\hline
 H    &                0.00000000    &  0.00000000   &  -8.06870600 \\ 
 C    &                0.00000000    &  0.00000000   &  -7.00712700 \\ 
 C    &                0.00000000    &  0.00000000   &  -5.79877900 \\ 
 C    &                0.00000000    &  0.00000000   &  -4.44974900 \\ 
 C    &                0.00000000    &  0.00000000   &  -3.22901600 \\ 
 C    &                0.00000000    &  0.00000000   &  -1.89190900 \\ 
 C    &                0.00000000    &  0.00000000   &  -0.66724600 \\ 
 C    &                0.00000000    &  0.00000000   &   0.66723600 \\ 
 C    &                0.00000000    &  0.00000000   &   1.89189900 \\ 
 C    &                0.00000000    &  0.00000000   &   3.22900900 \\ 
 C    &                0.00000000    &  0.00000000   &   4.44973600 \\ 
 C    &                0.00000000    &  0.00000000   &   5.79879100 \\ 
 C    &                0.00000000    &  0.00000000   &   7.00715000 \\ 
 H    &                0.00000000    &  0.00000000   &   8.06872600 \\ 
\hline
    \end{tabular*}
    %%%%%%%%%%%%%%%%%%%%%%%%%%%%%%%%%%%%%%%%%%%%%%%%%%%%%%%%%%%%%%%%%%%%%%%%%%%%%%%%%%%%%%
    \caption{Geometry of a singlet \ce{HC12H} chain optimized at DFT/B3LYP/6-311++g(3df, 3pd) level. All coordinates in angstrom.}
    \label{table:hc12h-xyz-singlet}
  \end{center}
\end{table}
\begin{table}[htbp] % [h!]
  \begin{center}
    %%%%%%%%%%%%%%%%%%%%%%%%%%%%
    \begin{tabular*}{0.49\textwidth}{@{\extracolsep{\fill}}rrrr}
      \hline
Atom &                    X        &   Y           &     Z         \\
\hline
 H    &                0.01152900    & -8.06549200    &  0.00000000 \\ 
 C    &                0.01077900    & -7.00401500    &  0.00000000 \\ 
 C    &                0.00890100    & -5.78238000    &  0.00000000 \\ 
 C    &                0.00726400    & -4.45795200    &  0.00000000 \\ 
 C    &                0.00565600    & -3.20353000    &  0.00000000 \\ 
 C    &                0.00394400    & -1.91363900    &  0.00000000 \\ 
 C    &                0.00208400    & -0.63883700    &  0.00000000 \\ 
 C    &                0.00000000    &  0.63882000    &  0.00000000 \\ 
 C    &               -0.00225500    &  1.91367300    &  0.00000000 \\ 
 C    &               -0.00483300    &  3.20347200    &  0.00000000 \\ 
 C    &               -0.00748200    &  4.45798000    &  0.00000000 \\ 
 C    &               -0.01045400    &  5.78235100    &  0.00000000 \\ 
 C    &               -0.01292200    &  7.00404500    &  0.00000000 \\ 
 H    &               -0.01562000    &  8.06556500    &  0.00000000 \\ 
\hline
\end{tabular*}
    %%%%%%%%%%%%%%%%%%%%%%%%%%%%%%%%%%%%%%%%%%%%%%%%%%%%%%%%%%%%%%%%%%%%%%%%%%%%%%%%%%%%%%
    \caption{Geometry of a triplet \ce{HC12H} chain optimized at DFT/B3LYP/6-311++g(3df, 3pd) level. All coordinates in angstrom.}
    \label{table:hc12h-xyz-triplet}
  \end{center}
\end{table}
\begin{table}[htbp] % [h!]
  \begin{center}
    %%%%%%%%%%%%%%%%%%%%%%%%%%%%
    \begin{tabular*}{0.49\textwidth}{@{\extracolsep{\fill}}rrrr}
      \hline
Atom &                    X        &   Y           &     Z         \\
\hline
 H    &   0.000000    &   0.000000    &   0.000000 \\ 
 C    &   0.000000    &   0.000000    &   1.061990 \\ 
 C    &   0.000000    &   0.000000    &   2.281660 \\ 
 C    &   0.000000    &   0.000000    &   3.605300 \\ 
 C    &   0.000000    &   0.000000    &   4.860230 \\ 
 C    &   0.000000    &   0.000000    &   6.141510 \\ 
 C    &   0.000000    &   0.000000    &   7.433850 \\ 
 C    &   0.000000    &   0.000000    &   8.678190 \\ 
 C    &   0.000000    &   0.000000    &  10.017590 \\ 
 N    &   0.000000    &   0.000000    &  11.184290 \\ 
\hline
\end{tabular*}
    %%%%%%%%%%%%%%%%%%%%%%%%%%%%%%%%%%%%%%%%%%%%%%%%%%%%%%%%%%%%%%%%%%%%%%%%%%%%%%%%%%%%%%
    \caption{Geometry of a singlet \ce{HC8N} chain optimized at DFT/B3LYP/6-311++g(3df, 3pd) level. All coordinates in angstrom.}
    \label{table:hc8n-xyz-singlet}
  \end{center}
\end{table}
\begin{table}[htbp] % [h!]
  \begin{center}
    %%%%%%%%%%%%%%%%%%%%%%%%%%%%
    \begin{tabular*}{0.49\textwidth}{@{\extracolsep{\fill}}rrrr}
      \hline
Atom &                    X        &   Y           &     Z         \\
\hline                                              
 H    &  0.000000    &   0.000000    &   0.050499 \\ 
 C    &  0.000000    &   0.000000    &   1.112529 \\ 
 C    &  0.000000    &   0.000000    &   2.332695 \\ 
 C    &  0.000000    &   0.000000    &   3.657696 \\ 
 C    &  0.000000    &   0.000000    &   4.914141 \\ 
 C    &  0.000000    &   0.000000    &   6.197112 \\ 
 C    &  0.000000    &   0.000000    &   7.491184 \\ 
 C    &  0.000000    &   0.000000    &   8.736581 \\ 
 C    &  0.000000    &   0.000000    &  10.077249 \\ 
 N    &  0.000000    &   0.000000    &  11.243844 \\ 
\hline
\end{tabular*}
    %%%%%%%%%%%%%%%%%%%%%%%%%%%%%%%%%%%%%%%%%%%%%%%%%%%%%%%%%%%%%%%%%%%%%%%%%%%%%%%%%%%%%%
    \caption{Geometry of a triplet \ce{HC8N} chain optimized at DFT/B3LYP/6-311++g(3df, 3pd) level. All coordinates in angstrom.}
    \label{table:hc8n-xyz-triplet}
  \end{center}
\end{table}
\begin{table}[htbp] % [h!]
  \begin{center}
    %%%%%%%%%%%%%%%%%%%%%%%%%%%%
    \begin{tabular*}{0.49\textwidth}{@{\extracolsep{\fill}}rrrr}
      \hline
Atom &                    X        &   Y           &     Z         \\
\hline
 H    &     0.000000    &   0.000000  &  0.008778 \\ 
 C    &     0.000000    &   1.061980  &  0.786078 \\ 
 C    &     0.000000    &   2.269590  & -0.127671 \\ 
 C    &     0.000000    &   3.618640  & -0.345776 \\ 
 C    &     0.000000    &   4.838020  & -0.158290 \\ 
 C    &     0.000000    &   6.176670  & -0.035931 \\ 
 C    &     0.000000    &   7.398190  & -0.589738 \\ 
 C    &     0.000000    &   8.737740  & -0.384900 \\ 
 C    &     0.000000    &   9.954280  &  1.087742 \\ 
 C    &     0.000000    &  11.310900  &  0.917371 \\ 
 N    &     0.000000    &  12.469730  & -1.157929 \\ 
\hline
\end{tabular*}
    %%%%%%%%%%%%%%%%%%%%%%%%%%%%%%%%%%%%%%%%%%%%%%%%%%%%%%%%%%%%%%%%%%%%%%%%%%%%%%%%%%%%%%
    \caption{Geometry of a singlet \ce{HC9N} chain optimized at DFT/B3LYP/6-311++g(3df, 3pd) level. All coordinates in angstrom.}
    \label{table:hc9n-xyz-singlet}
  \end{center}
\end{table}
\begin{table}[htbp] % [h!]
  \begin{center}
    %%%%%%%%%%%%%%%%%%%%%%%%%%%%
    \begin{tabular*}{0.49\textwidth}{@{\extracolsep{\fill}}rrrr}
      \hline
Atom &                    X        &   Y           &     Z         \\
\hline
 H    &                0.00000000    &  0.00000000   &  -6.76924400 \\ 
 C    &                0.00000000    &  0.00000000   &  -5.70728200 \\ 
 C    &                0.00000000    &  0.00000000   &  -4.48002200 \\ 
 C    &                0.00000000    &  0.00000000   &  -3.16567800 \\ 
 C    &                0.00000000    &  0.00000000   &  -1.89896700 \\ 
 C    &                0.00000000    &  0.00000000   &  -0.61966900 \\ 
 C    &                0.00000000    &  0.00000000   &   0.66054500 \\ 
 C    &                0.00000000    &  0.00000000   &   1.94595700 \\ 
 C    &                0.00000000    &  0.00000000   &   3.20223100 \\ 
 C    &                0.00000000    &  0.00000000   &   4.53430900 \\ 
 N    &                0.00000000    &  0.00000000   &   5.70581400 \\ 
\hline
\end{tabular*}
    %%%%%%%%%%%%%%%%%%%%%%%%%%%%%%%%%%%%%%%%%%%%%%%%%%%%%%%%%%%%%%%%%%%%%%%%%%%%%%%%%%%%%%
    \caption{Geometry of a triplet \ce{HC9N} chain optimized at DFT/B3LYP/6-311++g(3df, 3pd) level. All coordinates in angstrom.}
    \label{table:hc9n-xyz-triplet}
  \end{center}
\end{table}
\begin{table}[htbp] % [h!]
  \begin{center}
    %%%%%%%%%%%%%%%%%%%%%%%%%%%%
    \begin{tabular*}{0.49\textwidth}{@{\extracolsep{\fill}}rrrr}
      \hline
Atom &                    X        &   Y           &     Z         \\
\hline
 S   &    0.006105    &  0.000000    &  -0.019989 \\ 
 C   &   -0.017097    &  0.000000    &   1.538551 \\ 
 C   &   -0.036489    &  0.000000    &   2.812844 \\ 
 C   &   -0.056990    &  0.000000    &   4.092023 \\ 
 C   &   -0.078528    &  0.000000    &   5.361258 \\ 
 C   &   -0.101160    &  0.000000    &   6.651218 \\ 
 C   &   -0.124958    &  0.000000    &   7.937546 \\ 
\hline
\end{tabular*}
    %%%%%%%%%%%%%%%%%%%%%%%%%%%%%%%%%%%%%%%%%%%%%%%%%%%%%%%%%%%%%%%%%%%%%%%%%%%%%%%%%%%%%%
    \caption{Geometry of a singlet \ce{C6S} chain optimized at DFT/B3LYP/6-311++g(3df, 3pd) level. All coordinates in angstrom.}
    \label{table:c6s-xyz-singlet}
  \end{center}
\end{table}
\begin{table}[htbp] % [h!]
  \begin{center}
    %%%%%%%%%%%%%%%%%%%%%%%%%%%%
    \begin{tabular*}{0.49\textwidth}{@{\extracolsep{\fill}}rrrr}
      \hline
Atom &                    X        &   Y           &     Z         \\
\hline
 S   &    0.018284    &  0.000000    &  -0.218962 \\ 
 C   &   -0.015319    &  0.000000    &   1.339322 \\ 
 C   &   -0.042451    &  0.000000    &   2.612196 \\ 
 C   &   -0.064440    &  0.000000    &   3.890524 \\ 
 C   &   -0.084996    &  0.000000    &   5.159442 \\ 
 C   &   -0.102917    &  0.000000    &   6.447847 \\ 
 C   &   -0.117813    &  0.000000    &   7.736118 \\ 
\hline
\end{tabular*}
    %%%%%%%%%%%%%%%%%%%%%%%%%%%%%%%%%%%%%%%%%%%%%%%%%%%%%%%%%%%%%%%%%%%%%%%%%%%%%%%%%%%%%%
    \caption{Geometry of a triplet \ce{C6S} chain optimized at DFT/B3LYP/6-311++g(3df, 3pd) level. All coordinates in angstrom.}
    \label{table:c6s-xyz-triplet}
  \end{center}
\end{table}
\begin{table}[htbp] % [h!]
  \begin{center}
    %%%%%%%%%%%%%%%%%%%%%%%%%%%%
    \begin{tabular*}{0.49\textwidth}{@{\extracolsep{\fill}}rrrr}
      \hline
Atom &                    X        &   Y           &     Z         \\
\hline
 S    &               -0.28823100    & -3.87957200   &   0.00000000 \\ 
 C    &               -0.18370000    & -2.33341100   &   0.00000000 \\ 
 C    &               -0.09512400    & -1.05962500   &   0.00000000 \\ 
 C    &                0.00000000    &  0.20359900   &   0.00000000 \\ 
 C    &                0.09921500    &  1.47934800   &   0.00000000 \\ 
 C    &                0.20525100    &  2.73702600   &   0.00000000 \\ 
 C    &                0.31500700    &  4.02367400   &   0.00000000 \\ 
 C    &                0.42796800    &  5.29491500   &   0.00000000 \\ 
\hline
\end{tabular*}
    %%%%%%%%%%%%%%%%%%%%%%%%%%%%%%%%%%%%%%%%%%%%%%%%%%%%%%%%%%%%%%%%%%%%%%%%%%%%%%%%%%%%%%
    \caption{Geometry of a singlet \ce{C7S} chain optimized at DFT/B3LYP/6-311++g(3df, 3pd) level. All coordinates in angstrom.}
    \label{table:c7s-xyz-singlet}
  \end{center}
\end{table}
\begin{table}[htbp] % [h!]
  \begin{center}
    %%%%%%%%%%%%%%%%%%%%%%%%%%%%
    \begin{tabular*}{0.49\textwidth}{@{\extracolsep{\fill}}rrrr}
      \hline
Atom &                    X        &   Y           &     Z         \\
\hline
 S     &              -0.31032700    & -3.89879800    &  0.00000000 \\ 
 C     &              -0.19554500    & -2.33684200    &  0.00000000 \\ 
 C     &              -0.09954800    & -1.06757700    &  0.00000000 \\ 
 C     &               0.00000000    &  0.21110900    &  0.00000000 \\ 
 C     &               0.10794000    &  1.47994100    &  0.00000000 \\ 
 C     &               0.21939900    &  2.75289600    &  0.00000000 \\ 
 C     &               0.33716200    &  4.03332100    &  0.00000000 \\ 
 C     &               0.45812900    &  5.32394700    &  0.00000000 \\ 
\hline
\end{tabular*}
    %%%%%%%%%%%%%%%%%%%%%%%%%%%%%%%%%%%%%%%%%%%%%%%%%%%%%%%%%%%%%%%%%%%%%%%%%%%%%%%%%%%%%%
    \caption{Geometry of a triplet \ce{C7S} chain optimized at DFT/B3LYP/6-311++g(3df, 3pd) level. All coordinates in angstrom.}
    \label{table:c7s-xyz-triplet}
  \end{center}
\end{table}

\begin{table}[htbp] % [h!]
  \begin{center}
    %%%%%%%%%%%%%%%%%%%%%%%%%%%%
    \begin{tabular*}{0.49\textwidth}{@{\extracolsep{\fill}}rrrr}
      \hline
Atom &                    X        &   Y           &     Z         \\
\hline
 O    &   0.002680    &  0.000000    &   0.114732 \\ 
 C    &  -0.015651    &  0.000000    &   1.276918 \\ 
 C    &  -0.035540    &  0.000000    &   2.558251 \\ 
 C    &  -0.056544    &  0.000000    &   3.832815 \\ 
 C    &  -0.078556    &  0.000000    &   5.105768 \\ 
 C    &  -0.101140    &  0.000000    &   6.393987 \\ 
 C    &  -0.124900    &  0.000000    &   7.684016 \\ 
\hline
\end{tabular*}
    %%%%%%%%%%%%%%%%%%%%%%%%%%%%%%%%%%%%%%%%%%%%%%%%%%%%%%%%%%%%%%%%%%%%%%%%%%%%%%%%%%%%%%
    \caption{Geometry of a singlet \ce{C6O} chain optimized at DFT/B3LYP/6-311++g(3df, 3pd) level. All coordinates in angstrom.}
    \label{table:c6o-xyz-singlet}
  \end{center}
\end{table}
\begin{table}[htbp] % [h!]
  \begin{center}
    %%%%%%%%%%%%%%%%%%%%%%%%%%%%
    \begin{tabular*}{0.49\textwidth}{@{\extracolsep{\fill}}rrrr}
      \hline
Atom &                    X        &   Y           &     Z         \\
\hline                                              
 O    &   0.002724     & 0.000000     &  0.117263  \\  
 C    &  -0.015703     & 0.000000     &  1.278675  \\ 
 C    &  -0.035436     & 0.000000     &  2.559020  \\ 
 C    &  -0.056574     & 0.000000     &  3.832259  \\ 
 C    &  -0.078895     & 0.000000     &  5.104970  \\ 
 C    &  -0.101211     & 0.000000     &  6.390915  \\ 
 C    &  -0.124557     & 0.000000     &  7.683385  \\ 
\hline
\end{tabular*}
    %%%%%%%%%%%%%%%%%%%%%%%%%%%%%%%%%%%%%%%%%%%%%%%%%%%%%%%%%%%%%%%%%%%%%%%%%%%%%%%%%%%%%%
    \caption{Geometry of a triplet \ce{C6O} chain optimized at DFT/B3LYP/6-311++g(3df, 3pd) level. All coordinates in angstrom.}
    \label{table:c6o-xyz-triplet}
  \end{center}
\end{table}
\begin{table}[htbp] % [h!]
  \begin{center}
    %%%%%%%%%%%%%%%%%%%%%%%%%%%%
    \begin{tabular*}{0.49\textwidth}{@{\extracolsep{\fill}}rrrr}
      \hline
Atom &                    X        &   Y           &     Z         \\
\hline
   O &     -0.001655  &    0.000000  &     0.141663 \\
   C &     -0.015703  &    0.000000  &     1.298554 \\
   C &     -0.031158  &    0.000000  &     2.579403 \\
   C &     -0.053075  &    0.000000  &     3.842831 \\
   C &     -0.077053  &    0.000000  &     5.121938 \\
   C &     -0.106141  &    0.000000  &     6.383257 \\
   C &     -0.137515  &    0.000000  &     7.673639 \\
   C &     -0.171432  &    0.000000  &     8.948957 \\
\hline
\end{tabular*}
    %%%%%%%%%%%%%%%%%%%%%%%%%%%%%%%%%%%%%%%%%%%%%%%%%%%%%%%%%%%%%%%%%%%%%%%%%%%%%%%%%%%%%%
    \caption{Geometry of a singlet \ce{C7O} chain optimized at DFT/B3LYP/6-311++g(3df, 3pd) level. All coordinates in angstrom.}
    \label{table:c7o-xyz-singlet}
  \end{center}
\end{table}
\begin{table}[htbp] % [h!]
  \begin{center}
    %%%%%%%%%%%%%%%%%%%%%%%%%%%%
    \begin{tabular*}{0.49\textwidth}{@{\extracolsep{\fill}}rrrr}
      \hline
Atom &                    X        &   Y           &     Z         \\
\hline
 O    &  -0.000261    &  0.000000    &   0.116937 \\ 
 C    &  -0.015705    &  0.000000    &   1.282603 \\ 
 C    &  -0.032391    &  0.000000    &   2.564654 \\ 
 C    &  -0.052477    &  0.000000    &   3.841826 \\ 
 C    &  -0.077757    &  0.000000    &   5.119130 \\ 
 C    &  -0.104794    &  0.000000    &   6.397322 \\ 
 C    &  -0.137602    &  0.000000    &   7.683077 \\ 
 C    &  -0.172746    &  0.000000    &   8.984692 \\ 
\hline
\end{tabular*}
    %%%%%%%%%%%%%%%%%%%%%%%%%%%%%%%%%%%%%%%%%%%%%%%%%%%%%%%%%%%%%%%%%%%%%%%%%%%%%%%%%%%%%%
    \caption{Geometry of a triplet \ce{C7O} chain optimized at DFT/B3LYP/6-311++g(3df, 3pd) level. All coordinates in angstrom.}
    \label{table:c7o-xyz-triplet}
  \end{center}
\end{table}
\begin{table}[htbp] % [h!]
  \begin{center}
    %%%%%%%%%%%%%%%%%%%%%%%%%%%%
    \begin{tabular*}{0.49\textwidth}{@{\extracolsep{\fill}}rrrr}
      \hline
Atom &                    X        &   Y           &     Z         \\
\hline
 O    &   0.000000    &   0.000000    &   0.001750 \\ 
 C    &   0.000000    &   0.000000    &   1.169089 \\ 
 C    &   0.000000    &   0.000000    &   2.447533 \\ 
 C    &   0.000000    &   0.000000    &   3.725728 \\ 
 C    &   0.000000    &   0.000000    &   5.000070 \\ 
 C    &   0.000000    &   0.000000    &   6.278262 \\ 
 C    &   0.000000    &   0.000000    &   7.556708 \\ 
 O    &   0.000000    &   0.000000    &   8.724060 \\ 
\hline
\end{tabular*}
    %%%%%%%%%%%%%%%%%%%%%%%%%%%%%%%%%%%%%%%%%%%%%%%%%%%%%%%%%%%%%%%%%%%%%%%%%%%%%%%%%%%%%%
    \caption{Geometry of a singlet \ce{OC6O} chain optimized at DFT/B3LYP/6-311++g(3df, 3pd) level. All coordinates in angstrom.}
    \label{table:oc6o-xyz-singlet}
  \end{center}
\end{table}
\begin{table}[htbp] % [h!]
  \begin{center}
    %%%%%%%%%%%%%%%%%%%%%%%%%%%%
    \begin{tabular*}{0.49\textwidth}{@{\extracolsep{\fill}}rrrr}
      \hline
Atom &                    X        &   Y           &     Z         \\
\hline
 O    &   0.002724    &  0.000000    &   0.117263 \\ 
 C    &  -0.015703    &  0.000000    &   1.278675 \\ 
 C    &  -0.035436    &  0.000000    &   2.559020 \\ 
 C    &  -0.056574    &  0.000000    &   3.832259 \\ 
 C    &  -0.078895    &  0.000000    &   5.104970 \\ 
 C    &  -0.101211    &  0.000000    &   6.390915 \\ 
 C    &  -0.124557    &  0.000000    &   7.683385 \\ 
\hline
\end{tabular*}
    %%%%%%%%%%%%%%%%%%%%%%%%%%%%%%%%%%%%%%%%%%%%%%%%%%%%%%%%%%%%%%%%%%%%%%%%%%%%%%%%%%%%%%
    \caption{Geometry of a triplet \ce{OC6O} chain optimized at DFT/B3LYP/6-311++g(3df, 3pd) level. All coordinates in angstrom.}
    \label{table:oc6o-xyz-triplet}
  \end{center}
\end{table}
\begin{table}[htbp] % [h!]
  \begin{center}
    %%%%%%%%%%%%%%%%%%%%%%%%%%%%
    \begin{tabular*}{0.49\textwidth}{@{\extracolsep{\fill}}rrrr}
      \hline
Atom &                    X        &   Y           &     Z         \\
\hline
 O    &  -0.003872    &  0.000000    &   0.028377 \\ 
 C    &  -0.015441    &  0.000000    &   1.190634 \\ 
 C    &  -0.027672    &  0.000000    &   2.466546 \\ 
 C    &  -0.050153    &  0.000000    &   3.737453 \\ 
 C    &  -0.073579    &  0.000000    &   5.009382 \\ 
 C    &  -0.107730    &  0.000000    &   6.281028 \\ 
 C    &  -0.142310    &  0.000000    &   7.551702 \\ 
 C    &  -0.187463    &  0.000000    &   8.826846 \\ 
 O    &  -0.229527    &  0.000000    &   9.988413 \\ 
\hline
\end{tabular*}
    %%%%%%%%%%%%%%%%%%%%%%%%%%%%%%%%%%%%%%%%%%%%%%%%%%%%%%%%%%%%%%%%%%%%%%%%%%%%%%%%%%%%%%
    \caption{Geometry of a singlet \ce{OC7O} chain optimized at DFT/B3LYP/6-311++g(3df, 3pd) level. All coordinates in angstrom.}
    \label{table:oc7o-xyz-singlet}
  \end{center}
\end{table}
\begin{table}[htbp] % [h!]
  \begin{center}
    %%%%%%%%%%%%%%%%%%%%%%%%%%%%
    \begin{tabular*}{0.49\textwidth}{@{\extracolsep{\fill}}rrrr}
      \hline
Atom &                    X        &   Y           &     Z         \\
\hline
 O    &  -0.002686    &  0.000000    &  -0.000267 \\ 
 C    &  -0.015854    &  0.000000    &   1.170323 \\ 
 C    &  -0.029398    &  0.000000    &   2.450417 \\ 
 C    &  -0.047195    &  0.000000    &   3.730650 \\ 
 C    &  -0.074833    &  0.000000    &   5.009334 \\ 
 C    &  -0.104782    &  0.000000    &   6.287989 \\ 
 C    &  -0.144252    &  0.000000    &   7.567748 \\ 
 C    &  -0.188919    &  0.000000    &   8.847124 \\ 
 O    &  -0.229828    &  0.000000    &  10.017062 \\ 
\hline
\end{tabular*}
    %%%%%%%%%%%%%%%%%%%%%%%%%%%%%%%%%%%%%%%%%%%%%%%%%%%%%%%%%%%%%%%%%%%%%%%%%%%%%%%%%%%%%%
    \caption{Geometry of a triplet \ce{OC7O} chain optimized at DFT/B3LYP/6-311++g(3df, 3pd) level. All coordinates in angstrom.}
    \label{table:oc7o-xyz-triplet}
  \end{center}
\end{table}
\begin{figure*} % {hbtp}
  \centerline{\includegraphics[width=0.45\textwidth,angle=0]{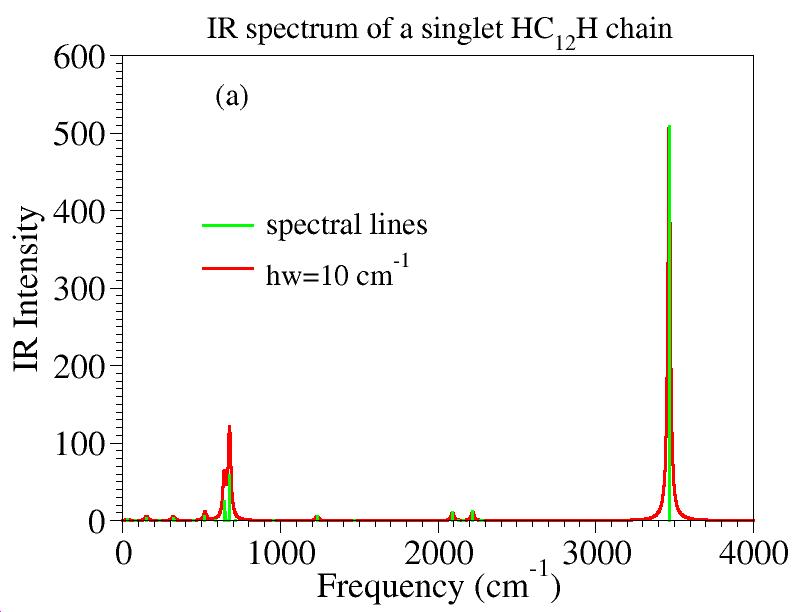}
  \includegraphics[width=0.45\textwidth,angle=0]{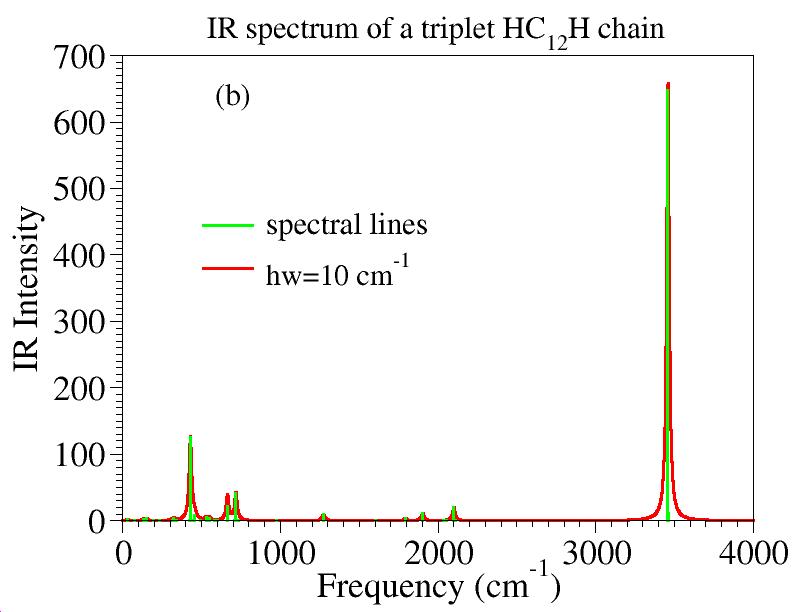}}
  \caption{Infrared spectra of (a) singlet and (b) triplet \ce{HC12H} chains.
    The solid red lines were deduced by convoluting the spectral lines (depicted in green)
    computed within a DFT/B3LYP/6-311++g(3df, 3pd) approach by using Lorentzian distributions
    whose halfwidth (hw) is indicated in the legend.
  }
  \label{fig:ir-hc12h}
\end{figure*}
\begin{figure*} % {hbtp}
  \centerline{\includegraphics[width=0.45\textwidth,angle=0]{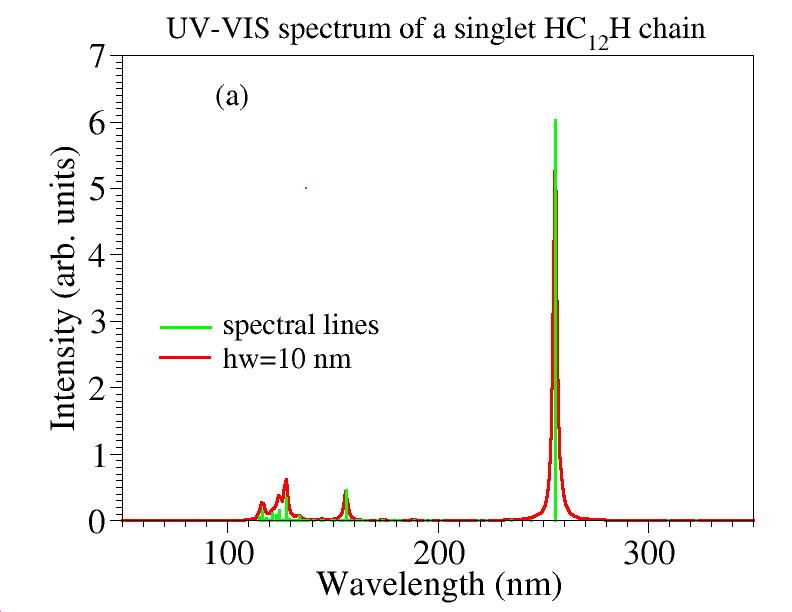}
  \includegraphics[width=0.45\textwidth,angle=0]{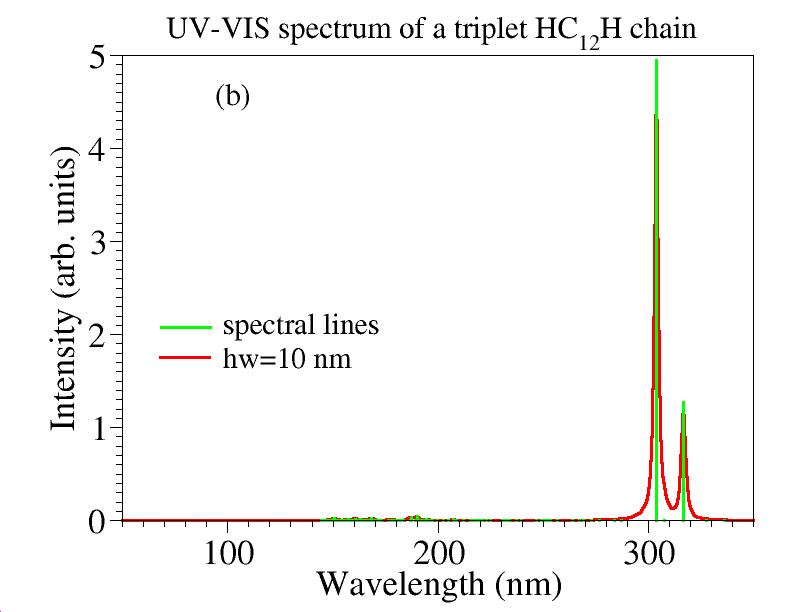}}
  \caption{UV-visible absorption spectra of (a) singlet and (b) triplet \ce{HC12H} chains.
     The solid red lines were deduced by convoluting the spectral lines (depicted in green)
    computed within a TD-DFT/CAM-B3LYP/6-311++g(3df, 3pd) approach by using Lorentzian distributions
    whose halfwidth (hw) is indicated in the legend.
  }
  \label{fig:uv-vis-hc12h}
\end{figure*}


\begin{mcitethebibliography}{75}
\providecommand*\natexlab[1]{#1}
\providecommand*\mciteSetBstSublistMode[1]{}
\providecommand*\mciteSetBstMaxWidthForm[2]{}
\providecommand*\mciteBstWouldAddEndPuncttrue
  {\def\EndOfBibitem{\unskip.}}
\providecommand*\mciteBstWouldAddEndPunctfalse
  {\let\EndOfBibitem\relax}
\providecommand*\mciteSetBstMidEndSepPunct[3]{}
\providecommand*\mciteSetBstSublistLabelBeginEnd[3]{}
\providecommand*\EndOfBibitem{}
\mciteSetBstSublistMode{f}
\mciteSetBstMaxWidthForm{subitem}{(\alph{mcitesubitemcount})}
\mciteSetBstSublistLabelBeginEnd
  {\mcitemaxwidthsubitemform\space}
  {\relax}
  {\relax}

\bibitem[Datta(2005)]{Datta:05}
Datta,~S. \emph{Quantum Transport: Atom to Transistor}; Cambridge Univ. Press:
  Cambridge, 2005\relax
\mciteBstWouldAddEndPuncttrue
\mciteSetBstMidEndSepPunct{\mcitedefaultmidpunct}
{\mcitedefaultendpunct}{\mcitedefaultseppunct}\relax
\EndOfBibitem
\bibitem[Cuevas and Scheer(2017)Cuevas, and Scheer]{CuevasScheer:17}
Cuevas,~J.~C.; Scheer,~E. \emph{Molecular Electronics: An Introduction to
  Theory and Experiment}, 2nd ed.; World Scientific, 2017; World Scientific
  Series in Nanoscience and Nanotechnology: Vol. 15\relax
\mciteBstWouldAddEndPuncttrue
\mciteSetBstMidEndSepPunct{\mcitedefaultmidpunct}
{\mcitedefaultendpunct}{\mcitedefaultseppunct}\relax
\EndOfBibitem
\bibitem[B\^aldea(2015)]{Baldea:2015e}
B\^aldea,~I., Ed. \emph{Molecular Electronics: An Experimental and Theoretical
  Approach}; Pan Stanford, 2015\relax
\mciteBstWouldAddEndPuncttrue
\mciteSetBstMidEndSepPunct{\mcitedefaultmidpunct}
{\mcitedefaultendpunct}{\mcitedefaultseppunct}\relax
\EndOfBibitem
\bibitem[Fazzi and Vozzi(2016)Fazzi, and Vozzi]{Fazzi:16}
Fazzi,~D.; Vozzi,~C. In \emph{Carbon Nanomaterials Sourcebook: Nanofibers,
  Nanoporous Structures, and Nanocomposites}; Sattler,~K.~D., Ed.; CRC Press,
  Boca Raton, FL, USA, 2016; Vol.~2; Chapter 2. Linear Carbon Chains, pp
  27--47\relax
\mciteBstWouldAddEndPuncttrue
\mciteSetBstMidEndSepPunct{\mcitedefaultmidpunct}
{\mcitedefaultendpunct}{\mcitedefaultseppunct}\relax
\EndOfBibitem
\bibitem[Goulay \latin{et~al.}(2009)Goulay, Trevitt, Meloni, Selby, Osborn,
  Taatjes, Vereecken, and Leone]{Goulay:09}
Goulay,~F.; Trevitt,~A.~J.; Meloni,~G.; Selby,~T.~M.; Osborn,~D.~L.;
  Taatjes,~C.~A.; Vereecken,~L.; Leone,~S.~R. Cyclic Versus Linear Isomers
  Produced by Reaction of the Methylidyne Radical (CH) with Small Unsaturated
  Hydrocarbons. \emph{J. Am. Chem. Soc.} \textbf{2009}, \emph{131}, 993--1005,
  PMID: 19123915\relax
\mciteBstWouldAddEndPuncttrue
\mciteSetBstMidEndSepPunct{\mcitedefaultmidpunct}
{\mcitedefaultendpunct}{\mcitedefaultseppunct}\relax
\EndOfBibitem
\bibitem[{Smith} and {Stecher}(1971){Smith}, and {Stecher}]{Smith:71}
{Smith},~A.~M.; {Stecher},~T.~P. Carbon Monoxide in the Interstellar Spectrum
  of Zeta Ophiuchi. \emph{Astrophys. J. Lett.} \textbf{1971}, \emph{164},
  L43\relax
\mciteBstWouldAddEndPuncttrue
\mciteSetBstMidEndSepPunct{\mcitedefaultmidpunct}
{\mcitedefaultendpunct}{\mcitedefaultseppunct}\relax
\EndOfBibitem
\bibitem[{Snyder} and {Buhl}(1971){Snyder}, and {Buhl}]{Snyder:71}
{Snyder},~L.~E.; {Buhl},~D. {Observations of Radio Emission from Interstellar
  Hydrogen Cyanide}. \emph{Astrophys. J. Lett.} \textbf{1971}, \emph{163},
  L47\relax
\mciteBstWouldAddEndPuncttrue
\mciteSetBstMidEndSepPunct{\mcitedefaultmidpunct}
{\mcitedefaultendpunct}{\mcitedefaultseppunct}\relax
\EndOfBibitem
\bibitem[Turner(1971)]{Turner:71}
Turner,~B.~E. Detection of Interstellar Cyanoacetylene. \emph{Astrophys. J.
  Lett.} \textbf{1971}, \emph{163}, L35\relax
\mciteBstWouldAddEndPuncttrue
\mciteSetBstMidEndSepPunct{\mcitedefaultmidpunct}
{\mcitedefaultendpunct}{\mcitedefaultseppunct}\relax
\EndOfBibitem
\bibitem[{Avery} \latin{et~al.}(1976){Avery}, {Broten}, {MacLeod}, {Oka}, and
  {Kroto}]{Avery:76}
{Avery},~L.~W.; {Broten},~N.~W.; {MacLeod},~J.~M.; {Oka},~T.; {Kroto},~H.~W.
  {Detection of the Heavy Interstellar Molecule Cyanodiacetylene}.
  \emph{Astrophys. J. Lett.} \textbf{1976}, \emph{205}, L173--L175\relax
\mciteBstWouldAddEndPuncttrue
\mciteSetBstMidEndSepPunct{\mcitedefaultmidpunct}
{\mcitedefaultendpunct}{\mcitedefaultseppunct}\relax
\EndOfBibitem
\bibitem[{Souza} and {Lutz}(1977){Souza}, and {Lutz}]{Souza:77}
{Souza},~S.~P.; {Lutz},~B.~L. Detection of \ce{C2} in the Interstellar Spectrum
  of Cygnus OB2 Number 12/VI Cygni Number 12/. \emph{Astrophys. J. Lett.}
  \textbf{1977}, \emph{216}, L49--L51\relax
\mciteBstWouldAddEndPuncttrue
\mciteSetBstMidEndSepPunct{\mcitedefaultmidpunct}
{\mcitedefaultendpunct}{\mcitedefaultseppunct}\relax
\EndOfBibitem
\bibitem[{Kroto} \latin{et~al.}(1978){Kroto}, {Kirby}, {Walton}, {Avery},
  {Broten}, {MacLeod}, and {Oka}]{Kroto:78}
{Kroto},~H.~W.; {Kirby},~C.; {Walton},~D.~R.~M.; {Avery},~L.~W.;
  {Broten},~N.~W.; {MacLeod},~J.~M.; {Oka},~T. The Detection of
  Cyanohexatriyne, H(C{$\equiv $} C)$_{3}$CN, in Heile's Cloud 2.
  \emph{Astrophys. J. Lett.} \textbf{1978}, \emph{219}, L133--L137\relax
\mciteBstWouldAddEndPuncttrue
\mciteSetBstMidEndSepPunct{\mcitedefaultmidpunct}
{\mcitedefaultendpunct}{\mcitedefaultseppunct}\relax
\EndOfBibitem
\bibitem[{Broten} \latin{et~al.}(1978){Broten}, {Oka}, {Avery}, {MacLeod}, and
  {Kroto}]{Broten:78}
{Broten},~N.~W.; {Oka},~T.; {Avery},~L.~W.; {MacLeod},~J.~M.; {Kroto},~H.~W.
  The Detection of \ce{HC9N} in Interstellar Space. \emph{Astrophys. J. Lett.}
  \textbf{1978}, \emph{223}, L105--L107\relax
\mciteBstWouldAddEndPuncttrue
\mciteSetBstMidEndSepPunct{\mcitedefaultmidpunct}
{\mcitedefaultendpunct}{\mcitedefaultseppunct}\relax
\EndOfBibitem
\bibitem[Matthews \latin{et~al.}(1984)Matthews, Irvine, Friberg, Brown, and
  Godfrey]{Matthews:84}
Matthews,~H.~E.; Irvine,~W.~M.; Friberg,~P.; Brown,~R.~D.; Godfrey,~P.~D. A New
  Interstellar Molecule: Triearbon Monoxide. \emph{Nature} \textbf{1984},
  \emph{310}, 125--126\relax
\mciteBstWouldAddEndPuncttrue
\mciteSetBstMidEndSepPunct{\mcitedefaultmidpunct}
{\mcitedefaultendpunct}{\mcitedefaultseppunct}\relax
\EndOfBibitem
\bibitem[Hinkle \latin{et~al.}(1988)Hinkle, Keady, and Bernath]{Hinkle:88}
Hinkle,~K.~W.; Keady,~J.~J.; Bernath,~P.~F. Detection of \ce{C3} in the
  Circumstellar Shell of IRC+10216. \emph{Science} \textbf{1988}, \emph{241},
  1319--1322\relax
\mciteBstWouldAddEndPuncttrue
\mciteSetBstMidEndSepPunct{\mcitedefaultmidpunct}
{\mcitedefaultendpunct}{\mcitedefaultseppunct}\relax
\EndOfBibitem
\bibitem[Bernath \latin{et~al.}(1989)Bernath, Hinkle, and Keady]{Bernath:89}
Bernath,~P.~F.; Hinkle,~K.~H.; Keady,~J.~J. Detection of C5 in the
  Circumstellar Shell of IRC+10216. \emph{Science} \textbf{1989}, \emph{244},
  562--564\relax
\mciteBstWouldAddEndPuncttrue
\mciteSetBstMidEndSepPunct{\mcitedefaultmidpunct}
{\mcitedefaultendpunct}{\mcitedefaultseppunct}\relax
\EndOfBibitem
\bibitem[{Ohishi} \latin{et~al.}(1991){Ohishi}, {Suzuki}, {Ishikawa}, {Yamada},
  {Kanamori}, {Irvine}, {Brown}, {Godfrey}, and {Kaifu}]{Ohishi:91}
{Ohishi},~M.; {Suzuki},~H.; {Ishikawa},~S.-I.; {Yamada},~C.; {Kanamori},~H.;
  {Irvine},~W.~M.; {Brown},~R.~D.; {Godfrey},~P.~D.; {Kaifu},~N. Detection of a
  New Carbon-Chain Molecule, CCO. \emph{Astrophys. J. Lett.} \textbf{1991},
  \emph{380}, L39--L42\relax
\mciteBstWouldAddEndPuncttrue
\mciteSetBstMidEndSepPunct{\mcitedefaultmidpunct}
{\mcitedefaultendpunct}{\mcitedefaultseppunct}\relax
\EndOfBibitem
\bibitem[{Gu\'elin} and {Cernicharo}(1991){Gu\'elin}, and
  {Cernicharo}]{Guelin:91}
{Gu\'elin},~M.; {Cernicharo},~J. {Astronomical Detection of the HCCN Radical -
  Toward a New Family of Carbon-Chain Molecules?} \emph{Astron. Astrophys.}
  \textbf{1991}, \emph{244}, L21--L24\relax
\mciteBstWouldAddEndPuncttrue
\mciteSetBstMidEndSepPunct{\mcitedefaultmidpunct}
{\mcitedefaultendpunct}{\mcitedefaultseppunct}\relax
\EndOfBibitem
\bibitem[Bell \latin{et~al.}(1997)Bell, Feldman, Travers, McCarthy, Gottlieb,
  and Thaddeus]{Bell:97}
Bell,~M.~B.; Feldman,~P.~A.; Travers,~M.~J.; McCarthy,~M.~C.; Gottlieb,~C.~A.;
  Thaddeus,~P. Detection of \ce{HC11N} in the Cold Dust Cloud TMC-1.
  \emph{Astrophys. J. Lett.} \textbf{1997}, \emph{483}, L61, See also
  ref.~\citenum{Travers:96b}\relax
\mciteBstWouldAddEndPuncttrue
\mciteSetBstMidEndSepPunct{\mcitedefaultmidpunct}
{\mcitedefaultendpunct}{\mcitedefaultseppunct}\relax
\EndOfBibitem
\bibitem[Cernicharo \latin{et~al.}(2000)Cernicharo, Goicoechea, and
  Caux]{Cernicharo:00}
Cernicharo,~J.; Goicoechea,~J.~R.; Caux,~E. Far-Infrared Detection of \ce{C3}
  in Sagittarius B2 and IRC +10216. \emph{Astrophys. J. Lett.} \textbf{2000},
  \emph{534}, L199\relax
\mciteBstWouldAddEndPuncttrue
\mciteSetBstMidEndSepPunct{\mcitedefaultmidpunct}
{\mcitedefaultendpunct}{\mcitedefaultseppunct}\relax
\EndOfBibitem
\bibitem[Cernicharo \latin{et~al.}(2004)Cernicharo, Gu\'elin, and
  Pardo]{Cernicharo:04}
Cernicharo,~J.; Gu\'elin,~M.; Pardo,~J.~R. Detection of the Linear Radical
  \ce{HC4N} in IRC +10216. \emph{Astrophys. J. Lett.} \textbf{2004},
  \emph{615}, L145\relax
\mciteBstWouldAddEndPuncttrue
\mciteSetBstMidEndSepPunct{\mcitedefaultmidpunct}
{\mcitedefaultendpunct}{\mcitedefaultseppunct}\relax
\EndOfBibitem
\bibitem[Graupner \latin{et~al.}(2008)Graupner, Field, and
  Saunders]{Graupner:08}
Graupner,~K.; Field,~T.~A.; Saunders,~G.~C. Experimental Evidence for Radiative
  Attachment in Astrochemistry from Electron Attachment to NCCCCN.
  \emph{Astrophys. J. Lett.} \textbf{2008}, \emph{685}, L95\relax
\mciteBstWouldAddEndPuncttrue
\mciteSetBstMidEndSepPunct{\mcitedefaultmidpunct}
{\mcitedefaultendpunct}{\mcitedefaultseppunct}\relax
\EndOfBibitem
\bibitem[Etim \latin{et~al.}(2016)Etim, Gorai, Das, Chakrabarti, and
  Arunan]{Etim:16a}
Etim,~E.~E.; Gorai,~P.; Das,~A.; Chakrabarti,~S.~K.; Arunan,~E. Systematic
  Theoretical Study on the Interstellar Carbon Chain Molecules.
  \emph{Astrophys. J.} \textbf{2016}, \emph{832}, 144\relax
\mciteBstWouldAddEndPuncttrue
\mciteSetBstMidEndSepPunct{\mcitedefaultmidpunct}
{\mcitedefaultendpunct}{\mcitedefaultseppunct}\relax
\EndOfBibitem
\bibitem[Eti()]{Etim16aImplicitACS}
Ref.~\citenum{Etim:16b} contains an explicit (and incorrect) assertion on the
  type of ground state (namely, singlet, \emph{cf.}~Table~1 of
  ref.~\citenum{Etim:16b}). Ref.~\citenum{Etim:16a}, a work done by the same
  group, does not explicitly stated what is the spin multiplicity of the state
  of the molecules considered but tacitly admitted singlet ground states of all
  cases where our results indicate triplet ground states. For more details the
  reader is referred to the {\si}.\relax
\mciteBstWouldAddEndPunctfalse
\mciteSetBstMidEndSepPunct{\mcitedefaultmidpunct}
{}{\mcitedefaultseppunct}\relax
\EndOfBibitem
\bibitem[Etim and Arunan(2016)Etim, and Arunan]{Etim:16b}
Etim,~E.~E.; Arunan,~E. Accurate Rotational Constants for Linear Interstellar
  Carbon Chains: Achieving Experimental Accuracy. \emph{Astrophys. Space Sci.}
  \textbf{2016}, \emph{362}, 4\relax
\mciteBstWouldAddEndPuncttrue
\mciteSetBstMidEndSepPunct{\mcitedefaultmidpunct}
{\mcitedefaultendpunct}{\mcitedefaultseppunct}\relax
\EndOfBibitem
\bibitem[Fan and Pfeiffer(1989)Fan, and Pfeiffer]{Fan:89}
Fan,~Q.; Pfeiffer,~G.~V. Theoretical Study of Linear \ce{C_n} (n=6-10) and
  \ce{HC_n H} (n=2-10) Molecules. \emph{Chem. Phys. Lett.} \textbf{1989},
  \emph{162}, 472 -- 478\relax
\mciteBstWouldAddEndPuncttrue
\mciteSetBstMidEndSepPunct{\mcitedefaultmidpunct}
{\mcitedefaultendpunct}{\mcitedefaultseppunct}\relax
\EndOfBibitem
\bibitem[Wu \latin{et~al.}(2016)Wu, Xu, Lu, Khamoshi, Liu, Han, Wu, Lin, Long,
  He, Cai, Yao, Zhang, and Wang]{Wu:16}
Wu,~Z.; Xu,~S.; Lu,~H.; Khamoshi,~A.; Liu,~G.-B.; Han,~T.; Wu,~Y.; Lin,~J.;
  Long,~G.; He,~Y.; Cai,~Y.; Yao,~Y.; Zhang,~F.; Wang,~N. Even-Odd
  Layer-Dependent Magnetotransport of High-Mobility Q-Valley Electrons in
  Transition Metal Disulfides. \emph{Nat. Commun.} \textbf{2016}, \emph{7},
  12955\relax
\mciteBstWouldAddEndPuncttrue
\mciteSetBstMidEndSepPunct{\mcitedefaultmidpunct}
{\mcitedefaultendpunct}{\mcitedefaultseppunct}\relax
\EndOfBibitem
\bibitem[Tao and Bernasek(2007)Tao, and Bernasek]{Tao:07}
Tao,~F.; Bernasek,~S.~L. Understanding Odd-Even Effects in Organic
  Self-Assembled Monolayers. \emph{Chem. Rev.} \textbf{2007}, \emph{107},
  1408--1453, PMID: 17439290\relax
\mciteBstWouldAddEndPuncttrue
\mciteSetBstMidEndSepPunct{\mcitedefaultmidpunct}
{\mcitedefaultendpunct}{\mcitedefaultseppunct}\relax
\EndOfBibitem
\bibitem[B\^aldea \latin{et~al.}(1999)B\^aldea, K\"oppel, and
  Cederbaum]{Baldea:99a}
B\^aldea,~I.; K\"oppel,~H.; Cederbaum,~L.~S. Structural and Magnetic
  Transitions in Ensembles of Mesoscopic Peierls Rings in a Magnetic Flux.
  \emph{Phys. Rev. B} \textbf{1999}, \emph{60}, 6646--6654\relax
\mciteBstWouldAddEndPuncttrue
\mciteSetBstMidEndSepPunct{\mcitedefaultmidpunct}
{\mcitedefaultendpunct}{\mcitedefaultseppunct}\relax
\EndOfBibitem
\bibitem[B\^aldea \latin{et~al.}(1999)B\^aldea, K\"oppel, and
  Cederbaum]{Baldea:99b}
B\^aldea,~I.; K\"oppel,~H.; Cederbaum,~L.~S. Quantum Phonon Fluctuations in
  Mesoscopic Dimerized Systems. \emph{J. Phys. Soc. Jpn.} \textbf{1999},
  \emph{68}, 1954--1962\relax
\mciteBstWouldAddEndPuncttrue
\mciteSetBstMidEndSepPunct{\mcitedefaultmidpunct}
{\mcitedefaultendpunct}{\mcitedefaultseppunct}\relax
\EndOfBibitem
\bibitem[B\^aldea \latin{et~al.}(2001)B\^aldea, K\"oppel, and
  Cederbaum]{Baldea:2001a}
B\^aldea,~I.; K\"oppel,~H.; Cederbaum,~L.~S. Collective Quantum Tunneling of
  Strongly Correlated Electrons in Commensurate Mesoscopic Rings. \emph{Eur.
  Phys. J. B} \textbf{2001}, \emph{20}, 289--299\relax
\mciteBstWouldAddEndPuncttrue
\mciteSetBstMidEndSepPunct{\mcitedefaultmidpunct}
{\mcitedefaultendpunct}{\mcitedefaultseppunct}\relax
\EndOfBibitem
\bibitem[B\^aldea and Cederbaum(2002)B\^aldea, and Cederbaum]{Baldea:2002}
B\^aldea,~I.; Cederbaum,~L.~S. Orbital Picture of Ionization and Its Breakdown
  in Nanoarrays of Quantum Dots. \emph{Phys. Rev. Lett.} \textbf{2002},
  \emph{89}, 133003, selected for Virtual Journal of Nanoscale Science \&
  Technology, \textbf{6} (12) (14 Sept, 2002)\relax
\mciteBstWouldAddEndPuncttrue
\mciteSetBstMidEndSepPunct{\mcitedefaultmidpunct}
{\mcitedefaultendpunct}{\mcitedefaultseppunct}\relax
\EndOfBibitem
\bibitem[H{\"u}ckel(1931)]{Hueckel:31a}
H{\"u}ckel,~E. Quantentheoretische Beitr{\"a}ge zum Benzolproblem. \emph{Z.
  Phys.} \textbf{1931}, \emph{70}, 204--286\relax
\mciteBstWouldAddEndPuncttrue
\mciteSetBstMidEndSepPunct{\mcitedefaultmidpunct}
{\mcitedefaultendpunct}{\mcitedefaultseppunct}\relax
\EndOfBibitem
\bibitem[H{\"u}ckel(1931)]{Hueckel:31b}
H{\"u}ckel,~E. Quanstentheoretische Beitr{\"a}ge zum Benzolproblem. \emph{Z.
  Phys.} \textbf{1931}, \emph{72}, 310--337\relax
\mciteBstWouldAddEndPuncttrue
\mciteSetBstMidEndSepPunct{\mcitedefaultmidpunct}
{\mcitedefaultendpunct}{\mcitedefaultseppunct}\relax
\EndOfBibitem
\bibitem[H{\"u}ckel(1932)]{Hueckel:32}
H{\"u}ckel,~E. Quantentheoretische Beitr{\"a}ge zum Problem der aromatischen
  und unges{\"a}ttigten Verbindungen. III. \emph{Z. Phys.} \textbf{1932},
  \emph{76}, 628--648\relax
\mciteBstWouldAddEndPuncttrue
\mciteSetBstMidEndSepPunct{\mcitedefaultmidpunct}
{\mcitedefaultendpunct}{\mcitedefaultseppunct}\relax
\EndOfBibitem
\bibitem[London(1937)]{London:37}
London,~F. Th\'eorie Quantique des Courants Interatomiques dans les
  Combinaisons Aromatiques. \emph{J. Phys. Radium} \textbf{1937}, \emph{8},
  397--409\relax
\mciteBstWouldAddEndPuncttrue
\mciteSetBstMidEndSepPunct{\mcitedefaultmidpunct}
{\mcitedefaultendpunct}{\mcitedefaultseppunct}\relax
\EndOfBibitem
\bibitem[Frisch \latin{et~al.}(2016)Frisch, Trucks, Schlegel, Scuseria, Robb,
  Cheeseman, Scalmani, Barone, Petersson, Nakatsuji, Li, Caricato, Marenich,
  Bloino, Janesko, Gomperts, Mennucci, Hratchian, Ortiz, Izmaylov, Sonnenberg,
  Williams-Young, Ding, Lipparini, Egidi, Goings, Peng, Petrone, Henderson,
  Ranasinghe, Zakrzewski, Gao, Rega, Zheng, Liang, Hada, Ehara, Toyota, Fukuda,
  Hasegawa, Ishida, Nakajima, Honda, Kitao, Nakai, Vreven, Throssell,
  J.~A.~Montgomery, Peralta, Ogliaro, Bearpark, Heyd, Brothers, Kudin,
  Staroverov, Keith, Kobayashi, Normand, Raghavachari, Rendell, Burant,
  Iyengar, Tomasi, Cossi, Millam, Klene, Adamo, Cammi, Ochterski, Martin,
  Morokuma, Farkas, Foresman, and Fox]{g16}
Frisch,~M.~J.; Trucks,~G.~W.; Schlegel,~H.~B.; Scuseria,~G.~E.; Robb,~M.~A.;
  Cheeseman,~J.~R.; Scalmani,~G.; Barone,~V.; Petersson,~G.~A.; Nakatsuji,~H.;
  Li,~X.; Caricato,~M.; Marenich,~A.~V.; Bloino,~J.; Janesko,~B.~G.;
  Gomperts,~R.; Mennucci,~B.; Hratchian,~H.~P.; Ortiz,~J.~V.; Izmaylov,~A.~F.
  \latin{et~al.}  Gaussian, Inc., Wallingford CT, Gaussian 16, Revision B.01.
  2016; \url{www.gaussian.com}\relax
\mciteBstWouldAddEndPuncttrue
\mciteSetBstMidEndSepPunct{\mcitedefaultmidpunct}
{\mcitedefaultendpunct}{\mcitedefaultseppunct}\relax
\EndOfBibitem
\bibitem[bwHPC(2013)]{bwHPC}
bwHPC, bwHPC program supported by the State of Baden-W\"{u}rttemberg and the
  German Research Foundation (DFG) through grant no INST 40/467-1 FUGG. 2013;
  \url{https://www.bwhpc.de/bwhpc-c5.html}\relax
\mciteBstWouldAddEndPuncttrue
\mciteSetBstMidEndSepPunct{\mcitedefaultmidpunct}
{\mcitedefaultendpunct}{\mcitedefaultseppunct}\relax
\EndOfBibitem
\bibitem[Bartlett and Purvis(1978)Bartlett, and Purvis]{Bartlett:78}
Bartlett,~R.~J.; Purvis,~G.~D. Many-Body Perturbation Theory, Coupled-Pair
  Many-Electron Theory, and the Importance of Quadruple Excitations for the
  Correlation Problem. \emph{Int. J. Quantum Chem.} \textbf{1978}, \emph{14},
  561--581\relax
\mciteBstWouldAddEndPuncttrue
\mciteSetBstMidEndSepPunct{\mcitedefaultmidpunct}
{\mcitedefaultendpunct}{\mcitedefaultseppunct}\relax
\EndOfBibitem
\bibitem[Ochterski(2000)]{Ochterski:00}
Ochterski,~J.~W. Thermochemistry in Gaussian. 2000;
  \url{https://gaussian.com/wp-content/uploads/dl/thermo.pdf}, Pittsburg, PA:
  Gaussian Inc., url:
  http://gaussian.com/wp-content/uploads/dl/thermo.pdf\relax
\mciteBstWouldAddEndPuncttrue
\mciteSetBstMidEndSepPunct{\mcitedefaultmidpunct}
{\mcitedefaultendpunct}{\mcitedefaultseppunct}\relax
\EndOfBibitem
\bibitem[Doney \latin{et~al.}(2018)Doney, Zhao, Stanton, and
  Linnartz]{Doney:18}
Doney,~K.~D.; Zhao,~D.; Stanton,~J.~F.; Linnartz,~H. Theoretical Investigation
  of the Infrared Spectrum of Small Polyynes. \emph{Phys. Chem. Chem. Phys.}
  \textbf{2018}, \emph{20}, 5501--5508\relax
\mciteBstWouldAddEndPuncttrue
\mciteSetBstMidEndSepPunct{\mcitedefaultmidpunct}
{\mcitedefaultendpunct}{\mcitedefaultseppunct}\relax
\EndOfBibitem
\bibitem[Abe(2013)]{Abe:13}
Abe,~M. Diradicals. \emph{Chem. Rev.} \textbf{2013}, \emph{113}, 7011--7088,
  PMID: 23883325\relax
\mciteBstWouldAddEndPuncttrue
\mciteSetBstMidEndSepPunct{\mcitedefaultmidpunct}
{\mcitedefaultendpunct}{\mcitedefaultseppunct}\relax
\EndOfBibitem
\bibitem[Cernicharo \latin{et~al.}(2001)Cernicharo, Heras, Tielens, Pardo,
  Herpin, Gu\'elin, and Waters]{Cernicharo:01}
Cernicharo,~J.; Heras,~A.~M.; Tielens,~A. G. G.~M.; Pardo,~J.~R.; Herpin,~F.;
  Gu\'elin,~M.; Waters,~L. B. F.~M. Infrared Space Observatory's Discovery of
  \ce{C4H2}, \ce{C6H2}, and Benzene in CRL 618. \emph{Astrophys. J. Lett.}
  \textbf{2001}, \emph{546}, L123--L126\relax
\mciteBstWouldAddEndPuncttrue
\mciteSetBstMidEndSepPunct{\mcitedefaultmidpunct}
{\mcitedefaultendpunct}{\mcitedefaultseppunct}\relax
\EndOfBibitem
\bibitem[Sommerfeld and Knecht(2005)Sommerfeld, and Knecht]{Sommerfeld:05}
Sommerfeld,~T.; Knecht,~S. Electronic Interaction Between Valence and
  Dipole-Bound States of the Cyanoacetylene Anion. \emph{Eur. Phys. J. D}
  \textbf{2005}, \emph{35}, 207--216\relax
\mciteBstWouldAddEndPuncttrue
\mciteSetBstMidEndSepPunct{\mcitedefaultmidpunct}
{\mcitedefaultendpunct}{\mcitedefaultseppunct}\relax
\EndOfBibitem
\bibitem[Graupner \latin{et~al.}(2006)Graupner, Merrigan, Field, Youngs, and
  Marr]{Graupner:06}
Graupner,~K.; Merrigan,~T.~L.; Field,~T.~A.; Youngs,~T. G.~A.; Marr,~P.~C.
  Dissociative electron attachment to HCCCN. \emph{New J. Phys.} \textbf{2006},
  \emph{8}, 117\relax
\mciteBstWouldAddEndPuncttrue
\mciteSetBstMidEndSepPunct{\mcitedefaultmidpunct}
{\mcitedefaultendpunct}{\mcitedefaultseppunct}\relax
\EndOfBibitem
\bibitem[B\^aldea(2019)]{Baldea:2019e}
B\^aldea,~I. Long Carbon-Based Chains of Interstellar Medium Can Have a Triplet
  Ground State. Why Is This Important for Astrochemistry? \emph{ACS Earth Space
  Chem.} \textbf{2019}, \emph{3}, 863--872\relax
\mciteBstWouldAddEndPuncttrue
\mciteSetBstMidEndSepPunct{\mcitedefaultmidpunct}
{\mcitedefaultendpunct}{\mcitedefaultseppunct}\relax
\EndOfBibitem
\bibitem[B\^aldea(2019)]{Baldea:2019g}
B\^aldea,~I. Alternation of Singlet and Triplet States in Carbon-Based Chain
  Molecules and Its Astrochemical Implications: Results of an Extensive
  Theoretical Study. \emph{Adv. Theor. Simul.} \textbf{2019}, \emph{2},
  1900084\relax
\mciteBstWouldAddEndPuncttrue
\mciteSetBstMidEndSepPunct{\mcitedefaultmidpunct}
{\mcitedefaultendpunct}{\mcitedefaultseppunct}\relax
\EndOfBibitem
\bibitem[{Cernicharo} \latin{et~al.}(1987){Cernicharo}, {Gu\'elin}, {Hein}, and
  {Kahane}]{Cernicharo:87}
{Cernicharo},~J.; {Gu\'elin},~M.; {Hein},~H.; {Kahane},~C. Sulfur in IRC +
  10216. \emph{Astron. Astrophys.} \textbf{1987}, \emph{181}, L9--L12\relax
\mciteBstWouldAddEndPuncttrue
\mciteSetBstMidEndSepPunct{\mcitedefaultmidpunct}
{\mcitedefaultendpunct}{\mcitedefaultseppunct}\relax
\EndOfBibitem
\bibitem[Ag\'undez \latin{et~al.}(2014)Ag\'undez, Cernicharo, and
  Gu\'elin]{Agundez:14}
Ag\'undez,~M.; Cernicharo,~J.; Gu\'elin,~M. New Molecules in IRC +10216:
  Confirmation of \ce{C5S} and Tentative Identification of MgCCH, NCCP, and
  \ce{SiH3CN}. \emph{Astron. Astrophys.} \textbf{2014}, \emph{570}, A45\relax
\mciteBstWouldAddEndPuncttrue
\mciteSetBstMidEndSepPunct{\mcitedefaultmidpunct}
{\mcitedefaultendpunct}{\mcitedefaultseppunct}\relax
\EndOfBibitem
\bibitem[Pople and Untch(1966)Pople, and Untch]{Pople:66}
Pople,~J.~A.; Untch,~K.~G. Induced Paramagnetic Ring Currents. \emph{J. Am.
  Chem. Soc.} \textbf{1966}, \emph{88}, 4811--4815\relax
\mciteBstWouldAddEndPuncttrue
\mciteSetBstMidEndSepPunct{\mcitedefaultmidpunct}
{\mcitedefaultendpunct}{\mcitedefaultseppunct}\relax
\EndOfBibitem
\bibitem[Jones and Gunnarsson(1989)Jones, and Gunnarsson]{Gunnarson:89}
Jones,~R.~O.; Gunnarsson,~O. The Density Functional Formalism, Its Applications
  and Prospects. \emph{Rev. Mod. Phys.} \textbf{1989}, \emph{61},
  689--746\relax
\mciteBstWouldAddEndPuncttrue
\mciteSetBstMidEndSepPunct{\mcitedefaultmidpunct}
{\mcitedefaultendpunct}{\mcitedefaultseppunct}\relax
\EndOfBibitem
\bibitem[B\^aldea(2012)]{Baldea:2012i}
B\^aldea,~I. Extending the Newns-Anderson Model to Allow Nanotransport Studies
  Through Molecules with Floppy Degrees of Freedom. \emph{Europhys. Lett.}
  \textbf{2012}, \emph{99}, 47002\relax
\mciteBstWouldAddEndPuncttrue
\mciteSetBstMidEndSepPunct{\mcitedefaultmidpunct}
{\mcitedefaultendpunct}{\mcitedefaultseppunct}\relax
\EndOfBibitem
\bibitem[B\^aldea(2013)]{Baldea:2013b}
B\^aldea,~I. Transition Voltage Spectroscopy Reveals Significant Solvent
  Effects on Molecular Transport and Settles an Important Issue in
  Bipyridine-Based Junctions. \emph{Nanoscale} \textbf{2013}, \emph{5},
  9222--9230\relax
\mciteBstWouldAddEndPuncttrue
\mciteSetBstMidEndSepPunct{\mcitedefaultmidpunct}
{\mcitedefaultendpunct}{\mcitedefaultseppunct}\relax
\EndOfBibitem
\bibitem[B\^aldea(2014)]{Baldea:2014c}
B\^aldea,~I. A Quantum Chemical Study from a Molecular Transport Perspective:
  Ionization and Electron Attachment Energies for Species Often Used to
  Fabricate Single-Molecule Junctions. \emph{Faraday Discuss.} \textbf{2014},
  \emph{174}, 37--56\relax
\mciteBstWouldAddEndPuncttrue
\mciteSetBstMidEndSepPunct{\mcitedefaultmidpunct}
{\mcitedefaultendpunct}{\mcitedefaultseppunct}\relax
\EndOfBibitem
\bibitem[B\^aldea(2014)]{Baldea:2014e}
B\^aldea,~I. Quantifying the Relative Molecular Orbital Alignment for Molecular
  Junctions with Similar Chemical Linkage to Electrodes. \emph{Nanotechnology}
  \textbf{2014}, \emph{25}, 455202\relax
\mciteBstWouldAddEndPuncttrue
\mciteSetBstMidEndSepPunct{\mcitedefaultmidpunct}
{\mcitedefaultendpunct}{\mcitedefaultseppunct}\relax
\EndOfBibitem
\bibitem[Xie \latin{et~al.}(2017)Xie, B\^aldea, Demissie, Smith, Wu, Haugstad,
  and Frisbie]{Baldea:2017e}
Xie,~Z.; B\^aldea,~I.; Demissie,~A.~T.; Smith,~C.~E.; Wu,~Y.; Haugstad,~G.;
  Frisbie,~C.~D. Exceptionally Small Statistical Variations in the Transport
  Properties of Metal-Molecule-Metal Junctions Composed of 80 Oligophenylene
  Dithiol Molecules. \emph{J. Am. Chem. Soc.} \textbf{2017}, \emph{139},
  5696--5699, PMID: 28394596\relax
\mciteBstWouldAddEndPuncttrue
\mciteSetBstMidEndSepPunct{\mcitedefaultmidpunct}
{\mcitedefaultendpunct}{\mcitedefaultseppunct}\relax
\EndOfBibitem
\bibitem[Hansen \latin{et~al.}(2008)Hansen, Klippenstein, Westmoreland, Kasper,
  Kohse-H\"oinghaus, Wang, and Cool]{Hansen:08}
Hansen,~N.; Klippenstein,~S.~J.; Westmoreland,~P.~R.; Kasper,~T.;
  Kohse-H\"oinghaus,~K.; Wang,~J.; Cool,~T.~A. A Combined Ab Initio and
  Photoionization Mass Spectrometric Study of Polyynes in Fuel-Rich Flames.
  \emph{Phys. Chem. Chem. Phys.} \textbf{2008}, \emph{10}, 366--374\relax
\mciteBstWouldAddEndPuncttrue
\mciteSetBstMidEndSepPunct{\mcitedefaultmidpunct}
{\mcitedefaultendpunct}{\mcitedefaultseppunct}\relax
\EndOfBibitem
\bibitem[Li \latin{et~al.}(2009)Li, Zhang, Tian, Yuan, Zhang, Yang, and
  Qi]{Li:09}
Li,~Y.; Zhang,~L.; Tian,~Z.; Yuan,~T.; Zhang,~K.; Yang,~B.; Qi,~F.
  Investigation of the Rich Premixed Laminar Acetylene/Oxygen/Argon Flame:
  Comprehensive Flame Structure and Special Concerns of Polyynes. \emph{Proc.
  Combust. Inst.} \textbf{2009}, \emph{32}, 1293 -- 1300\relax
\mciteBstWouldAddEndPuncttrue
\mciteSetBstMidEndSepPunct{\mcitedefaultmidpunct}
{\mcitedefaultendpunct}{\mcitedefaultseppunct}\relax
\EndOfBibitem
\bibitem[Ehara \latin{et~al.}(2005)Ehara, Hasegawa, and Nakatsuji]{Ehara:05}
Ehara,~M.; Hasegawa,~J.; Nakatsuji,~H. In \emph{Theory and Applications of
  Computational Chemistry}; Dykstra,~C.~E., Frenking,~G., Kim,~K.~S.,
  Scuseria,~G.~E., Eds.; Elsevier: Amsterdam, 2005; pp 1099 -- 1141\relax
\mciteBstWouldAddEndPuncttrue
\mciteSetBstMidEndSepPunct{\mcitedefaultmidpunct}
{\mcitedefaultendpunct}{\mcitedefaultseppunct}\relax
\EndOfBibitem
\bibitem[Christiansen(2006)]{Christiansen:06}
Christiansen,~O. Coupled Cluster Theory with Emphasis on Selected New
  Developments. \emph{Theor. Chem. Acc.} \textbf{2006}, \emph{116},
  106--123\relax
\mciteBstWouldAddEndPuncttrue
\mciteSetBstMidEndSepPunct{\mcitedefaultmidpunct}
{\mcitedefaultendpunct}{\mcitedefaultseppunct}\relax
\EndOfBibitem
\bibitem[White and Martin(1999)White, and Martin]{White:99}
White,~S.~R.; Martin,~R.~L. Ab Initio Quantum Chemistry Using the Density
  Matrix Renormalization Group. \emph{J. Chem. Phys.} \textbf{1999},
  \emph{110}, 4127--4130\relax
\mciteBstWouldAddEndPuncttrue
\mciteSetBstMidEndSepPunct{\mcitedefaultmidpunct}
{\mcitedefaultendpunct}{\mcitedefaultseppunct}\relax
\EndOfBibitem
\bibitem[Chan and Head-Gordon(2002)Chan, and Head-Gordon]{Chan:02}
Chan,~G. K.-L.; Head-Gordon,~M. Highly Correlated Calculations with a
  Polynomial Cost Algorithm: A Study of the Density Matrix Renormalization
  Group. \emph{J. Chem. Phys.} \textbf{2002}, \emph{116}, 4462--4476\relax
\mciteBstWouldAddEndPuncttrue
\mciteSetBstMidEndSepPunct{\mcitedefaultmidpunct}
{\mcitedefaultendpunct}{\mcitedefaultseppunct}\relax
\EndOfBibitem
\bibitem[Chan and Zgid(2009)Chan, and Zgid]{Chan:09}
Chan,~G. K.-L.; Zgid,~D. In \emph{Annu. Rep. Comput. Chem.}; Wheeler,~R.~A.,
  Ed.; Elsevier, 2009; Vol.~5; pp 149 -- 162\relax
\mciteBstWouldAddEndPuncttrue
\mciteSetBstMidEndSepPunct{\mcitedefaultmidpunct}
{\mcitedefaultendpunct}{\mcitedefaultseppunct}\relax
\EndOfBibitem
\bibitem[hc1()]{hc10n}
The linear \ce{HC10N} molecule is of interest on its own that needs a separate
  discussion \cite{Baldea:2019e}\relax
\mciteBstWouldAddEndPuncttrue
\mciteSetBstMidEndSepPunct{\mcitedefaultmidpunct}
{\mcitedefaultendpunct}{\mcitedefaultseppunct}\relax
\EndOfBibitem
\bibitem[Ball \latin{et~al.}(2000)Ball, McCarthy, and Thaddeus]{Ball:00}
Ball,~C.~D.; McCarthy,~M.~C.; Thaddeus,~P. Cavity Ringdown Spectroscopy of the
  Linear Carbon Chains \ce{HC7H}, \ce{HC9H}, \ce{HC_{11}H}, and \ce{HC_{13}H}.
  \emph{J. Chem. Phys.} \textbf{2000}, \emph{112}, 10149--10155\relax
\mciteBstWouldAddEndPuncttrue
\mciteSetBstMidEndSepPunct{\mcitedefaultmidpunct}
{\mcitedefaultendpunct}{\mcitedefaultseppunct}\relax
\EndOfBibitem
\bibitem[Chen \latin{et~al.}(2014)Chen, Steglich, Gupta, Rice, and
  Maier]{Maier:14}
Chen,~X.; Steglich,~M.; Gupta,~V.; Rice,~C.~A.; Maier,~J.~P. Gas Phase
  Electronic Spectra of Carbon Chains \ce{C_n} (n = 6--9). \emph{Phys. Chem.
  Chem. Phys.} \textbf{2014}, \emph{16}, 1161--1165\relax
\mciteBstWouldAddEndPuncttrue
\mciteSetBstMidEndSepPunct{\mcitedefaultmidpunct}
{\mcitedefaultendpunct}{\mcitedefaultseppunct}\relax
\EndOfBibitem
\bibitem[Aoki and Ikuta(1993)Aoki, and Ikuta]{Aoki:93}
Aoki,~K.; Ikuta,~S. Is a Triplet \ce{HC4N} Molecule Linear? \emph{J. Chem.
  Phys.} \textbf{1993}, \emph{98}, 7661--7662\relax
\mciteBstWouldAddEndPuncttrue
\mciteSetBstMidEndSepPunct{\mcitedefaultmidpunct}
{\mcitedefaultendpunct}{\mcitedefaultseppunct}\relax
\EndOfBibitem
\bibitem[Aoki \latin{et~al.}(1993)Aoki, Ikuta, and Murakami]{Aoki:94}
Aoki,~K.; Ikuta,~S.; Murakami,~A. Most Stable Isomer and Singlet-Triplet Energy
  Separation in the \ce{HC4N} molecule. \emph{Chem. Phys. Lett.} \textbf{1993},
  \emph{209}, 211 -- 215\relax
\mciteBstWouldAddEndPuncttrue
\mciteSetBstMidEndSepPunct{\mcitedefaultmidpunct}
{\mcitedefaultendpunct}{\mcitedefaultseppunct}\relax
\EndOfBibitem
\bibitem[Gutowski \latin{et~al.}(1996)Gutowski, Skurski, Boldyrev, Simons, and
  Jordan]{Gutowski:96}
Gutowski,~M.; Skurski,~P.; Boldyrev,~A.~I.; Simons,~J.; Jordan,~K.~D.
  Contribution of Electron Correlation to the Stability of Dipole-Bound Anionic
  States. \emph{Phys. Rev. A} \textbf{1996}, \emph{54}, 1906--1909\relax
\mciteBstWouldAddEndPuncttrue
\mciteSetBstMidEndSepPunct{\mcitedefaultmidpunct}
{\mcitedefaultendpunct}{\mcitedefaultseppunct}\relax
\EndOfBibitem
\bibitem[Kim \latin{et~al.}(1997)Kim, Lee, Nordlander, and Tom\'anek]{Kim:97}
Kim,~S.~G.; Lee,~Y.~H.; Nordlander,~P.; Tom\'anek,~D. Disintegration of Finite
  Carbon Chains in Electric Eields. \emph{Chem. Phys. Lett.} \textbf{1997},
  \emph{264}, 345 -- 350\relax
\mciteBstWouldAddEndPuncttrue
\mciteSetBstMidEndSepPunct{\mcitedefaultmidpunct}
{\mcitedefaultendpunct}{\mcitedefaultseppunct}\relax
\EndOfBibitem
\bibitem[Smith \latin{et~al.}(1994)Smith, Engel, Thoma, Schallmoser, Wurfel,
  and Bondybey]{Smith:94b}
Smith,~A.~M.; Engel,~C.; Thoma,~A.; Schallmoser,~G.; Wurfel,~B.~E.;
  Bondybey,~V.~E. Tentative Identification of \ce{C5N2} in Rare Gas Matrices.
  \emph{Chem. Phys.} \textbf{1994}, \emph{184}, 233 -- 245\relax
\mciteBstWouldAddEndPuncttrue
\mciteSetBstMidEndSepPunct{\mcitedefaultmidpunct}
{\mcitedefaultendpunct}{\mcitedefaultseppunct}\relax
\EndOfBibitem
\bibitem[Ohshima \latin{et~al.}(1995)Ohshima, Endo, and Ogata]{Ohshima:95}
Ohshima,~Y.; Endo,~Y.; Ogata,~T. Fourier-Transform Microwave Spectroscopy of
  Triplet Carbon Monoxides, C2O, C4O, C6O, and C8O. \emph{J. Chem. Phys.}
  \textbf{1995}, \emph{102}, 1493--1500\relax
\mciteBstWouldAddEndPuncttrue
\mciteSetBstMidEndSepPunct{\mcitedefaultmidpunct}
{\mcitedefaultendpunct}{\mcitedefaultseppunct}\relax
\EndOfBibitem
\bibitem[McGuire \latin{et~al.}(2018)McGuire, Martin-Drumel, Lee, Stanton,
  Gottlieb, and McCarthy]{McGuire:18b}
McGuire,~B.~A.; Martin-Drumel,~M.-A.; Lee,~K. L.~K.; Stanton,~J.~F.;
  Gottlieb,~C.~A.; McCarthy,~M.~C. Vibrational satellites of C2S, C3S, and C4S:
  Microwave Spectral Taxonomy as a Stepping Stone to the Millimeter-Wave Band.
  \emph{Phys. Chem. Chem. Phys.} \textbf{2018}, \emph{20}, 13870--13889\relax
\mciteBstWouldAddEndPuncttrue
\mciteSetBstMidEndSepPunct{\mcitedefaultmidpunct}
{\mcitedefaultendpunct}{\mcitedefaultseppunct}\relax
\EndOfBibitem
\bibitem[Vaizert \latin{et~al.}(2001)Vaizert, Motylewski, Wyss, Riaplov,
  Linnartz, and Maier]{Maier:01}
Vaizert,~O.; Motylewski,~T.; Wyss,~M.; Riaplov,~E.; Linnartz,~H.; Maier,~J.~P.
  The $A ^{3}\Sigma^{-}-X ^{3}\Sigma^{-}$ Electronic Transition of \ce{HC6N}.
  \emph{J. Chem. Phys.} \textbf{2001}, \emph{114}, 7918--7922\relax
\mciteBstWouldAddEndPuncttrue
\mciteSetBstMidEndSepPunct{\mcitedefaultmidpunct}
{\mcitedefaultendpunct}{\mcitedefaultseppunct}\relax
\EndOfBibitem
\bibitem[Travers \latin{et~al.}(1996)Travers, McCarthy, Kalmus, Gottlieb, and
  Thaddeus]{Travers:96b}
Travers,~M.~J.; McCarthy,~M.~C.; Kalmus,~P.; Gottlieb,~C.~A.; Thaddeus,~P.
  Laboratory Detection of the Linear Cyanopolyyne \ce{HC11N}. \emph{Astrophys.
  J.} \textbf{1996}, \emph{469}, L65--L68\relax
\mciteBstWouldAddEndPuncttrue
\mciteSetBstMidEndSepPunct{\mcitedefaultmidpunct}
{\mcitedefaultendpunct}{\mcitedefaultseppunct}\relax
\EndOfBibitem
\end{mcitethebibliography}
\end{document}